\documentclass{aa} 
\pdfoutput=1 

\usepackage{graphicx}
\usepackage{txfonts}
\usepackage{hyperref}
\begin{document} 

   \title{New database for a sample of optically bright lensed quasars in the northern hemisphere\thanks{Tables 
    4$-$6, 8$-$11, and 13$-$16 are only available in electronic form at the CDS via anonymous ftp to 
    cdsarc.u-strasbg.fr (130.79.128.5) or via http://cdsweb.u-strasbg.fr/cgi-bin/qcat?J/A+A/vol/page}}
   \titlerunning{New database of lensed quasars} 

   \author{R. Gil-Merino\inst{1}, L. J. Goicoechea\inst{1}, V. N. Shalyapin\inst{1,2} \and A. Oscoz\inst{3,4}}
   \authorrunning{R. Gil-Merino et al.}

   \institute{Departamento de F\'\i sica Moderna, Universidad de Cantabria (UC), 
             Avda. de Los Castros s/n, E-39005 Santander, Spain\\
             \email{r.gilmerino@gmail.com;goicol@unican.es;vshal@ukr.net}
             \and
             Institute for Radiophysics and Electronics, National Academy of 
             Sciences of Ukraine, 12 Proskura St., UA-61085 Kharkov, Ukraine
             \and
             Instituto de Astrof\'\i sica de Canarias (IAC), c/ V\'\i a L\'actea 
             s/n, E-38205 La Laguna, Spain
             \and
             Departamento de Astrof\'\i sica, Universidad de La Laguna, E-38200 
             La Laguna, Spain}


\abstract{In the framework of the Gravitational LENses and DArk MAtter (GLENDAMA) project, we present 
a database of nine gravitationally lensed quasars (GLQs) that have two or four images brighter than $r$ 
= 20 mag and are located in the northern hemisphere. This new database consists of a rich variety of 
follow-up observations included in the GLENDAMA global archive, which is publicly available online 
and contains 6557 processed astronomical frames of the nine lens systems over the period 1999$-$2016. 
In addition to the GLQs, our archive also incorporates binary quasars, accretion-dominated radio-loud 
quasars, and other objects, where about 50\% of the non-GLQs were observed as part of a campaign to 
identify GLQ candidates. Most observations of GLQs correspond to an ongoing long-term macro-programme 
with 2$-$10 m telescopes at the Roque de los Muchachos Observatory, and these data provide information 
on the distribution of dark matter at all scales. We outline some previous results from the database, 
and we additionally obtain new results for several GLQs that update the potential of the tool for 
astrophysical studies.} 

   \keywords{astronomical databases: miscellaneous --
                gravitational lensing: strong -- 
                gravitational lensing: micro --
                galaxies: general --
                quasars: general --
                cosmological parameters}

   \maketitle
%

\section{Introduction}
\label{sec:intro}

A quasar is a distant active galactic nucleus (AGN) of high luminosity powered by accretion 
into a super-massive black hole \citep[e.g.][]{Ree84}. The UV thermal emission is generated by hot 
gas orbiting the central black hole: the continuum comes from tiny sources and shows variability over 
several timescales, while broad emission lines are produced in regions around the continuum sources 
\citep[e.g.][]{Pet97,Kro99}. Only rarely is the same quasar seen
at different positions on the sky.
These positions are close together, and they are located around a massive galaxy acting as a lens. The 
gravitational field of the foreground galaxy bends the light from the background quasar and often 
produces two or four images of the distant AGN. Although a gravitationally lensed quasar (GLQ) is a 
rare phenomenon, observations of GLQs provide very valuable information about the structure of 
accretion flows, the distribution of mass in lensing galaxies, and the physical properties of the 
Universe as a whole \citep[e.g.][]{Schn92,Schn06}. 

A significant part of the UV emission of quasars at redshift $z >$ 1 is observed at 
optical wavelengths, and thus optical photometric monitoring of GLQs revealed a wide diversity of 
intrinsic flux variations. These variations were used, among other things, to determine accurate time 
delays between quasar images, which in turn led to constraints on the Hubble constant and the dark 
components of the Universe \citep[e.g.][]{Ogu07,Ser14,Wei14,Rat15,Yua15,Pan16,Bon17}, as well as on 
lensing mass distributions \citep[e.g.][]{Goi10}. Stars in lensing galaxies are also responsible for 
microlensing effects in optical light curves and spectra of GLQs, and the observed extrinsic 
variations and spectral distortions constrained the size of continuum and broad-line sources, the 
structure of emitting regions, the mass of super-massive black holes, and the composition of 
intervening galaxies \citep[e.g.][]{Sha02,Koc04,Ric04,Mor08,Slu12,Gue13,Mot17}. Deep imaging and 
spectroscopy of GLQs are also key tools to discuss the distribution of mass, dust, and gas in lensing 
objects \citep[e.g.][]{Schn06}. In addition, optical polarimetry may help to better understand the 
physical scenarios \citep[e.g.][]{Wil80,Cha01,Hut10}.

Since 1998, the Gravitational LENses and DArk MAtter (GLENDAMA) project is planning, conducting, and 
analysing (mainly) optical observations of GLQs and related objects. In the first decade of the 
current century, the advent of a robotic 2m telescope \citep{Ste04} to the Roque de los Muchachos  
Observatory (RMO) represented a revolution on the observational side of GLQs. A main advantage is the 
possibility of a rapid reaction in observations scheduling with a variety of available instruments. 
Along with the installation of the robotic telescope, the start of the scientific 
operational phase of a 10m telescope \citep{Alv06} paved the way to ambitious gravitational lensing 
programmes at the RMO. We thus focused on the construction of a comprehensive database for a sample 
of ten GLQs with bright images ($r <$ 20 mag) at $1 < z < 3$. The selected lens systems have different 
morphologies and angular separations between their images. In this paper, we introduce the current 
version of the database, including ready-to-use (processed) frames of nine targets. This astronomical 
material has been collected over 17 years, using facilities at the RMO, the Teide Observatory (TO), 
and space observatories 
\citep[{\it Swift} and {\it Chandra} monitoring campaigns of the first lensed quasar;][]{Gil12}. Our 
tenth and last target has been discovered in 2017 \citep[\object{PS J0147+4630};][]{Ber17,Lee17,Rub17}, 
and we are starting to observe this GLQ, in which three out of its four images are arranged in an 
arc-like configuration. We wish to perform an accurate follow-up of each target over 10$-$30 years, 
since observations on 10- $ \text{to }$30-year timescales are crucial to detect significant microlensing effects 
in practically all objects in the sample \citep{Mos11}. 

In addition to thousands of astronomical frames in a well-structured datastore that is publicly available 
online, the website of the GLENDAMA project offers high-level data products (light curves, calibrated 
spectra, polarisation measures, etc). We remark that the GLENDAMA observing programme does not only 
focus on imaging lens systems and light curves construction. The robotic telescope allows us to 
follow up the spectroscopic and polarimetric activity of some targets, and additionally, we obtain 
deep near-infrared (NIR) imaging with several 2$-$4m telescopes. Here, we present new results for six of the nine 
targets. Results for the other three lens systems have been published very recently. Despite 
of the existence of high-resolution spectra of some images of GLQs in the Sloan Digital Sky Survey 
(SDSS) database \citep[the SDSS spectroscopic database includes observations of the Baryon 
Oscillation Spectroscopic Survey $-$ BOSS;][]{Sme13}, we also conduct a programme with the very large 
telescope at the RMO to acquire spectra of unprecedented signal quality \citep[e.g.][]{Goi16,Sha17}. 
 
The paper is organised as follows: in Sect.~\ref{sec:obs}, we present an overview of the global 
archive and then describe the GLQ observations in detail. In Sect.~\ref{sec:results}, we review 
relevant intermediate results and discuss their astrophysical impact. New light curves, polarisations, 
and spectra at optical wavelengths (and deep NIR imaging of QSO B0957+561) are also presented and placed 
into perspective in Sect.~\ref{sec:results}. The summary and future prospects appear in 
Sect.~\ref{sec:final}.

\section{GLQ database in the GLENDAMA archive}
\label{sec:obs}

\subsection{Overview of the archive}
\label{sec:archive}

\begin{figure*}
\centering
\includegraphics[width=9cm]{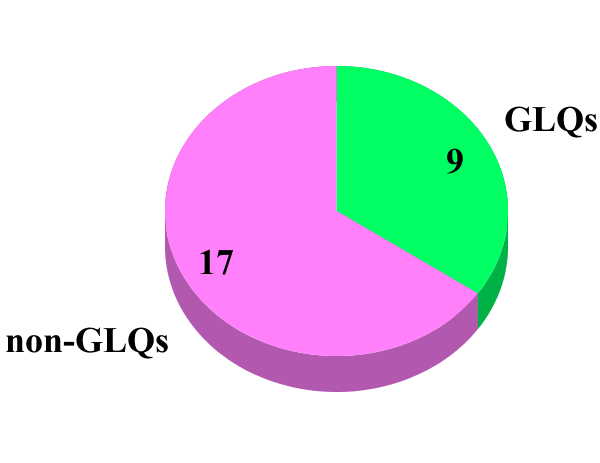}
\includegraphics[width=9cm]{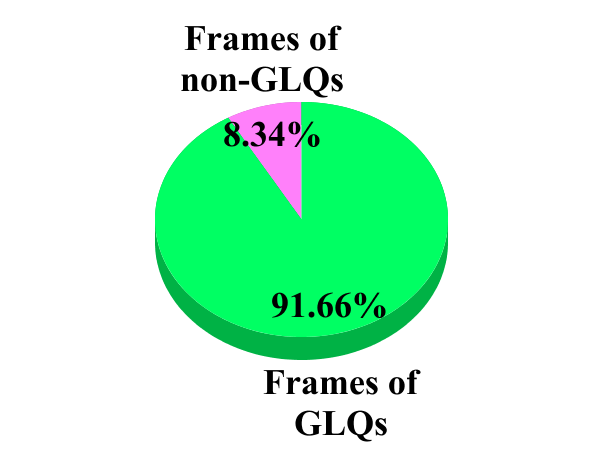}
\caption{Distribution of targets and frames in the GLENDAMA archive.}   
\label{fig:tarfra}
\end{figure*}

The global archive consists of a datastore of 40 GB in size, whose content is organised and 
visualised by using MySQL/PHP/JavaScript/HTML5 software\footnote{MySQL is a database management 
system that is developed, distributed and supported by Oracle Corporation. This software is available 
at \url{http://www.mysql.com/}. PHP is a general-purpose scripting language that is especially suited 
to web development, and is available at \url{http://php.net/}. JavaScript is an object-oriented 
computer programming language commonly used to create interactive effects within web browsers, 
developed by Mozilla Foundation at \url{https://developer.mozilla.org/en-US/docs/Web/JavaScript}. 
HTML5 
is the fifth version of the standard HTML markup language used for structuring and presenting web 
content, developed by the Word Wide Web Consortium at \url{https://www.w3.org/}}. A web user 
interface\footnote{\url{http://grupos.unican.es/glendama/database/}} (WUI) allows users to surf the 
archive, see all its content, and freely download any dataset. This interface is a three-step tool, where 
the first step is to select an object and then click the submit button to see the datasets 
available for the selected target. In this second screen, it is possible to select a dataset and 
press the retrieve button to view its details (telescope, instrument, file names, observation dates, 
exposure times, etc). In the third step of the WUI, the user can download the frames of 
interest\footnote{The size limit for each download (zip file), if any, is specified on the screen.}.
The GLENDAMA datastore incorporates more than 7000 ready-to-use astronomical 
frames of 26 targets falling into two classes: GLQs, and non-GLQs (binary quasars, accretion-dominated 
radio-loud quasars, and others). In spite of this, our observational effort was mainly concentrated on 
the construction 
of a GLQ database (see Fig.~\ref{fig:tarfra}). The full sample of GLQs and the bulk of data are 
described in detail in Sec.~\ref{sec:db}.

\begin{table*}
\centering
\caption{NUV-visible-NIR facilities.}
\begin{tabular}{lccc}
\hline\hline
Observatory & Telescope & Instrument & Observing modes \\
\hline
RMO & Gran Telescopio CANARIAS (GTC)\tablefootmark{a}     
                                        & OSIRIS      & LSS: R500B, R300R and R500R grisms\\
    & Isaac Newton Telescope (INT)\tablefootmark{b}                                 
                              & IDS         & LSS: R300V grating\\
    & Liverpool Telescope (LT)\tablefootmark{c}                            
                              & RATCam      & IMA: Sloan $griz$ filters\\
    &                         & IO:O      & IMA: Sloan $gri$ filters\\
    &                         & FRODOSpec   & IFS: blue and red gratings\\
    &                         & SPRAT       & LSS: BR grating configurations\\
    &                         & RINGO2      & POL: EMCCD with V+R filter\\
    &                         & RINGO3      & POL: BGR EMCCDs\\
    & Nordic Optical Telescope (NOT)\tablefootmark{d}                      
                              & StanCam     & IMA: Bessell $VR$ filters\\
    &                         & ALFOSC      & IMA: Bessel $R$ (\#76) and interference $i$ (\#12) filters\\
    &                         & ALFOSC      & LSS: grisms \#7, \#14 and \#18\\
    & Telescopio Nazionale Galileo (TNG)\tablefootmark{e} 
                              & NICS        & IMA: $JHK$ filters\\
    &                         & DOLORES     & LSS: LR-B grism\\
    & William Herschel Telescope (WHT)\tablefootmark{f}                 
                              & ISIS        & LSS: R300B and R316R gratings\\
TO  & IAC80 Telescope (IAC80)\tablefootmark{g}            
                              & optical CCD & IMA: Johnson-Bessell $BVRI$ filters\\
    & STELLA 1 Telescope (STELLA)\tablefootmark{h}                         
                              & WiFSIP      & IMA: Johnson-Cousins $UBV$ and Sloan $gr$ filters\\
{\it Swift} & UV and Optical Telescope (UVOT)\tablefootmark{i}                
                              & MIC         & IMA: $U$ filter\\
\hline
\end{tabular}
\tablefoot{
In the column for the observing modes, we use the acronyms imaging (IMA), integral-field 
spectroscopy (IFS), long-slit spectroscopy (LSS), and polarimetry (POL). The websites of the 
telescopes are\\
\tablefoottext{a}{\url{http://www.gtc.iac.es/}}\\
\tablefoottext{b}{\url{http://www.ing.iac.es/astronomy/telescopes/int/}}\\
\tablefoottext{c}{\url{http://telescope.livjm.ac.uk/}}\\
\tablefoottext{d}{\url{http://www.not.iac.es/}}\\
\tablefoottext{e}{\url{http://www.tng.iac.es/}}\\
\tablefoottext{f}{\url{http://www.ing.iac.es/astronomy/telescopes/wht/}}\\
\tablefoottext{g}{\url{http://www.iac.es/OOCC/iac-managed-telescopes/iac80/}}\\
\tablefoottext{h}{\url{http://www.aip.de/en/research/facilities/stella/instruments/}}\\
\tablefoottext{i}{\url{https://www.swift.psu.edu/uvot/}}
}
\label{tab:facil}
\end{table*}

The datastore includes many frames of non-GLQs. There are optical frames of four 
accretion-dominated radio-loud quasars in the sample of \citet{Lan08}: \object{RX J0254.6+3931}, 
\object{RX J2256.5+2618}, \object{RX J2318.5+3048,} and \object{1WGA J2347.6+0852}. Deep NIR imaging, 
optical spectroscopy, and a short-term $r$-band monitoring of the binary quasar \object{SDSS 
J1116+4118} \citep{Hen06,Hen10} are also available. This target and another two binary systems, 
\object{1WGA J1334.7+3757} and \object{QSO B2354+1840} \citep{Bor96,McH98,Zhd01}, were observed to 
discuss the physical scenario for widely separated pairs of quasars at similar redshift. In addition, 
we observed several systems that initially were selected as double-image quasar candidates through 
searches in the SDSS-III data releases \citep{Ahn12,Par12,Ahn14,Par14}: \object{SDSS J0240$-$0208} 
(quasar-quasar pair), \object{SDSS J0734+2733} (quasar-? pair)\footnote{This is not a GLQ, although 
one of the two sources is not yet completely identified.}, \object{SDSS J0735+2036} (quasar-star
pair), \object{SDSS J0755+1400} (quasar-? pair)$^4$, \object{SDSS J1617+3827} (GLQ?)\footnote{Although 
low-noise spectra (obtained when we were writing this article) confirm that the system is a true GLQ 
(Shalyapin et al. 2018, in preparation), it is considered as a doubtful object in this paper and 
included in the column "Others", which appears in the first step of the WUI.}, \object{SDSS J1642+3200} 
(quasar-AGN pair), \object{SDSS J1655+1948} (quasar-star pair), and \object{SDSS J2153+2732} (binary 
quasar). \object{PS1 J2241+4734} (star-galaxy pair) and \object{M87} also belong to the non-GLQ 
class. This last object is a well-known radio galaxy whose optical images can be used to analyse the 
isophotes and the jet emerging from its active nucleus.

\begin{table*}
\centering
\caption{Objects in the GLQ database.}
\begin{tabular}{lccccccccc}
\hline\hline
Object & $z$\tablefootmark{a} & N$_{\rm ima}$\tablefootmark{b} & $\Delta \theta$\tablefootmark{c} (\arcsec) &
$r$\tablefootmark{d} (mag) & Ref & $\Delta t$\tablefootmark{e} (d) & Ref & 
Lensing galaxy\tablefootmark{f} & Ref \\
\hline
\object{QSO B0909+532} & 1.38 & 2 & 1.1 & 16$-$17 & 1   & 50 $^{+2}_{-4}$ & 2 & E ($z$ = 0.83) & 3, 4, 5 \\
\object{FBQS J0951+2635} & 1.24 & 2 & 1.1 & 17$-$18 & 6 & 16 $\pm$ 2 & 7 & E ($z$ = 0.26) & 8, 9 \\
\object{QSO B0957+561} & 1.41 & 2 & 6.1 & 17 & 10, 11 & 416.5 $\pm$ 1.0\tablefootmark{$\star$} & 12 & 
E-cD ($z$ = 0.36) & 13, 14, 15 \\
\object{SDSS J1001+5027} & 1.84 & 2 & 2.9 & 17.5$-$18 & 16 & 119.3 $\pm$ 3.3 & 17 & E ($z$ = 0.41) & 18 \\
\object{SDSS J1339+1310} & 2.24 & 2 & 1.7 & 19 & 19 & 47 $^{+5}_{-6}$ & 20 & E ($z$ = 0.61) & 21 \\
\object{QSO B1413+117} & 2.55 & 4 & 1.4 & $\sim$ 18 & 22 & 23 $\pm$ 4 & 23 & 
? ($z$ = 1.88)\tablefootmark{$\star\star$} & 23 \\
\object{SDSS J1442+4055} & 2.57 & 2 & 2.1 & 18$-$19 & 24, 25 & 25 $^{+1}_{-2}$ & 26 & E ($z$ = 0.28) & 26 \\
\object{SDSS J1515+1511} & 2.05 & 2 & 2.0 & 18$-$19 & 27 & 211 $\pm$ 5 & 28 & S ($z$ = 0.74) & 27, 28, 29 \\
\object{QSO B2237+0305} & 1.69 & 4 & 1.8 & 17.5$-$18.5 & 30, 31 & 1.5 $\pm$ 2.0 & 32 & SBb ($z$ = 0.04) & 30, 31 \\
\hline
\end{tabular}
\tablefoot{
\tablefoottext{a}{Source redshift};
\tablefoottext{b}{number of quasar images};
\tablefoottext{c}{angular separation between images for double quasars or typical angular size for quadruple quasars};
\tablefoottext{d}{$r$-band magnitudes of quasar images (these values should be interpreted with caution, since we deal 
with variable objects)};
\tablefoottext{e}{measured time delay for double quasars or the longest of the measured delays for quadruple quasars 
(1$\sigma$ confidence interval)};
\tablefoottext{f}{classification (redshift)}. \\
\tablefoottext{$\star$}{Time delay in the $g$ band. There is evidence of chromaticity in the optical delay, since 
it is 420.6 $\pm$ 1.9 d in the $r$ band.} \\
\tablefoottext{$\star\star$}{Redshift from gravitational lensing data and a concordance cosmology. This measurement 
is in reasonable agreement with the photometric redshift of the secondary lensing galaxy and the most distant 
overdensity, as well as the redshift of one of the absorption systems}
}
\tablebib{
(1) \citet{Koc97}; (2) \citet{Hai13} \citep[see also][]{Goi08}; (3) \citet{Osc97}; (4) \citet{Leh00}; 
(5) \citet{Lub00}; (6) \citet{Sch98}; (7) \citet{Jak05}; (8) \citet{Koc00}; (9) \citet{Eig07};
(10) \citet{Wal79}; (11) \citet{Wey79}; (12) \citet{Sha12} \citep[see also][]{Kun97,Sha08}; 
(13) \citet{Sto80}; (14) \citet{You80}; (15) \citet{You81a}; (16) \citet{Ogu05}; (17) \citet{Rat13}; 
(18) \citet{Ina12}; (19) \citet{Ina09}; (20) \citet{Goi16}; (21) \citet{Sha14a}; (22) \citet{Mag88}; 
(23) \citet{Goi10} \citep[see also][]{Akh17}; (24) \citet{Ser16}; (25) \citet{Mor16};  
(26) Shalyapin \& Goicoechea (2018) (in preparation); (27) \citet{Ina14}; (28) \citet{Sha17}; (29) 
\citet{Rus16}; (30) \citet{Huc85}; (31) \citet{Yee88}; (32) \citet{Vak06}.
}
\label{tab:glqs}
\end{table*}

The GLENDAMA database covers the period 1999$-$2016 (it was updated on 1 October 2016), and we have 
used many telescopes and a varied instrumentation throughout the past 17 
years. In addition to an X-ray monitoring campaign of a lensed quasar in 2010 (see Sect.~\ref{sec:db}), 
the archive incorporates frames (imaging, polarimetry, and spectroscopy) that were taken with 
facilities operating in the near-ultraviolet (NUV)-visible-NIR spectral region. Such facilities and some 
additional details (filters, grisms, gratings, etc) are given in Table~\ref{tab:facil}. Users can 
also access information about air mass and seeing values (when available in file headers). Seeing 
values are 
not equally accurate through all the observations: for some instruments (e.g. RATCam, IO:O, RINGO2, 
and RINGO3), the full-width at half-maximum ($FWHM$) of the seeing disc is directly estimated from 
frames, and thus is a reliable reference. However, $FWHM$ values in FRODOSpec and SPRAT files are 
estimated before spectroscopic exposures, so these foreseen values may appreciably differ from true 
values. For spectroscopic observations, we offer frames of the science target and a calibration star. 
These files for the main target and the star have labels including the expressions 'obj' and 'std', 
respectively. 

\subsection{Sample of GLQs}
\label{sec:db}

We focused on nine GLQs in the northern hemisphere (see Table~\ref{tab:glqs}). Every GLQ in our sample 
has two or four images with $r <$ 20 mag. The source redshifts vary between 1.24 and 
2.57 ($\langle z \rangle \sim$ 1.9), and the sample includes five relatively compact systems and four 
wide separation double quasars. These last GLQs have two images separated by $\Delta \theta \geq$ 
2\arcsec. Over the first ten years of our follow-up observations (1999$-$2008), the target selections 
were based on the GLQs known in 1999. From 2009 on, we have also studied SDSS GLQs. Thus, we 
try to achieve a deeper knowledge of some classical targets, such as \object{QSO B0909+532}, \object{FBQS 
J0951+2635}, \object{QSO B0957+561}, \object{QSO B1413+117,} and \object{QSO B2237+0305}, and 
simultaneously, characterise other recently discovered systems (see Tables~\ref{tab:glqs} and 
\ref{tab:data}). We have also been involved in a search for new double quasars in the SDSS-III 
database with the purpose of "going the whole way": discovery, and subsequent characterisation 
\citep{Ser16}. After selecting three superb candidates through Maidanak Astronomical Observatory 
(MAO) deep imaging under good seeing conditions (i.e. quasar-companion pairs showing evidence 
for the existence of a near lensing galaxy, as well as parallel flux variations on a long timescale), 
\citet{Ser16} corfirmed the GLQ nature of \object{SDSS J1442+4055} \citep[see also][]{Mor16}. This 
object is being intensively observed to unveil its physical properties. Follow-up observations of the 
second superb candidate (\object{SDSS J1617+3827}) also led to the discovery of a faint double quasar 
(Shalyapin et al. 2018, in preparation). However, in this paper, the newly discovered GLQ is treated 
as an unconfirmed GLQ and incorporated into the non-GLQ class (see Sect.~\ref{sec:archive}). The third 
superb candidate (\object{SDSS J1642+3200}) turned out to be a system consisting of a quasar and a 
different AGN. 

\begin{table*}
\centering
\caption{GLQ frames in the NUV-visible-NIR spectral region.}
\begin{tabular}{lcccccc}
\hline\hline
Obs. Period & \multicolumn{4}{c}{Obs. Mode\tablefootmark{a}} & N$_{\rm frames}$ & Programme\\
\cline{2-5}
 & IMA & IFS & LSS & POL & & \\
\hline
\multicolumn{7}{c}{\object{QSO B0909+532}}\\
\hline 
2005$-$2007  & RATCam/$gr$                       &    &                &              &  451 & XCL04BL2\tablefootmark{b}\\
2010$-$2012  & RATCam/$r$                        &    &                &              &  119 & XCL04BL2\\
2012$-$2016  & IO:O/$gri$\tablefootmark{c}       &    &                &              &  345 & XCL04BL2\\
\hline
\multicolumn{7}{c}{\object{FBQS J0951+2635}}\\
\hline 
2007 Feb-May & RATCam/$i$                        &    &                &              &  259 & XCL04BL2\\
2009$-$2012  & RATCam/$r$                  	 &    &                &              &   29 & XCL04BL2\\
2013$-$2016  & IO:O/$r$                    	 &    &                &              &   43 & XCL04BL2\\
\hline
\multicolumn{7}{c}{\object{QSO B0957+561}\tablefootmark{d}}\\
\hline 
1999$-$2005  & IAC80-CCD/$BVRI$\tablefootmark{e} &    &                &              & 1108 & IAC-GLM\tablefootmark{f}\\
2000 Feb-Mar & StanCam/$VR$                      &    &                &              &   77 & GLITP\tablefootmark{g}\\
2005$-$2014  & RATCam/$griz$\tablefootmark{h}    &    &                &              & 1311 & XCL04BL2\\
2007 Dec     & NICS/$JHK$\tablefootmark{i}       &    &                &              &    3 & A16CAT128\\
2008 Mar     &                                   &    & IDS/R300V      &              &    2 & SST\tablefootmark{j}\\
2009$-$2013  &                                   &    & ALFOSC/\#7\#14 &              &    8 & SST\&NOT-SP\\
2010 Jan-Jun & MIC/$U$                           &    &                &              &   35 & TOO\#31567\\ 
2010$-$2014  &                                   & FRODOSpec/BR\tablefootmark{k} &  & &  122 & XCL04BL2\\
2011$-$2012  & & & &                                                       RINGO2/V+R &   32 & XCL04BL2\\
2012$-$2016  & IO:O/$gr$                         &    &                &              &  190 & XCL04BL2\\
2013$-$2016  & & & &                                                       RINGO3/BGR &  360 & XCL04BL2\\
2015         &                                   &    & SPRAT/R        &              &   13 & XCL04BL2\\
\hline
\multicolumn{7}{c}{\object{SDSS J1001+5027}}\\
\hline 
2010 Feb-May & RATCam/$g$                        &    &                &              &   46 & XCL04BL2\\
2013$-$2014  & IO:O/$gr$\tablefootmark{l}        & FRODOSpec/BR                  & &  &   50 & XCL04BL2\\
2014$-$2015  & & & &                                                       RINGO3/BGR &  120 & XCL04BL2\\
2015$-$2016  &                                   &    & SPRAT/B        &              &   11 & XCL04BL2\\
\hline
\multicolumn{7}{c}{\object{SDSS J1339+1310}}\\
\hline 
2009/2012    & RATCam/$r$                        &    &                &              &  293 & XCL04BL2\\
2010 Jun-Jul & RATCam/$i$                        &    &                &              &   20 & XCL04BL2\\
             & ALFOSC/$I$\tablefootmark{i}       &    &                &              &    1 & SST\\
2013 Apr     &                                   &    & OSIRIS/R500R   &              &    1 & GTC30$-$13A\\
2013$-$2016  & IO:O/$r$                          &    &                &              &  198 & XCL04BL2\\
2013$-$2014  & & & &                                                       RINGO3/BGR &   72 & XCL04BL2\\
2014 Mar/May &                                   &    & OSIRIS/R500BR  &              &    6 & GTC82$-$14A\\
\hline
\multicolumn{7}{c}{\object{QSO B1413+117}}\\
\hline 
2008 Feb-Jul & RATCam/$r$                        &    &                &              &   61 & XCL04BL2\\
2011 Mar/Jun &                                   &    & OSIRIS/R300R   &              &    6 & GTC35$-$11A\\ 
2013$-$2016  & IO:O/$r$                          &    &                &              &  125 & XCL04BL2\\
\hline
\multicolumn{7}{c}{\object{SDSS J1442+4055}}\\
\hline 
2015$-$2016  &                                   &    & SPRAT/BR       &              &   10 & XCL04BL2\\
2015$-$2016  & IO:O/$r$\tablefootmark{m}         &    &                &              &  144 & XCL04BL2\\
2016 Mar     &                                   &    & OSIRIS/R500BR  &              &    6 & GTC41$-$16A\\
\hline
\multicolumn{7}{c}{\object{SDSS J1515+1511}}\\
\hline 
2014$-$2016  & IO:O/$r$                          &    &                &              &  315 & XCL04BL2\\
2015 Apr     &                                   &    & OSIRIS/R500BR  &              &    4 & GTC29$-$15A\\
2015$-$2016  &                                   &    & SPRAT/BR       &              &   20 & XCL04BL2\\
\hline
\multicolumn{7}{c}{\object{QSO B2237+0305}}\\
\hline 
2007$-$2009  & RATCam/$gr$\tablefootmark{n}      &    &                &              &  174 & XCL04BL2\\
2013 Jun-Dec &                                   & FRODOSpec/BR        & & RINGO3/BGR &  204 & XCL04BL2\\
2013$-$2016  & IO:O/$gr$                         &    &                &              &  151 & XCL04BL2\\
\hline
\end{tabular}
\tablefoot{
\tablefoottext{a}{See Table~\ref{tab:facil} for details};
\tablefoottext{b}{this programme is mainly focused on a long-term optical monitoring of GLQs with the 
LT};
\tablefoottext{c}{no $i$-band data in 2014$-$2016};
\tablefoottext{d}{in addition to NUV-visible-NIR data, 12 X-ray (0.1$-$10 keV) frames were obtained as 
part of a monitoring campaign with {\it Chandra}/ACIS-S3 during the first semester of 2010 
(Programme: DDT\#10708333)};
\tablefoottext{e}{poorer sampling in the $BI$ bands};
\tablefoottext{f}{IAC Gravitational Lenses Monitoring};
\tablefoottext{g}{Gravitational Lenses International Time Project};
\tablefoottext{h}{only a few frames in 2013$-$2014. All $iz$-band frames were taken in 2010 and early 
2011};
\tablefoottext{i}{combined frames (deep imaging observations)};
\tablefoottext{j}{Spanish Service Time at the RMO};
\tablefoottext{k}{poorer sampling in 2010, 2013 and 2014};
\tablefoottext{l}{$\sim$ 90\% of frames in the $r$ band};
\tablefoottext{m}{only two frames in 2015};
\tablefoottext{n}{no data in 2007$-$2008 in the $g$-band.}
}
\label{tab:data}
\end{table*}

\begin{figure}
\centering
\includegraphics[width=9cm]{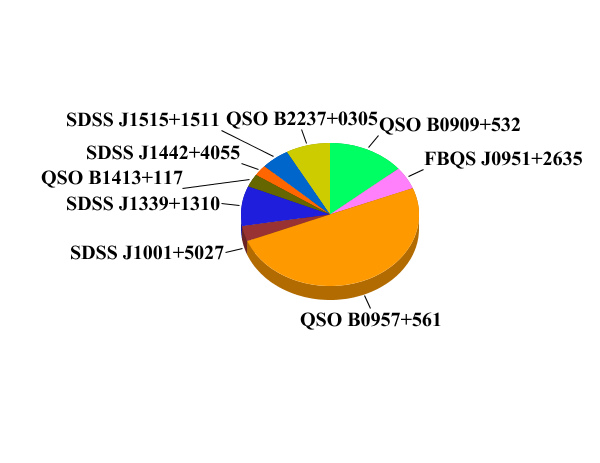}
\caption{Distribution of GLQ frames.}   
\label{fig:glqfra}
\end{figure}

After normal LT and GTC science operations at the RMO, most GLQ observations were carried out with 
these two telescopes. In a parallel effort, processing tools for some LT instruments (photometric 
pipelines for CCD imagers and specific software for FRODOSpec) and LT-GTC science products (several 
light curves and calibrated spectra) were made available to the community. However, the full set of 
NUV-visible-NIR frames of GLQs comes from a variety of observing programmes with the GTC, IAC80, INT, 
LT, NOT, TNG, and UVOT (see Table~\ref{tab:data}). In 2010, we also performed space-based observations 
of \object{QSO B0957+561} with the {\it Chandra} X-ray 
Observatory\footnote{\url{http://chandra.harvard.edu/}}. In this X-ray (0.1$-$10 keV) monitoring 
campaign, we used the Advanced CCD Imaging Spectrometer-S3 chip. The GLQ database consists of a total 
of 6557 processed frames, which are not homogeneously distributed among the nine objects (see 
Fig.~\ref{fig:glqfra}) for different reasons. For example, we used the first lensed quasar as a pilot 
target to check the performance of the majority of the instruments involved in the project, and therefore 
50\% of the GLQ frames correspond to \object{QSO B0957+561}. In general, dates of discovery, 
scientific aims, technical constraints, opportunities arising at some time periods, and decisions of 
time allocation committees are key factors to explain the pie chart in Fig.~\ref{fig:glqfra}. 
Unfortunately, the EOCA monitoring campaign of \object{QSO B0909+532} 
\citep{Ull06},
QuOC-Around-The-Clock observations of \object{QSO B0957+561} \citep{Col03} and the GLITP optical 
monitoring of \object{QSO B2237+0305} \citep{Alc02} could not be assembled in our disc-based storage for 
technical reasons.

All frames in Table~\ref{tab:data} are ready to use because they were processed with standard 
techniques (sometimes as part of specific pipelines) or more sophisticated reduction procedures. For 
example, the LT website (see notes to Table~\ref{tab:facil}) presents pipelines for RATCam, IO:O, 
FRODOSpec, SPRAT, RINGO2, and RINGO3, which contain basic instrumental reductions. We offer outputs 
from the LT pipelines for RATCam, IO:O, RINGO2, and RINGO3 without any extra processing. Thus, 
potential users should 
carefully consider whether supplementary tasks are required, for instance, cosmic ray cleaning, bad pixel 
mask, or defringing. We do not offer outputs from the standard L2 pipeline for the 2D spectrograph 
FRODOSpec (12$\times$12 square lenslets bonded to 144 optical fibres), but multi-extension FITS 
files, each consisting of four extensions: [1] $\equiv$ [L1] (output from the CCD processing pipeline 
L1, which performs bias subtraction, overscan trimming, and CCD flat fielding), [2] $\equiv$ [RSS] 
(144 row-stacked wavelength-calibrated spectra from the non-standard L2LENS 
software\footnote{\url{http://grupos.unican.es/glendama/LQLM_tools.htm}}), [3] $\equiv$ [CUBE] 
(spectral data cube giving the 2D flux in the 12$\times$12 spatial array at each wavelength pixel), 
and [4] $\equiv$ [COLCUBE] (datacube collapsed over its entire wavelength range). The main 
differences between the standard L2 pipeline \citep{Bar12} and the L2LENS reduction tool are 
described in \citet{Sha14b}. We remark that the LT is a unique facility for photometric, polarimetric, 
and spectroscopic monitoring campaigns of GLQs. However, taking into account the spatial resolution 
(pixel size of $\sim$ 0\farcs4$-$0\farcs5) of RINGO2, RINGO3, and SPRAT, we are currently tracking the 
evolution of broad-band fluxes for almost all systems, whereas we are only obtaining spectroscopic 
and/or polarimetric data of the wide separation double quasars: QSO B0957+561 (LSS \& POL), SDSS 
J1001+5027 (LSS), SDSS J1442+4055 (LSS), and SDSS J1515+1511 (LSS).  

The long-slit spectroscopy (SPRAT, OSIRIS, ALFOSC, and IDS) was processed using standard methods of 
bias subtraction, trimming, flat-fielding, cosmic-ray rejection, sky subtraction, and wavelength 
calibration. The reduction steps of the StanCam frames included bias subtraction and flat-fielding 
using sky flats, while the combined frames for deep-imaging observations with ALFOSC were obtained in 
a standard way. We also applied standard reduction procedures to the IAC80 original data, although 
only the final $VR$ datasets contain WCS information in the FITS headers. These most relevant data 
were also corrected for cosmic-ray hits on the CCD. The NICS frames were processed with the SNAP 
software\footnote{\url{http://www.arcetri.astro.it/~filippo/snap/}}, and different types of 
instrumental reductions are applied to {\it Swift} and {\it Chandra} observations before data are 
made available to users. These space-based observatories perform specific processing tasks that are 
outlined in dedicated websites. 

\section{Results from the GLQ database}
\label{sec:results}

\subsection{Previous results}
\label{sec:oldres}

Through frames in the database, as well as through those that could not be incorporated into the datastore 
for technical reasons (see Sect.~\ref{sec:db}), we obtained light curves and spectra that led to many
astrophysical outcomes. Most of these are grouped into Sect.~\ref{sec:accretion}, 
Sect.~\ref{sec:lensmass}, Sect.~\ref{sec:dustgas}, and Sect.~\ref{sec:cosmology}.       

\subsubsection{Quasar accretion flows}
\label{sec:accretion}

The RATCam/$r$ light curves of \object{QSO B0909+532} in 
2005$-$2006 indicated that symmetric triangular flares in an accretion disc (AD) is the best scenario 
(of those tested by us) to account for the variability of the MUV ($\sim$ 2600 \AA) continuum 
emission from the quasar \citep{Goi08}. In addition, combining RATCam/$gr$ and United States Naval 
Observatory (USNO)/$r$ data of \object{QSO B0909+532} over the period 2005$-$2012, \citet{Hai13} 
found prominent microlensing events and constrained the size of the MUV continuum source in the AD, 
deriving a typical half-light radius of $r_{1/2} \sim$ 20$-$50 Schwarzschild 
radii\footnote{Assuming 
a disc inclination of 60\degr \ and a central black hole with a mass ranging from 10$^{8.51} 
M_{\odot}$ (from the C\,{\sc iv} emission) to 10$^{8.9} M_{\odot}$ \citep[from the H$\beta$ and 
C\,{\sc iv} emissions;][]{Ass11}}. Regarding \object{QSO B0957+561}, old IAC80/$R$ data, the 
RATCam/$r$ brightness records spanning 2005$-$2007, and the USNO/$r$ dataset in 2008$-$2011 
were used by \citet{Hai12} to detect a microlensing event and measure the size of the continuum 
source emitting at $\sim$ 2600 \AA\ \citep[see, however,][]{Sha12}. Their 1$\sigma$ interval for the 
size ($r_{1/2}$) of this MUV source was $10^{16}-10^{17}$ cm (inclination of 60\degr). There is also 
strong evidence supporting the presence of a centrally irradiated AD in the heart of \object{QSO 
B0957+561}. A {\it Chandra}-UVOT-LT monitoring campaign from late 2009 to mid-2010 suggested that a 
central EUV source drives the variability of the first GLQ, so EUV flares originated in the immediate 
vicinity of the black hole are thermally reprocessed in the AD at 20$-$30 Schwarzschild radii from 
the dark object\footnote{We consider a black hole with a mass of 10$^{9.4} M_{\odot}$ \citep[average 
of estimates through the C\,{\sc iv} and Mg\,{\sc ii} emission lines;][]{Pen06} instead of 10$^{9} 
M_{\odot}$ \citep[C\,{\sc iv} emission-line estimate of][]{Ass11}} \citep{Gil12,Goi12}. Interpreting 
the reverberation-based size of the 2600 \AA\ source as a flux-weighted emitting radius  
\citep[e.g.][]{Fau16}, we obtained $r_{1/2}$ = (1.1 $\pm$ 0.2) $\times$ 10$^{16}$ cm (1$\sigma$ 
interval), and thus the source size from the microlensing analysis of \citet{Hai12} is marginally
consistent with this measurement. Our accurate value of $r_{1/2}$ is in good agreement with the 
overlapping region between the Hainline et al. interval and the microlensing-based constraint 
obtained by \citet{Ref00}.        

RATCam-IO:O light curves and OSIRIS spectra of \object{SDSS J1339+1310} indicated that 
this system is likely the main microlensing factory discovered so far \citep{Sha14a,Goi16}. Thus, data 
of \object{SDSS J1339+1310} are very promising tools to reveal fine details of the structure of its 
accretion flow. In particular, we have shown how microlensing magnification ratios of the continuum 
can be used to check the structure of the AD, and we have reported some physical properties of broad 
line emitting regions: the Fe\,{\sc iii} region is more compact than the Fe\,{\sc ii} region, while the 
C\,{\sc iv} region has an anisotropic structure and a size probably not much larger than the AD. 
There is also clear evidence that high-ionisation regions have smaller sizes than low-ionisation 
regions. This was found using high-ionisation emission lines in OSIRIS spectra of \object{SDSS 
J1339+1310} \citep[Si\,{\sc iv}/O\,{\sc iv} and C\,{\sc iv};][]{Goi16} and \object{SDSS J1515+1511} 
\citep[C\,{\sc iv} and He\,{\sc ii};][]{Sha17}, in good agreement with the results in \citet{Gue13} 
from a sample of 16 GLQs. We also showed that the GLITP light curve of a microlensing 
high-magnification event in the A image of \object{QSO B2237+0305} \citep{Alc02}, alone or in 
conjunction with data from the OGLE collaboration \citep{Woz00}, can be used to prove the structure 
of the inner accretion flow in the distant quasar \citep{Sha02,Goi03,Gil06}. This accurate 
microlensing curve (from October 1999 to early February 2000) has been discussed by several other 
groups \citep[e.g.][]{Koc04,Bog04,Vak04,Mor05,Uda06,Kop07,Ale11,Abo12,Med15}.
 
\subsubsection{Lensing mass distributions}
\label{sec:lensmass}

Deep $I$-band imaging (ALFOSC) and spectroscopic (OSIRIS) 
observations of \object{SDSS J1339+1310} allowed us to reliably reconstruct the mass distribution  
acting as a strong gravitational lens \citep{Sha14a}. Using a singular isothermal ellipsoid (SIE) to
model the mass of the main lensing galaxy \citep[e.g.][]{Koo06}, we obtained an offset between light
and mass position angles. This misalignment suggests that \object{SDSS J1339+1310} is 
affected by external shear $\gamma$ arising from the environment of the main lens (early-type 
galaxy) at $z$ = 0.61 \citep[e.g.][]{Gav12}. We then considered an SIE + $\gamma$ mass model, where 
the SIE was aligned with the light distribution of the main lens. Although the uncertainty in the SIE 
mass scale was below 10\%, new observational constraints on the macrolens flux ratio and the time 
delay \citep[e.g.][]{Goi16} must produce a much more accurate SIE + $\gamma$ solution. A 
cross-correlation analysis of the RATCam light curves of the four images of \object{QSO B1413+117} 
yielded three independent delays, which were also used to improve the lens solution for this system 
and to estimate the previously unknown lens redshift \citep{Goi10}. The mass model consisted of an SIE 
(main lensing galaxy), a singular isothermal sphere (secondary lensing galaxy), and external shear 
\citep{Mac09}, and we derived a lens redshift of $z$ = 1.88$^{+0.09}_{-0.11}$ \citep[1$\sigma$ 
interval; see also][]{Akh17}. Additionally, from OSIRIS spectroscopy of field objects in the external 
shear direction, we identified an emission line galaxy at $z \sim$ 0.57 that is responsible for $<$ 
2\% of $\gamma \sim$ 0.1 \citep{Sha13}.  

Very recently, IO:O light curves and OSIRIS-SPRAT spectra of \object{SDSS J1515+1511} 
have been used to obtain strong constraints on its time delay and its macrolens flux ratio 
\citep{Sha17}. \citet{Ina14} tentatively associated the main lensing galaxy with an Fe/Mg absorption 
system at $z$ = 0.74 (intervening gas), therefore we have assumed the existence of intervening dust at this 
redshift to measure the macrolens flux ratio. Our observational constraints practically did not 
modify the previous SIE + $\gamma$ solution \citep{Rus16}. Moreover, the redshift of the lensing mass 
was found to be consistent with $z$ = 0.74, which confirmed the putative value of $z$ for the main 
lens (edge-on disc-like galaxy). From the OSIRIS data, we also extracted the spectrum of an object
that is $\sim$ 15\arcsec\ away from the quasar images. This early-type galaxy at $z$ = 0.54 may 
account for $<$ 10\% of the large external shear ($\gamma \sim$ 0.3).
 
\subsubsection{Dust and metals in main lensing galaxies}
\label{sec:dustgas}

We 
probed the intervening medium along the lines of sight towards the two images A and B of \object{QSO 
B0957+561} with great effort. The light rays associated with these images pass through two separate regions 
within a 
giant elliptical (lens) galaxy at $z$ = 0.36 (see Table~\ref{tab:glqs}). Although there is no 
evidence of Mg\,{\sc ii} absorption at $z$ = 0.36 \citep{You81b}, we studied the possible presence of 
dust in the cD galaxy during a long quiescent phase of microlensing activity \citep[e.g.][]{Sha12}. 
Using continuum delay-corrected flux ratios $B/A$ from {\it Hubble} Space Telescope (HST) spectra 
and GLITP/$VR$ photometric data in 1999$-$2001, we found a chromatic behaviour resembling extinction 
laws for galaxies in the Local Group \citep{Goi05}. While the macrolens flux ratio is 0.75 
\citep[e.g.][]{Gar94}, the continuum ratios were greater than 1, indicating that the A 
image is more affected by dust. We obtained a differential visual extinction $\Delta A_{\rm{AB}}(V) = 
A_{\rm{B}}(V) - A_{\rm{A}}(V) \sim$ $-$0.3 mag, which can be interpreted in different ways. For 
example, the simplest scenario is the presence of a dust cloud in front of the image A, at $\sim$ 26 
kpc from the centre of the cD galaxy. This cloud must be compact enough to produce a negligible 
extinction over the broad line emitting regions, since emission-line flux ratios agree reasonably 
well with $B/A \sim$ 0.75 \citep[e.g.][]{Sch91,Goi05}.  

Time-domain observations of \object{QSO B0957+561} were even more intriguing than those 
made in the spectral domain. RATCam/$gr$ light curves in 2008$-$2010 showed well-sampled, sharp 
intrinsic fluctuations with an unprecedentedly high signal-to-noise ratio. These allowed us to measure 
very accurate $g$-band and $r$-band time delays, which were inconsistent with each other: the $r$-band 
delay exceeded the 417-d delay in the $g$ band by about 3 d \citep{Sha12}. In two periods of 
violent activity, we also detected an increase in the continuum flux ratios $B/A$, as well as a 
correlation between $B/A$ values and flux level of B. This posed
the question whether the dust cloud affecting the A image might
be 
responsible for all these time-domain anomalies. \citet{Sha12} naively suggested that chromatic 
dispersion \citep[e.g.][]{Bor99} might account for a three-day lag between $g$-band and $r$-band 
signals crossing a dusty region. However, it is hard to reconcile an interband lag of some days with 
a structure belonging to the giant elliptical galaxy and containing standard dust. In addition, the 
increase in the flux ratios (diminution of A relative to B) during violent episodes was 
associated with highly polarised light passing through a dust-rich region with aligned elongated 
dust grains. This light may suffer from a higher extinction than that of weakly polarised light in 
periods of normal activity. In Sect.~\ref{sec:q0957} (Overview), we revise our previous crude 
interpretation for the chromatic time delay and the continuum flux ratios between the two images of 
\object{QSO B0957+561}.  

The main lens in \object{SDSS J1339+1310} is an early-type galaxy at $z$ = 0.61, and 
SDSS-OSIRIS spectra of both quasar images display Fe\,{\sc ii}, Mg\,{\sc ii,} and Mg\,{\sc i} 
absorption lines at the lens redshift. These metals are not uniformly distributed inside the 
galaxy, since the Mg\,{\sc ii} absorption is stronger in the A image. From OSIRIS spectra of A and B, 
we also inferred a typical value $\Delta A_{\rm{AB}}(V)$ = $-$0.27 mag for the differential visual 
extinction in the system \citep{Goi16}. Hence, we find that A is the most reddened image, supporting 
the notion that the more metal-rich the gas, the higher the dust content. \citet{Ina14} also carried 
out 
observations of the A and B images of \object{SDSS J1515+1511} with the DOLORES spectrograph on the 
TNG. Their data revealed the existence of strong Mg\,{\sc ii} absorption at the lens redshift in the 
spectrum of B. This finding was corroborated by our OSIRIS data of the B image, displaying Fe\,{\sc 
i}, Fe\,{\sc ii,} and Mg\,{\sc ii} absorption in the edge-on disc-like galaxy at $z$ = 0.74. Such 
absorption features were not detected in the OSIRIS spectrum of the A image. We consistently obtained 
that B is affected more by dust extinction than A, and $\Delta A_{\rm{AB}}(V)$ = 0.130 $\pm$ 0.013 
mag \citep[1$\sigma$ interval;][]{Sha17}. We also note that \citet{Eli06} studied the differential 
visual extinction in ten lensing galaxies at $z \leq$ 1, reporting many values ranging from 0.1 to 0.3 
mag.

\subsubsection{Cosmology}
\label{sec:cosmology}

A time delay of a GLQ enables us to measure the current expansion rate of the 
Universe (the so-called Hubble constant $H_0$), provided the lensing mass distribution and its 
redshift are reasonably well constrained through additional data \citep[e.g.][]{Jac15}. Thus, we 
obtained accurate time delays (with a mean error of 3$-$4 d) between the images of 5 GLQs: 
\object{QSO B0909+532}, \object{QSO B0957+561}, \object{SDSS J1339+1310}, \object{QSO B1413+117,} and 
\object{SDSS J1515+1511} (see Cols. 7$-$8 in Table~\ref{tab:glqs}), which can potentially be 
used to determine $H_0$. Our first time delay estimation of \object{QSO B0909+532} \citep{Ull06} was 
used by \citet{Ogu07} and \citet{Par10} to find $H_0$ values around 66$-$70 km s$^{-1}$ Mpc$^{-1}$ 
for a flat universe. They performed a simultaneous analysis of 16$-$18 GLQs, adopting a flat universe 
model with standard amounts of matter ($\Omega_M$) and dark energy ($\Omega_{\Lambda}$) that satisfy 
$\Omega_M + \Omega_{\Lambda} = 1$. \citet{Ser14} confirmed these $H_0$ values using weaker 
constraints on the matter and dark energy parameters, while \citet{Rat15} also inferred $H_0$ = 68.1 
$\pm$ 5.9 km s$^{-1}$ Mpc$^{-1}$ ($\Omega_M$ = 0.3, $\Omega_{\Lambda}$ = 0.7) from 10 GLQs with 
relative astrometry, lens redshift, and time delays sufficiently accurate, as well as with a simple 
lensing mass. This last study was partially based on the LT delays of \object{QSO B0909+532} and 
\object{QSO B0957+561}. In addition to the determination of $H_0$, our time delays have also been used 
to discuss different cosmological and gravity models \citep[e.g.][]{Tia13,Wei14,Yua15,Pan16}.  

Recently, we have determined the time delay in the two double quasars 
\object{SDSS J1339+1310} and \object{SDSS J1515+1511} (see Table~\ref{tab:glqs}), and it is easy to 
probe the impact of these delays on the estimation of $H_0$ via gravitational lensing. For example,
assuming a self-consistent lens redshift $z$ = 0.742 in \object{SDSS J1515+1511}, and a flat universe 
with $\Omega_M$ = 0.27 and $\Omega_{\Lambda}$ = 0.73, the best-fit value for $H_0$ was 72 km s$^{-1}$ 
Mpc$^{-1}$ \citep{Sha17}. Taking an external convergence $\kappa_{\rm{ext}} \sim$ 0.015 (due to a 
galaxy that is $\sim 15\arcsec$ away from the quasar images) and $H_0^{\rm{new}} = H_0^{\rm{old}} 
\times (1 - \kappa_{\rm{ext}})$ \citep[e.g.][]{Ser14} into account, the Hubble constant is decreased 
by $\sim$ 1.5\% until it reaches about 71 km s$^{-1}$ Mpc$^{-1}$. Therefore, the time delay of \object{SDSS 
J1515+1511} leads to $H_0$ 
values supporting previous estimates from other lens systems. Our delay database has a size and quality 
similar to those of the COSMOGRAIL collaboration \citep[e.g.][and references therein]{Bon16}, 
which is a complex effort involving several 1$-$2m telescopes at different sites. The LT monitoring 
offers a unique opportunity to obtain homogeneous and accurate light curves of GLQs, and thus time
delays with an uncertainty of a few days.  

\subsection{New photometric, polarimetric, and spectroscopic results}
\label{sec:newres}

In this section, we introduce new results for six objects in the GLQ database. For three objects that 
have been updated in recent papers, i.e. \object{SDSS J1339+1310}, \object{SDSS J1442+4055,} and 
\object{SDSS J1515+1511}, we do not include science data derived from frames in the database.

\subsubsection{QSO B0909+532}
\label{sec:q0909}

The RATCam photometry in the $r$ band over the period between January 2005 and June 2011 (198 epochs) 
has been published in \citet{Goi08} and \citet{Hai13}. Here, we present additional $r$-band 
photometric data of the two quasar images A and B, which were obtained from RATCam frames in 
February-April 2012 (18 epochs), as well as using the IO:O camera in the period spanning from October 
2012 to June 2016 (128 epochs; see Table~\ref{tab:data}). This new camera has a 
$10\arcmin\times10\arcmin$ field of view and a pixel scale of $\sim 0\farcs30$ (binning 2$\times$2),
and we set the exposure time to 200 or 150 s. After some initial processing tasks, including cosmic-ray 
removal and bad pixel masking, a crowded-field photometry pipeline produced magnitudes of A and B
for every IO:O frame. Our pipeline relies on IRAF\footnote{IRAF is distributed by the National 
Optical Astronomy Observatory, which is operated by the Association of Universities for Research in 
Astronomy (AURA) under cooperative agreement with the National Science Foundation. This software is 
available at \url{http://iraf.noao.edu/}} packages and the IMFITFITS software \citep{McL98}. As the 
lensing galaxy is not apparent in optical frames of \object{QSO B0909+532}, a simple photometric 
model can describe the crowded region associated with such GLQ. This model only consists of two close 
stellar-like sources, where each source is described by an empirical point-spread function (PSF). To
perform the PSF fitting of the double quasar, we mostly considered the 2D profile of the "b" field 
star as the PSF, after removing the local background to clean its distribution of instrumental flux. 
However, when this bright star was saturated in certain frames, the PSF was derived from the profile 
of the "c" field star \citep[e.g.][]{Koc97}.

\begin{figure}[!ht]
\centering
\includegraphics[width=9cm]{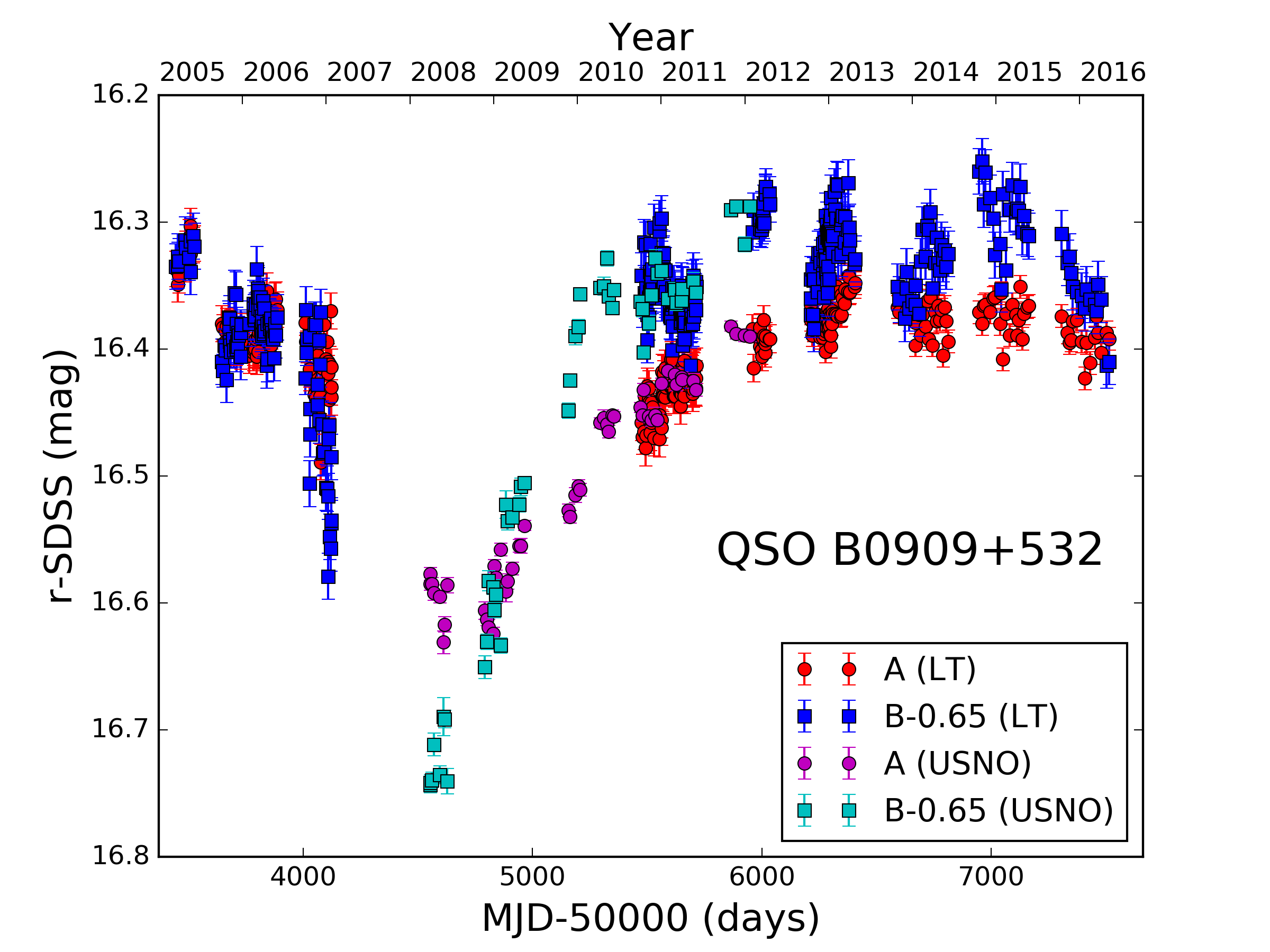}
\hspace*{-0.7cm}
\includegraphics[width=9cm]{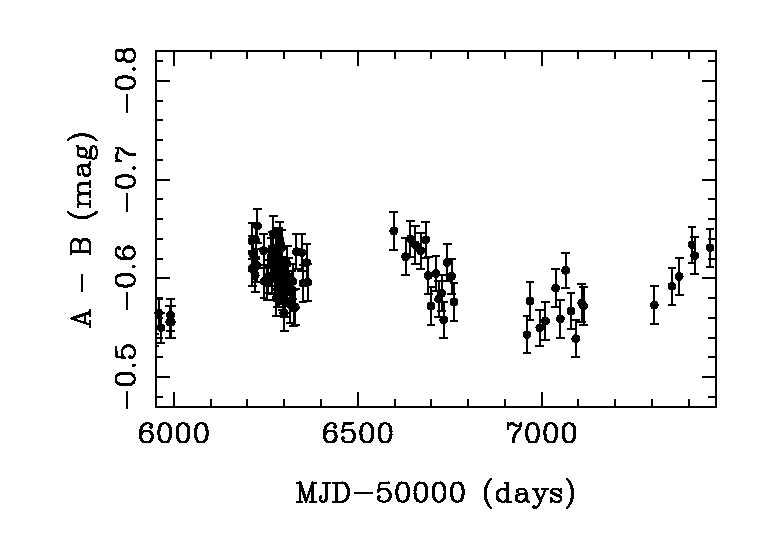}
\hspace*{-0.7cm}
\includegraphics[width=9cm]{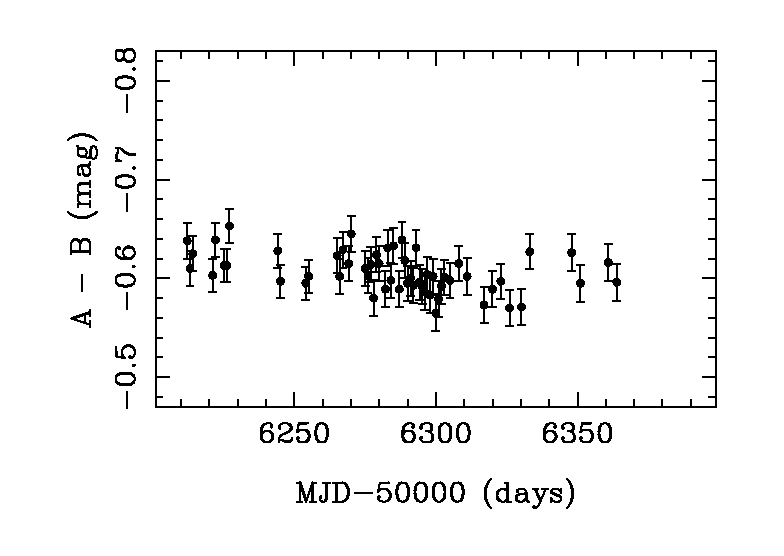}
\caption{Light curves of \object{QSO B0909+532} in the $r$ band. The top panel shows the LT-USNO 
brightness records of both quasar images. The brightness of B is offset by $-$0.65 mag to facilitate 
comparison, and the new LT data correspond to our monitoring in 2012$-$2016. The middle panel 
incorporates the DLC between 2012 and 2016, and the bottom panel displays the zoomed-in DLC 
around day 6300. To construct the DLC, the data of the A image have been shifted by $-$50 d (time 
delay), and then binned around the dates of the B image (using bins with a semisize of 10 d). Only
bins including two or more data have been taken into account to compute differences between A and B.}
\label{fig:q0909}
\end{figure}

To obtain the IO:O light curves, we removed magnitudes when the signal-to-noise ratio ($S/N$) of the 
"c" field star fell below 30. This star has a brightness close to that of the A image, and the $S/N$ 
was measured through an aperture of radius equal to twice the $FWHM$ of 
the seeing disc. By visual inspection of the pre-selected brightness records, we then found that the 
magnitudes of A and B at a few epochs strongly deviate from adjacent data. These outliers were also 
discarded. The whole selection procedure yielded a rejection rate of about 6\% (8 out of 136 epochs). 
In a last step, assuming the root-mean-square deviations 
between magnitudes on consecutive nights as errors, the uncertainties were 0.011 and 0.017 mag for A 
and B. The RATCam-IO:O light curves of A and B covering the period 2005 to 2016 are available in 
tabular format at the CDS\footnote{See \url{http://grupos.unican.es/glendama/LQLM_results.htm} for 
updated results}: Table 4 includes $r$-SDSS magnitudes and their errors at 344 epochs. Column 1 lists 
the observing date (MJD$-$50\,000), Cols. 2 and 3 give photometric data for the quasar image A, and 
Cols. 4 and 5 give photometric data for the quasar image B. Thus, we combined all our $r$-band 
measurements in a machine-readable ASCII file, using MJD$-$50\,000 dates instead of JD$-$2\,450\,000 
ones. Now, in all the GLENDAMA light curves, the origin of the time axis is MJD$-$50\,000. The 
$r$-band data collected by us and the USNO group during a 12-year period are also displayed in the 
top panel of Fig.~\ref{fig:q0909}. The new 146 epochs of magnitudes (after day 5959) lead to new 
microlensing variability in the difference light curve (DLC; see the middle and bottom panels of 
Fig.~\ref{fig:q0909}). Although this variability has an amplitude of $\sim$ 0.1 mag, is not as 
strong as in the previous period between days 4000 and 5400 \citep[see the bottom panel in Fig. 3 
of][]{Hai13}. The new extrinsic signal might better constrain the size of the continuum source 
emitting at $\sim$ 2600 \AA\ (see Sec.~\ref{sec:accretion}).

\subsubsection{FBQS J0951+2635}
\label{sec:q0951}
 
\citet{Jak05} monitored the double quasar \object{FBQS J0951+2635} soon after its discovery in 1998 
\citep{Sch98}, measuring a time delay of about 16 d and reporting evidence for microlensing in the 
period 1999$-$2001 \citep[see also][]{Par06}. We also presented $R$-band light curves of the two 
images of the GLQ \citep{Sha09}. These last records (37 epochs), based on Uzbekistan MAO observations 
between 2001 and 2006, indicated the existence of a long-timescale microlensing 
fluctuation. The MAO monitoring  programme was conducted by an international collaboration of 
astronomers from Russia, Ukraine, Uzbekistan, and other countries. Here, we analyse new LT photometric 
observations made during an 8-year period (2009$-$2016), which allow us to draw the evolution of the 
extrinsic variation over this century. Our database contains 72 frames in the $r$ band, divided into 
two groups (see Table~\ref{tab:data}): 29 RATCam frames in 2009$-$2012 (for each monitoring night, 
we usually obtained three consecutive 300 s exposures) and 43 IO:O frames in 2013$-$2016 
(typically, two consecutive 250 s exposures per monitoring night). To fill the LT gap in 2010, 
3$\times$300 s ALFOSC exposures of the lens system were taken with the Bessel $R$ filter on 8 
February 2010. We also analyse these frames, which are not included in the database. 

\begin{figure}
\centering
\includegraphics[width=9cm]{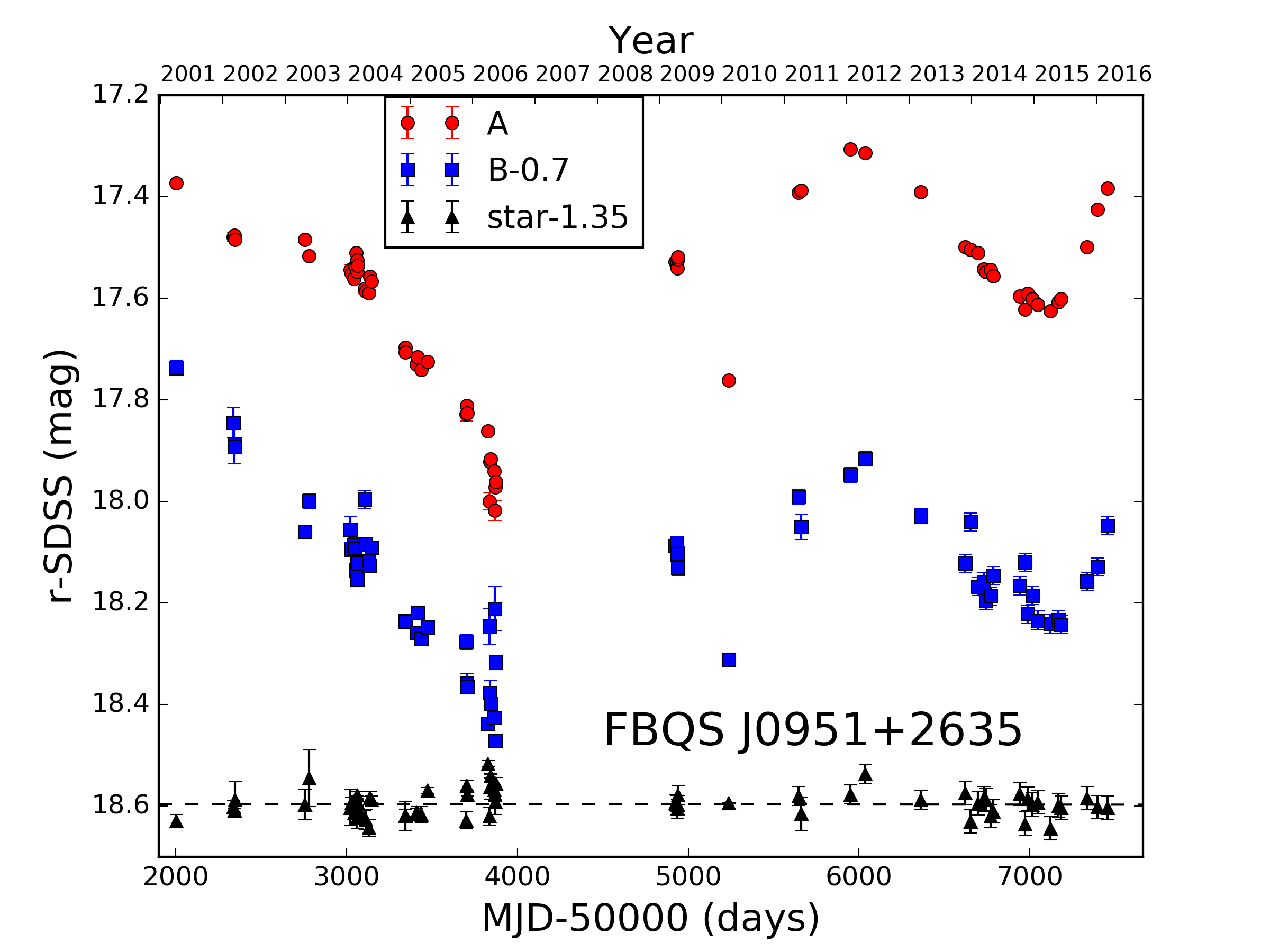}
\hspace*{-0.7cm}
\includegraphics[width=9cm]{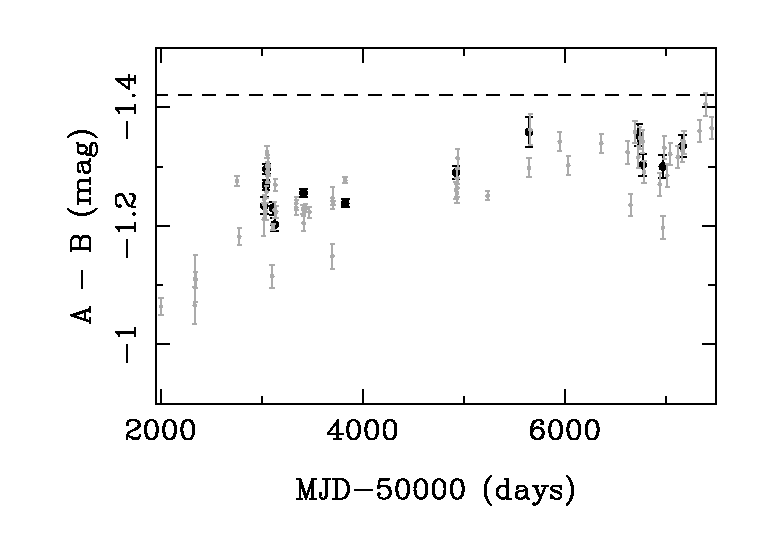}
\caption{Light curves of \object{FBQS J0951+2635} in the $r$ band. The top panel shows the LT-NOT-MAO  
brightness records of both quasar images and the faint star S3. There is a gap of about 1000 d 
between the MAO monitoring campaign (before day 4000) and the LT-NOT follow-up observations starting 
in 2009. The records of B and S3 are offset by $-$0.70 and $-$1.35 mag, respectively. The bottom 
panel displays the DLC (black data points), where the data of the B image have been shifted by $-$16  
d (time delay) and binned around the dates of the A image (using bins with a semisize of 4 d). We 
also show the single-epoch magnitude differences (grey data points), as well as the single-epoch flux
ratio of the Mg\,{\sc ii} emission line in magnitudes (horizontal dashed line).} 
\label{fig:q0951}
\end{figure}      

The lensing galaxy is too faint to be detected with a red filter, and thus the system was described 
as two stellar-like objects, that is, two PSFs. The S1 field star was used to estimate the PSF, whereas we 
considered the S3 field star to check the reliability of the quasar brightness fluctuations 
\citep[see the finding chart in Fig. 1 of][]{Sha09}. We used IMFITFITS to obtain PSF-fitting 
photometry for the two quasar images and the field stars. Most of the LT frames (64 of 72) led to 
reasonable photometric results, and these usable frames were then combined on a nightly basis to 
produce $r$-band magnitudes at 28 epochs. For each object, the typical photometric error for an 
individual exposure was determined from the intra-night scatter of the magnitude values measured on 
the individual frames. These intra-night scatters were 0.007 mag (A), 0.025 mag (B), and 0.033 mag 
(S3); B is fainter than A in $\sim$ 1.3 mag (and only 1\farcs1 away from the brightest image A), 
and S3 is fainter than B in $\sim$ 1.2 mag. The errors for combined frames were reduced by a factor 
of $N^{1/2}$, where $N$ = 2$-$3 is the number of individual exposures. After constructing the LT 
$r$-band brightness records, we merged this new dataset and the NOT $R$-band data at day 5236 
(MJD$-$50\,000; derived from the ALFOSC exposures on 8 February 2010) using an $r - R_{\rm{NOT}}$ 
offset of 0.153 mag. We also found an $r - R_{\rm{MAO}}$ offset of 0.489 mag, and merged the LT-NOT 
and the MAO data in 2001$-$2006. We remark that both magnitude offsets were calculated from the 
records of the non-variable star S3.   

The top panel of Fig.~\ref{fig:q0951} shows the LT-NOT-MAO $r$-band light curves of the double quasar 
and the comparison star S3. The brightness changes of A and B are significantly greater than the 
observational noise level in the record of S3, which is appreciably fainter than both quasar images. 
In addition, the almost parallel behaviour of A and B indicates the presence of intrinsic variations. 
In Table 5 at the CDS$^{12}$, using the same format as Table 4, we include the $r$-SDSS magnitudes of 
A and B (and their errors) at 66 epochs over the period 2001$-$2016. Column 1 contains the observing 
dates (MJD$-$50\,000), Cols. 2 and 3 give the magnitudes and magnitude errors of A, and Cols. 4 
and 5 give the magnitudes and magnitude errors of B. Regarding the extrinsic signal in the $r$ band,
the DLC and the single-epoch differences are shown in the bottom panel of Fig.~\ref{fig:q0951}. It is 
apparent that the DLC values basically agree with single-epoch differences close to them. However, it 
is not clear whether the microlensing variation that was observed in the period 1999$-$2006 
\citep{Jak05,Par06,Sha09} is completed or not. Although the DLC in 2009$-$2016 is roughly consistent 
with a quiescent phase of microlensing activity, the single-epoch differences suggest an active 
phase, in which the current $r$-band flux ratio could be similar to the single-epoch Mg\,{\sc ii} 
flux ratio as measured in 2001 by \citet{Jak05}.  

\subsubsection{QSO B0957+561}
\label{sec:q0957}

\paragraph{\underline{Optical photometry}}

\begin{figure}
\centering
\hspace*{-0.7cm}
\includegraphics[width=9cm]{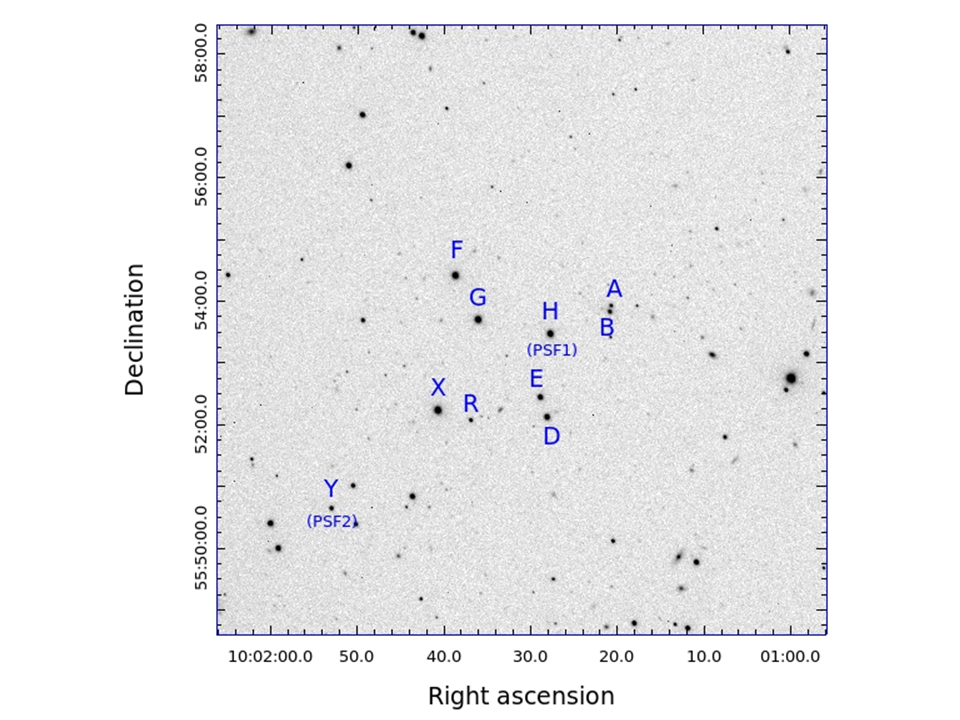}
\includegraphics[width=9cm]{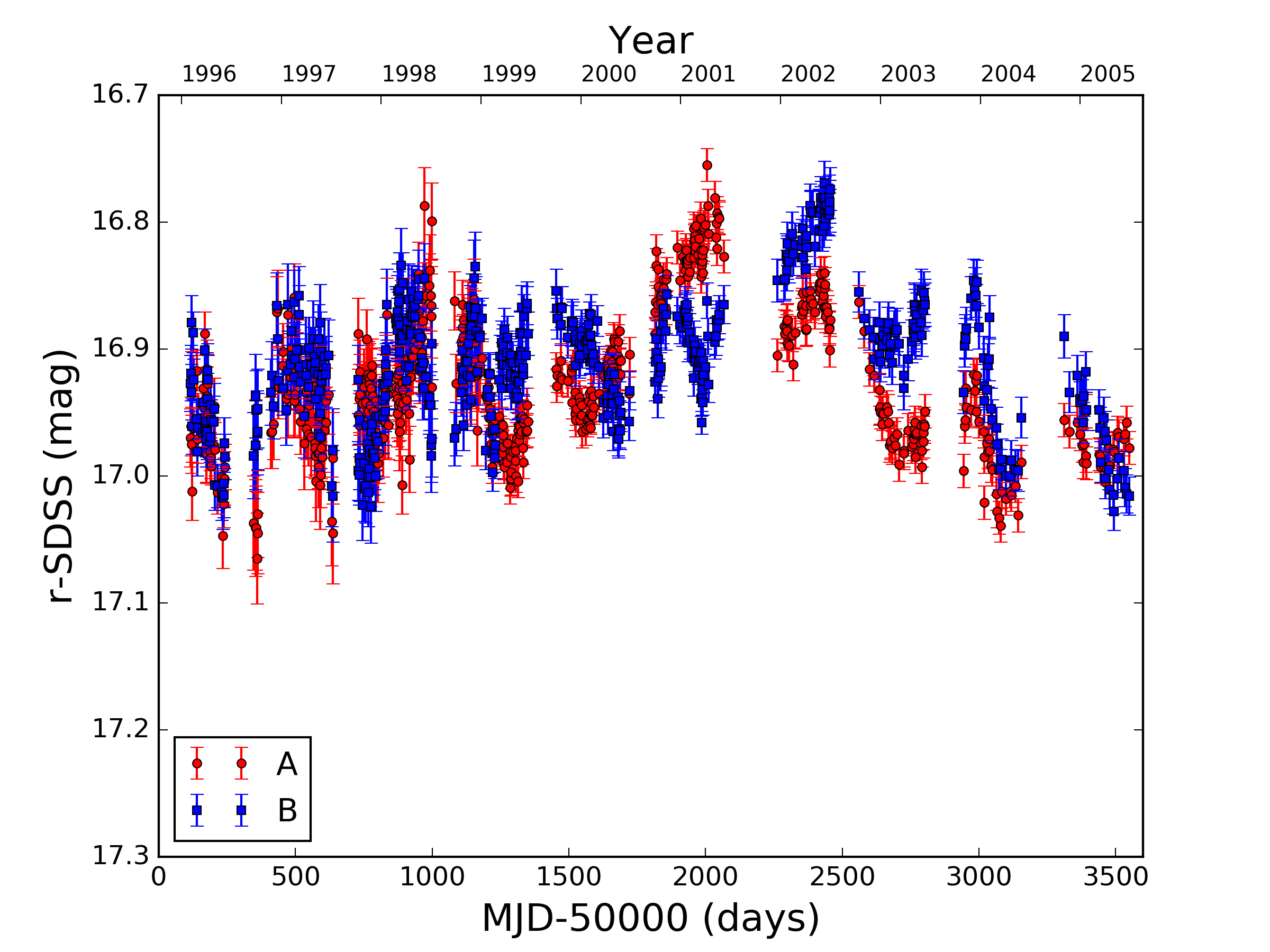}
\includegraphics[width=9cm]{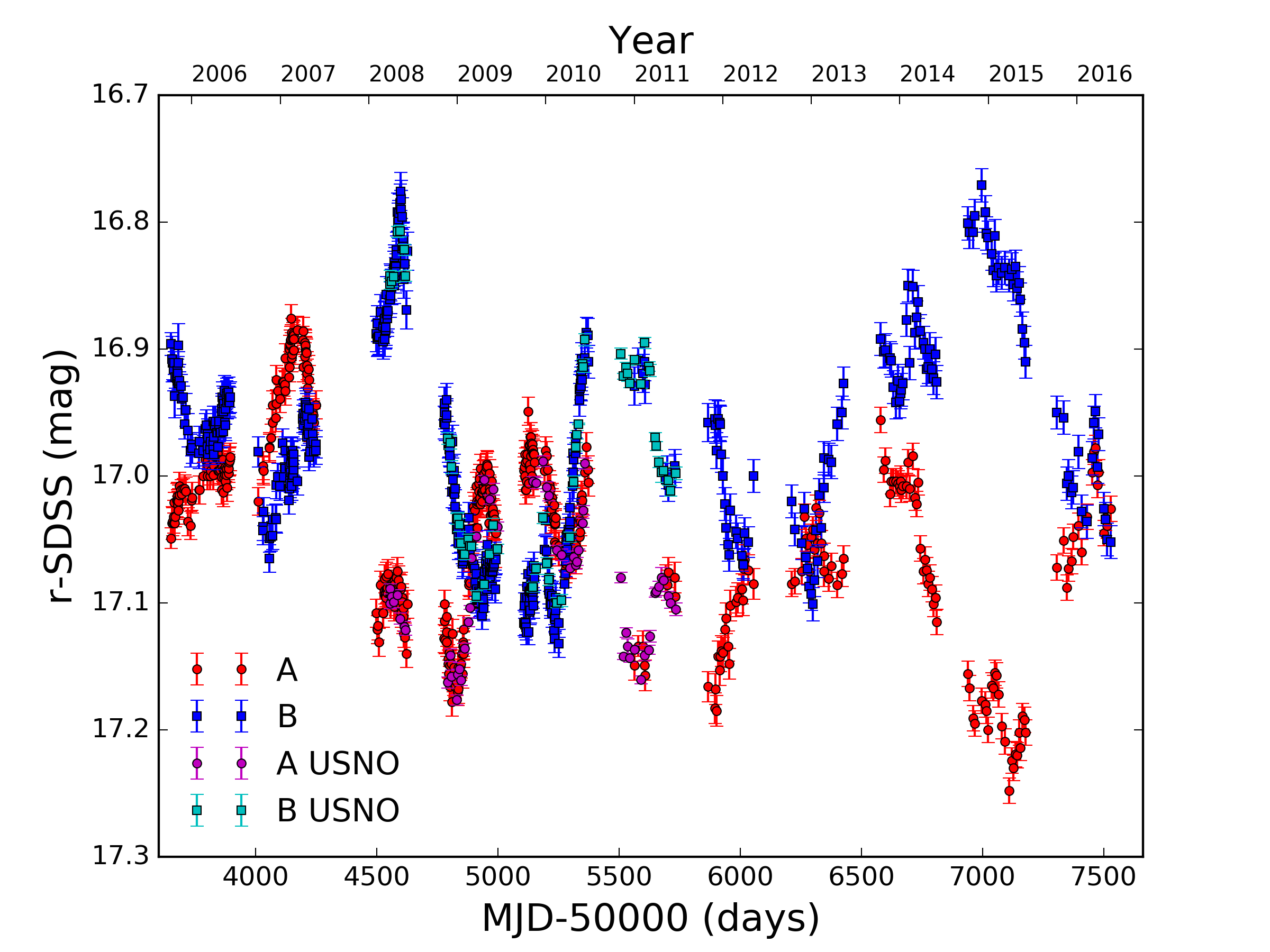}
\caption{Photometric monitoring of \object{QSO B0957+561}. The top panel displays the IO:O $r$-band 
frame taken on 21 January 2015. In this typical frame in terms of seeing ($FWHM$ = 1\farcs45), the 
two quasar images and several field stars are properly labelled \citep[these objects, except for star 
Y, are depicted on the finding chart of][]{Ova03}. The middle and bottom panels show the IAC80-LT 
light curves in the $r$ band in two different monitoring periods. The first period includes IAC80 
data (middle panel), while the vast majority of data in the second period correspond to LT 
observations (bottom panel). For comparison purposes, the bottom panel also contains the USNO 
brightness records in 2008$-$2011 using our own estimate for the $r_{\rm{LT}} - r_{\rm{USNO}}$ 
magnitude offset \citep[14.430 mag instead of the 14.455 mag offset used by][]{Hai12}.}
\label{fig:q0957r}
\end{figure}

We observed \object{QSO B0957+561} in red passbands from 1996 to 2016, that is, for 21 years. We used 
the IAC80 Telescope during the first observing period (1996$-$2005), while we monitored the double
quasar with the LT from 2005 to 2016. The IAC80-CCD frames in the $R$ band for the period 1996$-$2001 
were previously processed using the PHO2COM photometric task \citep{Ser99,Osc02}. Here, we focus on 
the most recent IAC80-CCD $R$-band frames taken between January 1999 and November 2005, which are 
included in our database (see Table~\ref{tab:data}). The CCD camera covered an area of about 
$7\arcmin\times7\arcmin$ on the sky, with a pixel scale of $\sim 0\farcs43$. This field of view 
allowed us to simultaneously image the AB components of the GLQ and the YXRFGEDH field stars (see the 
top panel of Fig.~\ref{fig:q0957r}). The typical exposure time was 300 s, although longer combined 
exposures of 900$-$1200 s were also used. We selected 515 $R$-band frames with a reasonable size of the 
seeing disc, and then performed PSF photometry on the field stars and lens system with the IMFITFITS 
software. As usual, the clean 2D profile of the H star was considered as the empirical PSF and the lens 
system was modelled as two stellar-like sources (i.e. two PSFs) plus a de Vaucouleurs profile convolved 
with the PSF. For these $R$-band frames, we found evidence of inhomogeneity over the field of view and 
carried out a frame-to-frame inhomogeneity correction based on the idea by \citet{Gil88}. When we performed 
photometry on frames of any lens system, we paid special attention to colour and inhomogeneity terms, as 
well as other atmospheric and instrumental effects \citep[e.g.][]{Sha08}. 

We thus obtained new photometric data of \object{QSO B0957+561} from 515 IAC80-CCD frames in the $R$ 
band over a seven-year period in 1999$-$2005. After an $S/N >$ 70 selection (the $S/N$ 
values were calculated within circles of 7-pixel radius centred on the quasar image A), the number of 
frames was reduced to 441. To estimate typical photometric errors, we computed magnitude differences 
between consecutive nights. These night-to-night brightness variations led to an uncertainty of 0.014 
mag for both A and B, meaning that photometry to 1.4\% was achieved for the lensed quasar. If there were 
several measurements on the same night, they were grouped to get more accurate light curves. This 
grouping produced 367 magnitudes for each quasar image. We also rejected outliers, so that the new IAC80 
dataset contains 347 epochs. We note that it is important to merge the old IAC80 data 
\citep[PHO2COM light curves in Fig. 1 of][]{Osc02} and the new IAC80 brightness records, as well as 
the new IAC80 data and the LT $r$-band light curves over 2005$-$2010 \citep{Sha08,Sha12}. Before  
constructing a global IAC80 database in the $R$ band, we applied an outlier detection, data cleaning, and 
intra-night grouping method to the old photometry. Thus, we found small $R_{\rm{new}} - R_{\rm{old}}$ 
offsets of 0.004 mag in A and 0.031 mag in B, and merged the old (1996$-$2001) and the new (1999$-$2005) 
data. The shifted old magnitudes in 1999$-$2001 were then replaced by our new photometric data in 
that period. In addition, to transform the $R$-band magnitudes ($R_{\rm{new}}$) into the $r$ band of 
the SDSS photometric system, we found similar $r - R_{\rm{new}}$ offsets of $-$0.236 mag in A and 
$-$0.233 mag in B.
 
After building the light curves in the $r$ band until 2010 (942 epochs), the next step was to 
incorporate all the available data for the period 2011$-$2016. We fully processed and analysed the 
RATCam $r$-band frames between 2011 and 2014 (RATCam was decommissioned at the end of February 2014), 
which have provided magnitudes in 34 additional nights. We also obtained photometric outputs from 95 
IO:O frames in the $r$ band. Because this optical camera has
a high sensitivity, the brightest stars 
in its field of view are often saturated. Hence, we used the H star to build the PSF in two-thirds of 
the frames, while the Y star was used in the remaining one-third of the frames (see the PSF1 and PSF2 stars 
in the top panel of Fig.~\ref{fig:q0957r}). Following the standard selection procedure of IO:O data, we 
removed quasar magnitudes for three frames in which the $S/N$ of image A was below 30. Fortunately, 
we did not detect any outlier, and thus added 91 new magnitude
epochs (two frames were taken 
on the same night). The IO:O monitoring is characterised by a mean sampling rate of about one frame 
every 7$-$10 d, and this precludes an estimation of the variability on consecutive nights. For a
given quasar image, we made trios of consecutive data within time intervals shorter than 14.5 d. 
We also performed a linear interpolation to the initial and final magnitudes in each trio to generate 
an interpolated magnitude at the same epoch as that of the central data point, and we then derived a 
typical photometric error by comparing measured and interpolated central magnitudes. This method 
produces reasonable errors of 0.010 mag in A and 0.013 mag in B, which are consistent with the 
uncertainty in the magnitude of the control star R (0.011 mag; A, B, and R have similar brightness).

\begin{figure}
\centering
\includegraphics[width=9cm]{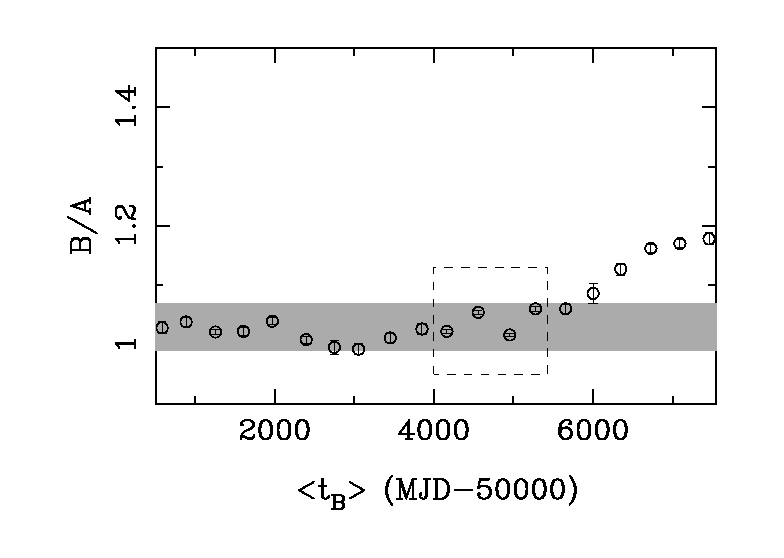}
\includegraphics[width=9cm]{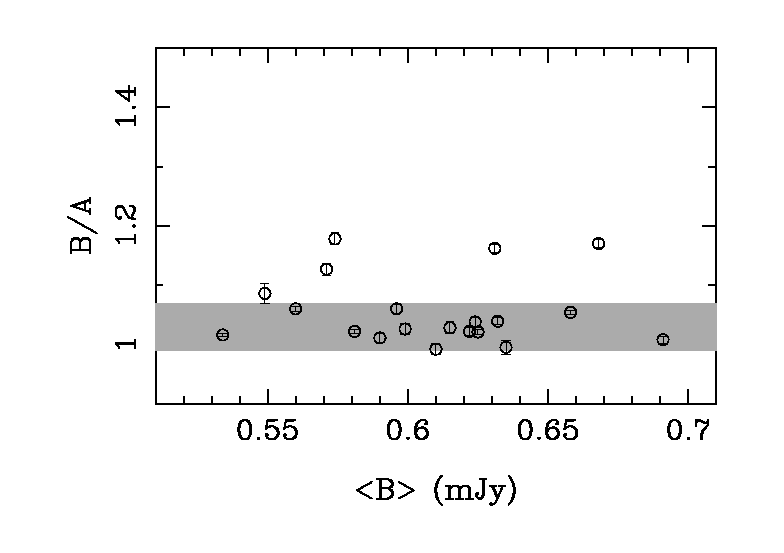}
\caption{Delay-corrected $r$-band flux ratio of \object{QSO B0957+561}. The 2$\sigma$ measurements of 
$B/A$ are obtained in the overlapping periods between A(+420 d) and B. Here, the $\langle 
t_{\rm{B}} \rangle$ and $\langle B \rangle$ values are the epochs and fluxes of B averaged over these 
periods. The grey highlighted rectangles illustrate the $r$-band flux ratio from HST spectra in 
1999$-$2000 (2$\sigma$ confidence interval), while the rectangle with dashed sides highlights the 
oscillating trend from LT photometric data in 2005$-$2010. The top panel shows the yearly variability
and the bottom panel displays the dependence of $B/A$ on $\langle B \rangle$.}
\label{fig:q0957frat}
\end{figure}

The IAC80-LT $r$-band light curves over the 21-year period 1996$-$2016 are available at the 
CDS$^{12}$: the format of Table 6 is similar to those of Tables 4 and 5, including $r$-SDSS magnitudes 
and errors of A and B at 1067 epochs (MJD$-$50\,000). These records are shown in the middle 
and bottom panels of Fig.~\ref{fig:q0957r}, which indicate that the variability of the GLQ has 
increased over the last data decade. While the quasi-constancy of the 
delay-corrected flux ratio $B/A$ in red passbands between 1987 and 2007 (dates of the trailing image 
B) is well known, for $B/A \sim$ 1.03 \citep[e.g.][]{Osc02,Sha08}, there is some controversy on the behaviour of 
the flux ratio in more recent years when higher variability occured \citep{Hai12,Sha12}. Using LT-USNO 
$r$-band data, \citet{Hai12} suggested the existence of a slow increase in $B/A$ over days 
4100$-$5700 (epochs of B). However, a reanalysis of LT-USNO data revealed an oscillating behaviour 
between days 4100 and 5400 (see the rectangle with dashed sides in the top panel of 
Fig.~\ref{fig:q0957frat}) that calls into question the presence of a microlensing gradient during 
these epochs \citep{Sha12}. To try to remedy this problem,  we derived the 
$r$-band flux ratio in 20 time segments of B covering the full photometric monitoring campaign from the
IAC80-LT light curves (see 
Appendix~\ref{sec:appa1}). 

In the top panel of Fig.~\ref{fig:q0957frat}, we present the long-term evolution of $B/A$ (see 
Table~\ref{tab:q0957frat}), where the $\langle t_{\rm{B}} \rangle$ values are the average epochs for 
the overlapping periods between the time-delay shifted flux record of A and the time segments of B. 
The error bars represent formal 2$\sigma$ confidence intervals, and the grey highlighted rectangle is 
the 2$\sigma$ measurement of the $r$-band $B/A$ from HST spectra in 1999$-$2000 \citep{Goi05}. 
Although we detect a low-amplitude variability over the first 5000 days of data, the values of $B/A$ 
are outside the HST band from day 6000. Hence, although the DLC in \citet{Hai12} likely has a biased 
shape, the new results support their claim that a microlensing event occurred in recent years. An 
extrinsic event (decrease in $B/A$) with similar amplitude has only been detected in the first years 
of the 1980s \citep[e.g.][]{Pel98}, so this new accurately measured fluctuation offers a unique 
opportunity to unveil physical properties of the source and the primary lensing galaxy. In the bottom 
panel of Fig.~\ref{fig:q0957frat}, we also show the lack of correlation between $B/A$ and the average 
flux $\langle B \rangle$. Based on four measurements of $B/A$ from LT observations in 2005$-$2010 
(see the rectangle with dashed sides in the top panel), we have previously found evidence of a 
correlation between flux ratio and variability of B. However, from this larger collection of data, we 
did not find any clear $B/A - \sigma_B$ relationship, where the $\sigma_B$ values are the standard 
deviations of the flux of B in the overlapping periods between A(+420 d) and B. 

\paragraph{\underline{Imaging polarimetry}}

As part of a pilot programme to probe the suitability of the
main instruments on the LT for studying 
GLQs, we also conducted polarimetric and spectroscopic monitorings of \object{QSO B0957+561} with the 
2m robotic telescope. For polarimetric follow-up observations, we used the imaging polarimeters 
RINGO2 and RINGO3. The basic idea behind these instruments is to take eight consecutive exposures of the
same duration for eight different rotor positions of a rapidly rotating polaroid. The data are then 
stacked for each rotor position to produce eight final frames in a given optical band 
\citep[e.g.][and references therein]{Jer16}. Combining photometric measurements in the eight frames 
allows determining the polarisation \citep{Cla02}. RINGO2 saw first light in June 2009 and was 
decommissioned in October 2012. This optical polarimeter used an EMCCD composed of 512$\times$512 
pixels (pixel scale of $\sim 0\farcs45$) and a hybrid V+R filter covering the wavelength range 460 to 
720 nm. We obtained 8$\times$200 s frames on each of the first two nights of observation (21 December 
2011 and 13 January 2012), and performed slightly longer imaging-polarimetry observations 
(8$\times$300 s frames) on 23 January 2012 and 26 March 2012. The data on 13 January 2012 are not 
usable because the PSF is very elongated along a specific direction as a result of tracking problems. 
The RINGO3 multicolour polarimeter was brought into service in January 2013, and it incorporates a 
pair of dichroic mirrors that split the light into three beams to simultaneously obtain exposures in 
three broad-bands using three different 512$\times$512 pixel EMCCDs \citep{Arn12}: B (350$-$640 nm), 
G (650$-$760 nm), and R (770$-$1000 nm). We obtained useful data (8$\times$300 s frames with each 
EMCCD) on 16 out of 18 observing nights over the period 2013$-$2017. As polarimetric observations 
have been interrupted in late February 2017, we analysed all available frames, even those not yet 
included in our GLQ database. 

\setcounter{table}{6}
\begin{table*}
\centering
\caption{Polarimetric results from RINGO2 and RINGO3 observations of \object{QSO B0957+561}.}
\begin{tabular}{lccccccc}
\hline\hline
Obs. Configuration & \multicolumn{3}{c}{A} & & \multicolumn{3}{c}{B} \\
\cline{2-4}
\cline{6-8}
 & $PD$ (\%) & $PD_{\rm{corr}}$ (\%) & $PA$ (\degr) & & $PD$ (\%) & $PD_{\rm{corr}}$ (\%) & $PA$ (\degr) \\
\hline 
RINGO2 (V+R) & 0.38 $\pm$ 0.42 & 0 & 14 $\pm$ 32 & & 0.47 $\pm$ 0.42 & 0 & $-$32 $\pm$ 25 \\ 
RINGO3/B (blue) & 0.43 $\pm$ 0.20 & 0.38 & $-$14 $\pm$ 13 & & 0.17 $\pm$ 0.20 & 0 & $-$12 $\pm$ 34 \\ 
RINGO3/G (green) & 0.52 $\pm$ 0.24 & 0.46 & 9 $\pm$ 13 & & 0.13 $\pm$ 0.24 & 0 & $-$9 $\pm$ 53 \\ 
RINGO3/R (red) & 0.70 $\pm$ 0.30 & 0.63 & $-$34 $\pm$ 12 & & 0.25 $\pm$ 0.30 & 0 & $-$1 $\pm$ 34 \\ 
\hline
\end{tabular}
\label{tab:q0957pol}
\end{table*}

In Appendix~\ref{sec:appa2}, we present details on the reduction of RINGO2 and RINGO3 observations of 
\object{QSO B0957+561}. After removing main instrumental biases, the Stokes parameters ($q_{\rm{A}}$, 
$u_{\rm{A}}$) and ($q_{\rm{B}}$, $u_{\rm{B}}$) at different observing epochs are depicted in 
Fig.~\ref{fig:q0957ringo2qu} and Fig.~\ref{fig:q0957ringo3qu}. Although the construction of 
polarisation curves of the two quasar images is a very attractive possibility, we should firstly 
check whether the scatters in these $q-u$ diagrams are caused by true variability. 
Accordingly, scatters in parameter distributions of the quasar images were compared to scatters in 
distributions of Stokes parameters for the non-variable field stars E and D (see the top panel of 
Fig.~\ref{fig:q0957r}). We concentrated on RINGO2 and RINGO3/B data, which are based on the best 
observations in terms of $S/N$, and deduced that deviations from the mean values in 
Fig.~\ref{fig:q0957ringo2qu} and the top panel of Fig.~\ref{fig:q0957ringo3qu} are essentially due to 
random noise. Thus, for a given observational configuration (polarimeter and optical band), the 
polarisation of each image is characterised by mean values ($\bar{q}$, $\bar{u}$) and standard errors 
($\sigma_{\bar{q}}$, $\sigma_{\bar{u}}$). The polarisation degree and polarisation angle were derived 
as $PD = (\bar{q}^2 + \bar{u}^2)^{1/2}$ and $PA = 0.5 \tan^{-1}(\bar{u}/\bar{q})$, respectively 
\citep[e.g.][]{Cla02}. We also estimated a common random error for $\bar{q}$ and $\bar{u}$ of both 
images ($\sigma_{\rm{pol}}$) through the average of the four standard errors, and obtained 
$\sigma_{PD} = \sigma_{\rm{pol}}$ and $\sigma_{PA} = 0.5\ (\sigma_{\rm{pol}}/PD)$ from a standard 
propagation of uncertainties. 

Table~\ref{tab:q0957pol} includes our main results for the two quasar images in the four 
observational configurations: RINGO2, RINGO3/B, RINGO3/G, and RINGO3/R. We 
note two important details: first, there are only three individual observations from RINGO2, and with 
just three data points ($q$, $u$) for each image, the $\sigma_{\rm{pol}}$ value is 50\% uncertain. To
account for this extra uncertainty, the errors in the first data row of Table~\ref{tab:q0957pol} are 
increased by 50\%. Second, as the $PD$ of weakly polarised sources is systematically overestimated 
\citep[e.g.][]{Sim85}, we also report the corrected polarisation degree ($PD_{\rm{corr}}$). For 
$PD/\sigma_{PD}$ lower than or similar to 1, the best estimate of the actual polarisation amplitude is 
zero: $PD_{\rm{corr}}$ = 0 \citep[e.g.][]{Sim85}. Otherwise, we use the estimator described by  
\citet{War74}. The results in Table~\ref{tab:q0957pol} indicates that the polarisation of \object{QSO 
B0957+561} has remained at low levels during the 5.2-year polarimetric follow-up. Contrary to what 
\citet{Sha12} proposed to explain certain time-domain observations of the first GLQ (see 
Sec.~\ref{sec:dustgas}), we have not found evidence for high-polarisation states in epochs of violent 
activity. While the RINGO3 data of B are consistent with zero polarisation (or $PD \leq$ 0.3\% from 
the weighted average over the three bands), the data of A suggest a polarisation amplitude of about 
0.5\%. The detection of this 0.5\% polarisation in A (which could depend on wavelength; see the
$PD_{\rm{corr}}$ values in the third column of Table~\ref{tab:q0957pol}) deserves more attention. 
Before this work, \citet{Wil80} conducted polarimetric observations of \object{QSO B0957+561}
using unfiltered white light. They reported $PD$ = 0.7 $\pm$ 0.4\% ($PD_{\rm{corr}}$ = 0.6\%) for A 
and $PD$ = 1.6 $\pm$ 0.4\% ($PD_{\rm{corr}}$ = 1.5\%) for B. Therefore, the current polarisation 
degree of image A agrees well with the 1980 value of Wills et al., while the current $PD$ of image B 
does not. \citet{Dol95} also studied the polarisation of A and B in the UV. However, their HST data 
led to large uncertainties of $\sim$ 1.5\% in the $PD$ of both images, and no reliable detection was 
obtained.            

\paragraph{\underline{Spectroscopy}}

Our spectroscopic monitoring of \object{QSO B0957+561} includes many observing epochs with FRODOSpec 
on the LT. This spectrograph is equipped with an integral field unit that consists of 12$\times$12 
lenslets each 0\farcs83 on sky, covering a field of view of 9\farcs84$\times$9\farcs84.
However, the FRODOSpec programme in 2010$-$2014 was not as successful as expected, and only $\sim$ 
25\% of the 2500$-$2700 s exposures led to usable spectra of both quasar images. The difficulty in 
placing in a robotic mode two sources separated by 6\farcs1 within a square of side $\sim$ 
10\arcsec\ was one of the main reasons for a relatively low efficiency of the IFS monitoring. We 
obtained 16 reasonably good individual exposures with FRODOSpec, and one of them (on 1 March 2011) 
was exhaustively analysed by \citet{Sha14b}. This paper addressed the whole processing 
method we used to obtain flux-calibrated spectra of sources in crowded fields from FRODOSpec 
observations\footnote{The associated L2LENS software is available at 
\url{http://grupos.unican.es/glendama/LQLM_tools.htm}}. For both the IFS and the LSS (in this
subsection and in Sect.~\ref{sec:q1001}), we 
almost always compared quasar spectral fluxes averaged over the $g$ and/or $r$ passbands with 
corresponding concurrent fluxes from RATCam/IO:O frames. This comparison has permitted us to check 
the initial calibration of spectra and recalibrate them when required. Here, we concentrated 
on the Mg\,{\sc ii} emission at 2800 \AA\ observed at red wavelengths because the red grating of the 
integral-field spectrograph provides the highest $S/N$ values. The red grating spectra of A, B, and 
the primary lensing galaxy (GAL) are available at the CDS$^{12}$: Table 8 includes wavelengths (\AA) 
along with fluxes of A, B, and GAL (10$^{-17}$ erg cm$^{-2}$ s$^{-1}$ \AA$^{-1}$) for each of the 16 
observing dates (yyyymmdd). 

\begin{figure}
\centering
\includegraphics[width=9cm]{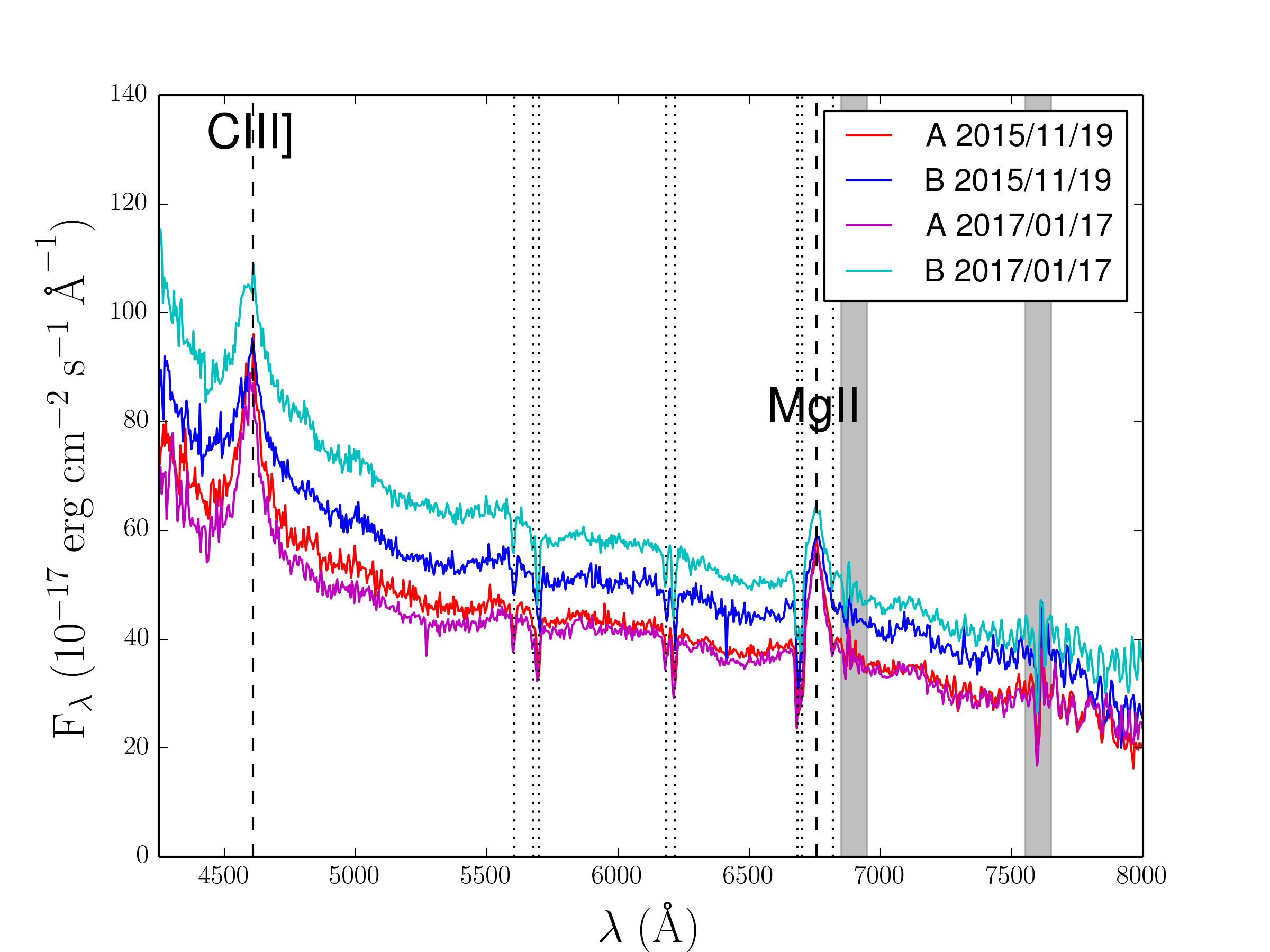}
\caption{SPRAT spectroscopy of \object{QSO B0957+561} on 19 November 2015 and 17 January 2017. 
Vertical dashed and dotted lines indicate emission and absorption features, while grey highlighted 
regions are associated with atmospheric artefacts.}
\label{fig:q0957sprat}
\end{figure}

We conducted additional LSS, with the long slit in the direction joining A
and B. At each wavelength bin, the spectroscopic data along the slit were fitted to an 1D model 
consisting of two Gaussian profiles with a fixed separation between them. This procedure provided 
spectra for A and B. As the data were not taken along the parallactic angle, differential atmospheric 
refraction (DAR) produced chromatic offsets of both quasar images across the slit \citep{Fil82}, and 
thus wavelength-dependent slit losses. We assumed that the two sources were exactly centred on the 
slit at $\sim$ 6200 \AA\ (acquisition frame in the $r$ band), and then derived DAR-induced slit 
losses and corrected original spectra. Using $g$-band and/or $r$-band fluxes from RATCam/IO:O frames 
(see above), we also accounted for weak spectral contaminations by GAL. Observations with SPRAT at 
five epochs between 2015 and 2017 (the last two are not included in the current version of the GLQ 
database) were used to study the Mg\,{\sc ii} line at 21 epochs. The SPRAT spectra show the Mg\,{\sc 
ii} and C\,{\sc iii}] emission lines, as well as several absorption features (see 
Fig.~\ref{fig:q0957sprat}). Table 9 at the CDS$^{12}$ is structured in the same manner as Table 8, 
but incorporating the fluxes of A and B from the observations with SPRAT. Although 
NOT/ALFOSC spectroscopic data in 2009$-$2013 (four epochs) contain Mg\,{\sc ii}, C\,{\sc iii}] and 
C\,{\sc iv} emission features, the Mg\,{\sc ii} line is near the red edge of the NOT/ALFOSC/\#7 
spectra. We were not able to accurately calibrate the NOT/ALFOSC spectroscopy on 29 January 2009, therefore 
we only extracted usable spectra at three epochs. These NOT/ALFOSC data are presented in Tables 10 (grism 
7) and 11 (grism 14) at the CDS$^{12}$, using the same format and units as Table 9. In addition, 
we were unfortunately unable to infer reliable results for any emission line from the INT/IDS 
spectroscopy on 31 March 2008 because of poor atmospheric seeing. 

\begin{figure}
\centering
\includegraphics[width=9cm]{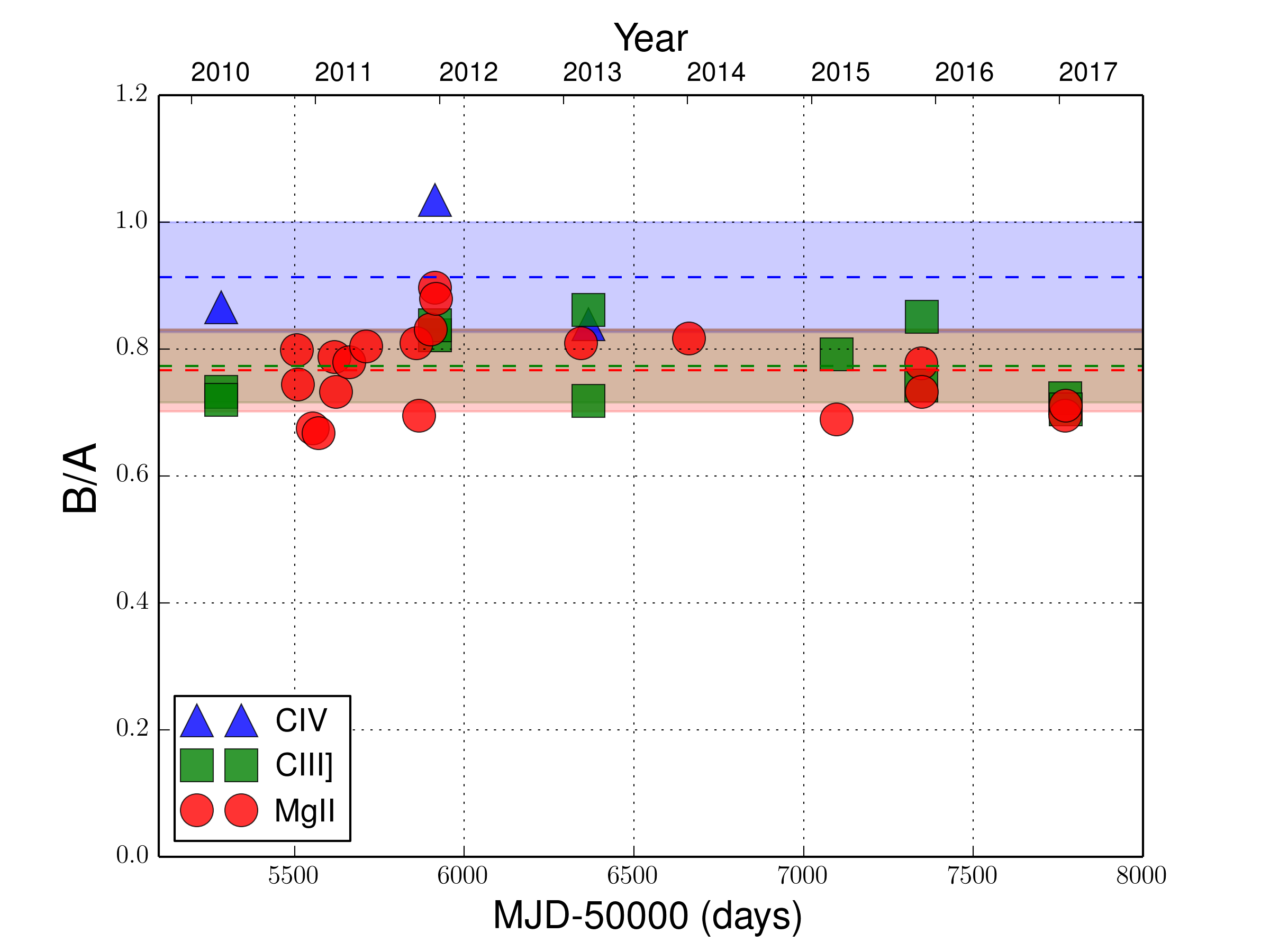}
\caption{Single-epoch flux ratios of emission lines. The blue triangles, green squares, and red circles 
represent the C\,{\sc iv}, 
C\,{\sc iii}], and Mg\,{\sc ii} flux ratios, respectively. 
Average values and standard deviations are marked by dashed lines and bands: blue (C\,{\sc iv}), 
green (C\,{\sc iii}]), and red (Mg\,{\sc ii}). The green and red bands overlap almost 
completely, resulting in a brown band.}   
\label{fig:q0957linefrat}
\end{figure}

In Appendix~\ref{sec:appa3}, we analyse the profiles, fluxes and single-epoch flux ratios of the 
Mg\,{\sc ii}, C\,{\sc iii}], and C\,{\sc iv} emission lines. The single-epoch Mg\,{\sc ii} flux ratios
are marked with red circles in Fig.~\ref{fig:q0957linefrat}. We also show their average value (dashed 
red line) and standard deviation (red band). The average flux ratio is $\langle B/A 
\rangle_{\rm{Mg\,II}}$ = 0.77 $\pm$ 0.02, in good agreement with the macrolens (radio core) flux 
ratio \citep[0.75 $\pm$ 0.02;][]{Gar94} and the first estimation of the delay-corrected Mg\,{\sc ii} 
flux ratio \citep[0.75 $\pm$ 0.02;][]{Sch91}. Although the red circles in 
Fig.~\ref{fig:q0957linefrat} come from fluxes of A and B that are not separated by the time delay 
between the two images, the distribution
of single-epoch flux ratios can be used to determine the delay-corrected value of $B/A$. 
The unaccounted line variability yields biases in both directions (underestimates and 
overestimates), and thus generates a random noise. As we only have three measures of the C\,{\sc 
iv} flux ratio (see the blue triangles in Fig.~\ref{fig:q0957linefrat}), $\langle B/A 
\rangle_{\rm{C\,IV}}$ = 0.91 $\pm$ 0.09 could be a biased estimator of the delay-corrected value. 
However, the statistical result based on$\text{}$ about ten observing epochs of the C\,{\sc 
iii}] line is noteworthy (see the green squares, the dashed green line, and the green band in 
Fig.~\ref{fig:q0957linefrat}): $\langle B/A \rangle_{\rm{C\,III]}}$ = 0.77 $\pm$ 0.02. 

From spectroscopic observations separated by $\sim$ 425 d, that
is, a time lag very close to the measured 
delays between A and B, we can obtain delay-corrected values of $(B/A)_{\rm{Mg\,II}}$ and 
$(B/A)_{\rm{C\,III]}}$ (see Table~\ref{tab:q0957MgIIana} and Table~\ref{tab:q0957CIII]ana}). Thus, 
the FRODOSpec spectra of A and B on 21 December 2011 and 20 February 2013, respectively, were used to 
calculate a typical value $(B/A)_{\rm{Mg\,II}}$ = 0.71 in the first half of this decade. More recent 
SPRAT spectra also allowed us to determine two delay-corrected flux ratios for each emission line: 
$(B/A)_{\rm{Mg\,II}}$ = 0.76 and $(B/A)_{\rm{C\,III]}}$ = 0.80 from the spectra in 
Fig.~\ref{fig:q0957sprat}, and $(B/A)_{\rm{Mg\,II}}$ = 0.73 and $(B/A)_{\rm{C\,III]}}$ = 0.69 from 
the spectrum of A on 21 November 2015 and the spectrum of B on 18 January 2017. A basic statistics 
leads to $\langle B/A \rangle_{\rm{Mg\,II}}$ = 0.73 $\pm$ 0.02 and $\langle B/A 
\rangle_{\rm{C\,III]}}$ = 0.75. These measures and the single-epoch flux ratios for both lines are 
consistent with previous estimates based on analyses of magnesium and carbon emissions 
\citep[e.g.][and $\langle B/A \rangle$ = 0.72 $\pm$ 0.04 from Fig. 13 of Motta et al. 
2012]{Sch91,Goi05}, as well as the macrolens flux ratio of $\sim$ 0.75. Therefore, although the 
behaviour of the C\,{\sc iv} flux ratio during the initial phase of the ongoing microlensing event in 
the continuum is not a clear matter (see above), there is strong evidence that the Mg\,{\sc ii} and 
C\,{\sc iii}] emitting regions have not been affected by dust extinction and microlensing during the 
past 30 years. 

\paragraph{\underline{Deep NIR imaging}}

\begin{figure}
\centering
\includegraphics[width=9cm]{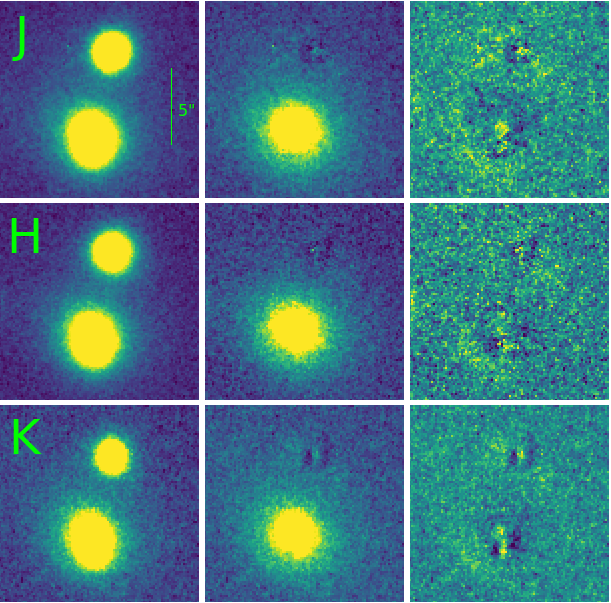}
\caption{TNG/NICS imaging of \object{QSO B0957+561} in subarcsec seeing conditions. The left panels 
show the lens system in the final combined frames with total exposure times of 3.12 ks ($J$ band), 
2.2 ks ($H$ band), and 4.5 ks ($K$ band). The middle panels show the residual signal after subtracting 
only the two quasar images (A+B), and they clearly display the main lensing galaxy. The right panels 
contain the residual light after subtracting the global photometric model (A+B+GAL). See main text 
for details.} 
\label{fig:q0957nir}
\end{figure}

NIR frames of \object{QSO B0957+561} were obtained on 28 December 2007 with the TNG using the 
instrument NICS in imaging mode. All frames were taken with the small field camera, which provides a 
pixel scale of 0\farcs13 and a field of view of 2\farcm2$\times$2\farcm2. We also used three 
different filters $JHK$ covering the spectral range of 1.27$-$2.20 $\mu$m, that is, $\sim$ 5260$-$9110 
\AA\ in the quasar rest frame, and combined individual exposures in each passband to produce final 
frames with subarcsec spatial resolution. The left panels of Fig.~\ref{fig:q0957nir} show the 
strong-lensing region encompassing the three science targets A, B, and GAL. The total exposure times are 
3120, 2200, and 4500 s in the $J$, $H,$ and $K$ bands, respectively. Each combined frame of \object{QSO 
B0957+561} incorporates the lens system and the bright reference star H (which does not appear in 
Fig.~\ref{fig:q0957nir}), so that the PSF can be finely sampled and an accurate PSF-fitting photometry of 
the lens system can be performed. 

\setcounter{table}{11}
\begin{table}
\centering
\caption{Best photometric solutions in the $JHK$ bands.}
\begin{tabular}{lccc}
\hline\hline
                                &       $J$        &       $H$       &       $K$     \\
\hline
$r_{\rm{eff}}$ (\arcsec)        &       4.78       &       3.84      &       3.26    \\
$e$                             &       0.22       &       0.30      &       0.25    \\
$\theta_e$ (\degr)              &       60         &       63        &       55      \\  
$GAL$ (mag)                     &       15.215     &       14.620    &       13.892  \\
$A$ (mag)                       &       16.199     &       15.381    &       15.035  \\
$B$ (mag)                       &       16.198     &       15.453    &       15.148  \\
\hline
\end{tabular}
\tablefoot{
Here, $r_{\rm{eff}}$, $e = 1 - b/a$ and $\theta_e$ are the effective radius, ellipticity, and 
position angle of the major axis (it is measured east of north) of the de Vaucouleurs profile of the
lensing galaxy, and $GAL$, $A,$ and $B$ are the calibrated brightnesses of the galaxy and the 
quasar images. 
} 
\label{tab:q0957nirphot}
\end{table}

As usual, the lens system was modelled as two PSFs (A and B) plus a de Vaucouleurs profile convolved 
with the PSF (GAL). We then determined the structure parameters of the galaxy by setting the 
positions of B and GAL (relative to A) to those derived from HST data in the $H$ band \citep{Kee00}. 
These IMFITFITS structure parameters are shown in Table~\ref{tab:q0957nirphot}. The $J$-band size 
($r_{\rm{eff}}$) is similar to the optical size \citep{Kee98}, and the galaxy is more compact at 
longer wavelengths. Additionally, the $e$ and $\theta_e$ values in the NIR almost coincide with 
previous optical estimates at isophotal radii $>$ 1\arcsec \citep{Ber97,Fad10}. We note that our 
solution in the $H$ band differs from the $H$-band photometric structure reported by \citet{Kee00}, 
since we obtain higher values of $r_{\rm{eff}}$, $e,$ and $\theta_e$. In Fig.~\ref{fig:q0957nir}, we 
display the residual instrumental fluxes after subtracting A+B (middle panels) and A+B+GAL (right 
panels), and the right panels contain arc-like residues resembling the host-galaxy light distribution 
from HST $H$-band observations. Using the $JHKs$ magnitudes of the H star in the Two Micron All Sky 
Survey \citep[2MASS;][]{Skr06} database, we also inferred magnitudes for the galaxy and the two 
quasar images (see the last three rows in Table~\ref{tab:q0957nirphot}). 

Unfortunately, we failed to obtain additional NIR data in early 2009, and thus delay-corrected flux 
ratios could not be probed. However, \citet{Kee00} measured a single-epoch flux ratio $(B/A)_H$ = 
0.93 from observations in 1998, and using data acquired about ten years later, we obtain $(B/A)_H$ = 
0.94 (see Table~\ref{tab:q0957nirphot}). This suggests that the single-epoch flux ratio in the $H$ 
band is stable on long timescales and might be a rough estimator of the delay-corrected value of 
$B/A$. Based on the magnitudes in Table~\ref{tab:q0957nirphot}, the NIR flux ratios of \object{QSO 
B0957+561} vary from 1 ($J$ band) to 0.9 ($K$ band), decreasing as the wavelength increases. For 
comparison purposes, we also analysed mid-IR (MIR) observations in the data archive of the {\it Spitzer} Space 
Telescope. We found two {\it Spitzer} combined frames including \object{QSO 
B0957+561}\footnote{Program Id: 20528, PI: Crystal L. Martin}, each corresponding to a 202 s exposure 
with MIPS at 24 $\mu$m, and the fluxes of the quasar images in both frames were extracted using the 
MOPEX
package\footnote{\url{http://irsa.ipac.caltech.edu/data/SPITZER/docs/dataanalysistools/tools/mopex/}}.
The average fluxes $\langle A \rangle$ = 12135 $\mu$Jy and $\langle B \rangle$ = 8932.5 $\mu$Jy lead  
to a 24 $\mu$m flux ratio of 0.74, in very good agreement with radio and emission-line flux ratios. 
We remark that the radiation observed at 24 $\mu$m is emitted at $\sim$ 10 $\mu$m from a dusty torus
surrounding the AD and broad line emitting regions \citep{Ant93}, and passes through the lens galaxy 
with a wavelength of 17.6 $\mu$m. Thus, this radiation is insensitive to extinction and microlensing. 

\paragraph{\underline{Overview}}

After accumulating data for many years, we are gaining a clearer perspective of the physical 
processes at work on \object{QSO B0957+561}. From a 5.5-year optical monitoring of the two quasar 
images, we found oscillating behaviours of the delay-corrected flux ratios in the $g$ and $r$ bands,
with maximum values of $B/A$ when the flux variations are greater \citep{Sha12}. This result was
crudely related to the presence of dust along the line of sight towards image A (within the 
lensing galaxy) and the emission of highly polarised light during episodes of violent activity. Here, 
we present $r$-band light curves covering 21 years of observations (from 1996 to 2016), which allow 
us to better understand the long-term evolution of $(B/A)_r$. Based on these longer-lasting brightness 
records, $(B/A)_r$ does not seem correlated with flux level or flux variation, meaning that violent 
activity is not the unique driver of changes in $(B/A)_g$ and $(B/A)_r$ (see below). In addition, a 
5.2-year optical polarimetric follow-up does not show any evidence for high polarisation degrees when 
large flux variations occur. In fact, $PD <$ 1\% during the entire monitoring period. 

\citet{Sha12} also reported a chromatic time delay between A and B. They used a standard 
cross-correlation in the $g$ and $r$ bands, and their results were interpreted as being due to 
chromatic dispersion of the A image light by a dusty region inside the lensing galaxy. However, while 
the chromaticity of the time delay from standard techniques does not seem arguable, its 
interpretation is very likely incorrect (see the second paragraph in Sec.~\ref{sec:dustgas}). Very 
recently, \citet{Tie18} proposed that measured delays of a GLQ may contain microlensing-induced 
contributions of a few days, which would depend on the position of the AD across microlensing 
magnificatiion maps. A microlensing-based interpretation for the $\text{approximately}$ three-day 
difference between the 
delays in the $g$ and $r$ bands is unlikely, however. The time delays of 417 d ($g$ band) and 420 d 
($r$ band) are consistent with data from two independent experiments separated by $\sim$ 15 years 
\citep{Kun97,Sha12}, and thus microlensing does not seem to play a relevant role. 

A more plausible scenario  may account for a few observed "anomalies" in \object{QSO 
B0957+561} without a need to invoke highly polarised emission phases, the existence of exotic dust, or 
complex microlensing effects that remain over decades. The UV-visible-NIR continuum observed in the 
quasar comes from the direct UV-visible emission of the AD and the diffuse UV-visible light emitted 
by broad-line clouds (BLC), and this last contribution could be relatively significant 
\citep[e.g.][and references therein]{Kor01}. The continuum of the BLC includes scattered (Rayleigh 
and Thomson) and thermal (Balmer and Paschen recombination) radiation, and high-density gas clouds 
are particularly efficient in producing a diffuse component \citep{Ree89}. From a wider perspective, 
heavily blended iron lines also produce a pseudo-continuum in quasar spectra 
\citep[e.g.][]{Wil85,Mao93}, and recent research has provided evidence for two different emitting 
regions in \object{QSO B1413+117} \citep{Slu15}. The compact emission of this quasar is probably 
scattered by electrons and/or dust in an extended region.

HST spectra of \object{QSO B0957+561} in April 1999 and June 2000 allowed us to construct 
delay-corrected continuum flux ratios $B/A$ at UV-visible-NIR wavelengths during a period of low 
quasar activity and microlensing quiescence \citep{Goi05}. These data and HST emission-line 
ratios are consistent with a simple picture: the direct light of the A image is affected by a compact 
dusty region in the intervening cD galaxy (see the first paragraph in Sect.~\ref{sec:dustgas} and new 
results in this section), which is adopted here for further discussion. As a result, regarding the 
continuum observed in the two quasar images, the diffuse contribution plays a more important role in 
A because its direct light is partially extinguished by dust. In the absence of extended diffuse 
light, the continuum flux ratio at the observed wavelength $\lambda$ is given by 
$B_{\lambda}(t)/A_{\lambda}(t - \Delta t) = 0.75/\epsilon_{\lambda}$, where 0.75 is the macrolens 
ratio and $\epsilon_{\lambda}$ is the dust extinction law. However, assuming that $T_{\lambda}$ is 
the light-travel time (in the observer's frame) between the direct compact source and the clouds 
emitting the observed radiation, as well as a diffuse-to-direct emission ratio $\delta_{\lambda} <<$ 
1, $\epsilon_{\lambda}$ should be replaced by an effective extinction $\epsilon_{\lambda}^{\rm{eff}} 
\approx \epsilon_{\lambda} + (1 - \epsilon_{\lambda})\delta_{\lambda} [B_{\lambda}(t - 
T_{\lambda})/B_{\lambda}(t)]$. During low activity periods, $B_{\lambda}(t - 
T_{\lambda})/B_{\lambda}(t) \sim$ 1 and spectral anomalies in $B/A$ are due to peaks in 
$\delta_{\lambda}$, i.e. $\epsilon_{\lambda}^{\rm{eff}} \sim \epsilon_{\lambda}(1 - \delta_{\lambda}) 
+ \delta_{\lambda}$. Thus, the apparent distortion of the 2175-\AA\ extinction bump (observed around 
2960 \AA) is reasonably related to the Ly$\alpha$ Rayleigh scattering feature, while the flattening 
at NIR wavelengths would be (at least partially) due to a large amount of diffuse emission at $\sim$ 
3000$-$4000 \AA\ \citep[around the Balmer jump; e.g.][]{Kor01}. 
           
The toy model outlined in the previous paragraph can also be used to discuss some time-domain 
anomalies. Considering a central epoch in the time segment TS4 (e.g. the day 5300 in such episode of 
violent variability; see Appendix~\ref{sec:appa1}), we estimated $B_r(t - T_r)/B_r(t) \sim$ 
0.90$-$0.95 if $T_r \sim$ 50$-$100 d \citep[e.g.][]{Gue13}. As the diffuse light contribution to 
$\epsilon_r^{\rm{eff}}$ decreases by $\sim$ 5$-$10\% (with respect to low activity periods), this may 
produce the 4\% increase observed in $(B/A)_r$ at day 5300. Moreover, the diffuse-light term in the
effective extinction decreases by about 9$-$17\% at the $r$-band flux peak, in reasonable agreement  
with the measured increase of a 9\% in the flux ratio at day 5370. Therefore, the model is also able to 
explain the flux ratio anomalies during sharp intrinsic variations of flux. Despite this sucess of 
the AD + BLC scenario in accounting for some local fluctuations in $(B/A)_r$, a microlensing-induced 
variation is taking place in recent years. The time-evolution of the flux ratios in the $gr$ bands is 
thus a powerful tool to constrain the sizes of the $g$- and $r$-band continuum sources, and compare
microlensing-based measures with reverberation mapping ones.  

A central EUV light source most likely drives the variability of \object{QSO B0957+561} 
\citep{Gil12}. Although this ionising radiation cannot be observed directly \citep[e.g.][]{Mic93}, 
its variations are thermally reprocessed within the AD to generate fluctuations in the UV  
emission that are observed in the visible continuum. These EUV and UV variations are also reprocessed 
in the BLC, but extended regions respond later and less coherently than compact ones. Even though 
detailed simulations are required to obtain a realistic description of the BLC emissivity \citep[e.g. 
using CLOUDY models;][]{Fer98,Fer03}, we again used the toy model to obtain some insights into expected 
delays in the AD + BLC scenario. For instance, in the $g$ band, the flux of A at $t - \Delta t$ can  
be related to the fluxes of B at $t$ and $t - T_g$, that is, $A_g(t - \Delta t) \approx 1.33\epsilon_g 
B_g(t) + 1.33(1 - \epsilon_g)\delta_g B_g(t - T_g)$. However, when estimating the time delay $\Delta 
t$ from a standard cross-correlation, we are implicitly assuming a linear relationship $A_g(t - 
\Delta t) = \alpha B_g(t) + \beta$. The key point is that the contamination by diffuse light from the 
gas clouds does not produce a constant added ($\beta$), but a variable one $\beta(t) 
\propto B_g(t - T_g)$. The linear law does not hold here, so we indeed derive an effective time delay 
$\Delta t_g = \Delta t - \tau_g$ instead of the true achromatic value. The amount of deviation 
($\tau_g > 0$) depends on the relative weight and the degree of variability of the delayed signal 
$\beta(t)$. As the relative extinction $(1 - \epsilon)/\epsilon$ and the variability of B increase 
toward shorter wavelengths, it is not at all surprising to measure a lag $\Delta t_r - \Delta t_g = 
\tau_g - \tau_r \sim$ 3 d.

\subsubsection{SDSS J1001+5027}
\label{sec:q1001}

\begin{figure}
\centering
\includegraphics[width=9cm]{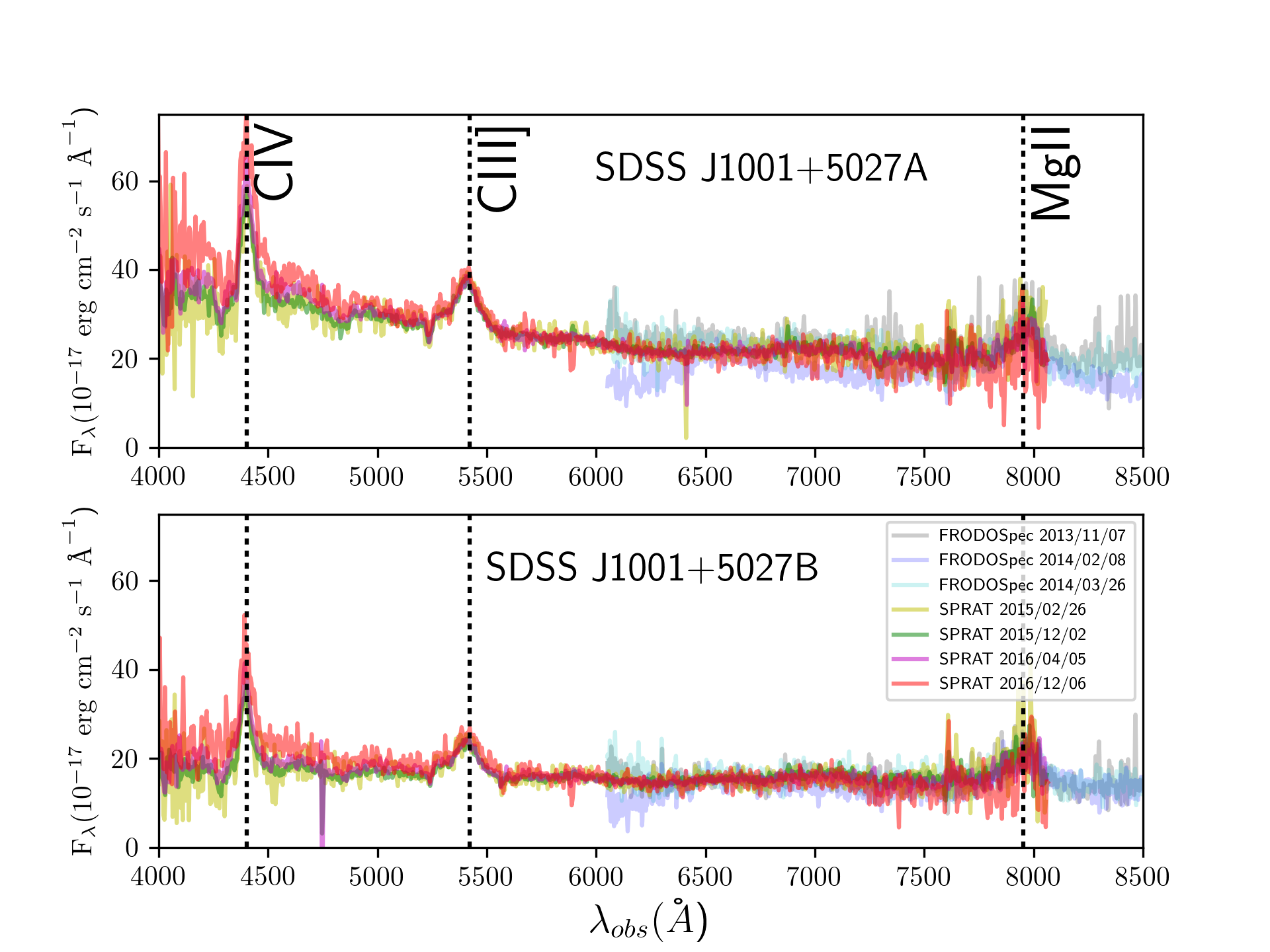}
\caption{FRODOSpec red-grating and SPRAT spectra of \object{SDSS J1001+5027} in 2013$-$2016. We 
display the FRODOSpec spectra after a three-point smoothing filter is applied. Our global dataset 
contains information on the C\,{\sc iv}, C\,{\sc iii}], and Mg\,{\sc ii} emissions at $z$ = 1.838 
\citep{Ogu05}.}
\label{fig:q1001spec}
\end{figure}

The COSMOGRAIL collaboration monitored the two images of \object{SDSS J1001+5027}, measuring a time 
delay of about four months \citep{Rat13}. We therefore interrupted the LT photometric monitoring 
of the GLQ (whose primary goal is determining the time delay) and started spectroscopic 
follow-up observations separated by four months. The first successful spectra of \object{SDSS 
J1001+5027} were obtained using FRODOSpec on 7 November 2013, 8 February 2014, and 26 March 2014. We 
did not obtain data in early June 2014, and thus, by means of FRODOSpec observations, only 
 one comparison between A and B is possible at the same emission time. We took a single 3000 s exposure 
on each of the three nights. The blue grating signal was very noisy, therefore only red grism data were 
considered of astrophysical interest. We also observed the GLQ with SPRAT on several nights. The 
1\farcs8 ($\sim$ 4 pixel) wide slit was oriented along the line joining A and B, and we used the blue 
grating mode. We initially checked the feasibility of the SPRAT programme by taking 3$\times$300 s 
exposures on 26 February 2015. After analysing these tentative data, we decided to use longer 
exposures (4$\times$600 s) per observing run. We then obtained spectroscopic frames at four 
additional epochs: 2 December 2015, 5 April 2016, 6 December 2016, and 5 April 2017, but data on the 
last epoch are not usable. Spectra of the two quasar images were extracted by setting its 
angular separation to that reported by \citet{Rus16} and fitting two 1D Gaussian profiles 
(corrections for DAR-induced spectral distortions and flux calibrations are outlined in the subsection 
on spectroscopy in Sect.~\ref{sec:q0957}). The 
FRODOSpec red-grating spectra can be downloaded from the CDS$^{12}$: Table 13 includes wavelengths 
(\AA) along with fluxes of A and B (10$^{-17}$ erg cm$^{-2}$ s$^{-1}$ \AA$^{-1}$) for each of the three 
observing dates (yyyymmdd). The SPRAT data appear in Table 14 at the CDS$^{12}$, using the same data 
structure as Table 13. All usable LT spectra at seven different epochs are also shown in 
Fig.~\ref{fig:q1001spec}, where the FRODOSpec spectral energy distributions are smoothed with a 
three-point filter to reproduce the 4.6-\AA\ bins of SPRAT.

\begin{figure}
\centering
\includegraphics[width=9cm]{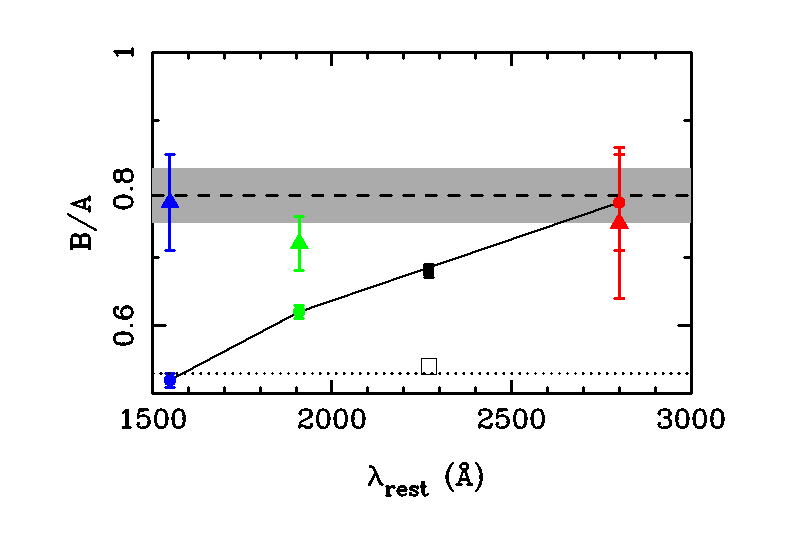}
\caption{Continuum and emission-line flux ratios of \object{SDSS J1001+5027}. The coloured circles 
and triangles describe the results for the continuum and the emission lines, respectively (see main 
text). Blue, green, and red are associated with the continuum (line) at 1549 \AA\ (C\,{\sc 
iv}), 1909 \AA\ (C\,{\sc iii}]), and 2800 \AA\ (Mg\,{\sc ii}). For comparison purposes, the black 
filled and open squares are results from the COSMOGRAIL monitoring in the $R$ band \citep{Rat13}, 
whereas the horizontal dashed line and the grey highlighted rectangle show the $K'$-band flux ratio 
(1$\sigma$ confidence interval) from Subaru Telescope adaptive optics observations \citep{Rus16}.}
\label{fig:q1001contlinefrat}
\end{figure}

Using previous simultaneous spectra of A and B, \citet{Ogu05} noted that the C\,{\sc iv} flux ratio 
$(B/A)_{\rm{C\,IV}}$ was significantly higher than the continuum flux ratio at 1549 \AA\ in the 
quasar rest frame. Moreover, the single-epoch continuum flux ratios were higher for the longer 
wavelengths. In Appendix~\ref{sec:appb}, we analyse the new spectroscopy of \object{SDSS J1001+5027}, 
highlighting the results on the Mg\,{\sc ii}, C\,{\sc iii}], and C\,{\sc iv} emissions, and the 
continuum fluxes at the rest-frame wavelengths of the three emission lines (see 
Tables~\ref{tab:q1001MgIIana}, \ref{tab:q1001CIII]ana}, and \ref{tab:q1001CIVana}). From the 
single-epoch flux ratios in Table~\ref{tab:q1001CIVana}, we infer $\langle B/A \rangle_{\rm{C\,IV}}$
= 0.78 $\pm$ 0.07 and $\langle B/A \rangle_{\rm{cont}}$ = 0.52 $\pm$ 0.01 at 1549 \AA. We also report 
that $\langle B/A \rangle_{\rm{cont}}$ grows from 0.52 at 1549 \AA\ to 0.78 at 2800 \AA, and this 
growing trend becomes clearly apparent in Fig.~\ref{fig:q1001contlinefrat} (coloured circles). The 
new continuum and emission-line flux ratios (coloured triangles represent the results for the 
emission lines) essentially agree with the findings of \citet{Ogu05}. When statistical studies are 
made with only a few data points, the standard deviation of the standard deviation (i.e. the 
uncertainty in the uncertainty) is large. Here, to account for this additional uncertainty when $N$ = 4 
(carbon lines) and $N$ = 3 (magnesium line), the standard deviations of the means were increased by 
40 and 50\%, respectively. These error bar enlargements are also useful to account for possible 
variability effects. Hence, we consider the coloured symbols (graph markers) in 
Fig.~\ref{fig:q1001contlinefrat} as a proxy to the delay-corrected flux ratios. At present, our 
observations on 7 November 2013, 26 March 2014, 2 December 2015, and 5 April 2016 lead to a single 
value for each delay-corrected ratio, which is not depicted in Fig.~\ref{fig:q1001contlinefrat} 
because it might be strongly biased. 

In Fig.~\ref{fig:q1001contlinefrat}, we display $(B/A)_R$ = 0.68 $\pm$ 0.01 \citep[black filled 
square;][]{Rat13} and $(B/A)_{K'}$ = 0.79 $\pm$ 0.04 \citep[dashed line and grey rectangle;][]{Rus16} 
along with the LT flux ratios for the continuum and the emission lines. The emission-line flux ratios 
are in reasonably accord with the continuum flux ratios at the longest wavelengths, that is, 
$(B/A)_{\rm{cont}}$ at 2800 \AA\ and $(B/A)_{K'}$, so that the macrolens flux ratio may lie within the 
grey rectangle depicted in this figure. For this scenario, the behaviour of the continuum flux 
ratios at shorter wavelengths (smaller emitting regions) requires a convincing explanation. The 
observed positive slope in $(B/A)_{\rm{cont}}$ (see the solid line in 
Fig.~\ref{fig:q1001contlinefrat}) could be primarily due to chromatic microlensing 
\citep[e.g.][]{Med09} or extinction by a compact dusty cloud along the line of sight to the B image. 
The relative constancy of the positive $(B/A)_{\rm{cont}}$ gradient over a timescale of $\sim$ 10 
years (see Fig. 2 of Oguri et al.) favours the extinction phenomenon. In addition, the BOSS spectrum 
of B on 1 March 2014 includes a Mg\,{\sc ii} absorption doublet (2796 and 2803 \AA) at the redshift 
of the main lensing galaxy \citep[$z$ = 0.415;][]{Ina12}, which is absent in the SDSS spectrum of A 
on 3 October 2003. This metal ion is most likely associated with dust (see 
Sect.~\ref{sec:dustgas}), suggesting the existence of significant extinction in the B image. 
However, \citet{Rat13} also indicated that $(B/A)_R$ = 0.54 (open square in 
Fig.~\ref{fig:q1001contlinefrat}) if the $R$-band magnitudes of the B image are contaminated by an 
unknown source with a non-variable flux. Then, assuming that contamination increases with 
wavelength, $(B/A)_{\rm{cont}}$ could be almost constant at 1500$-$3000 \AA\ (dotted horizontal 
line). The origin of the observed flux ratios of \object{SDSS J1001+5027} merits further study. 

\subsubsection{QSO B1413+117}
\label{sec:q1413}

The RATCam $r$-band photometry in February-July 2008 (33 epochs) was presented in Table 1 of 
\citet{Goi10}. In this section, we describe additional LT $r$-band observations, outline the most 
relevant processing tasks, and obtain new light curves of the four quasar images (A-D). The LT data 
archive includes usable frames for three observing nights with RATCam in May-June 
2006\footnote{Proposal ID: CL06A07, PI: Evencio Mediavilla}. For each of these nights, we found 6$-$9 
decent 100 s exposures. While the quasar images are bright ($r \sim$ 18 mag), the main lensing galaxy 
is a very faint source \citep[$r \geq$ 23;][]{Kne98}. Therefore, the fluxes in the 
crowded region of each individual frame were extracted using the IMFITFITS software and setting a simple 
photometric model consisting of four close PSFs. The empirical PSF was derived from the S45 field 
star, which was also taken as reference for differential photometry \citep[see the finding chart in 
Fig. 1 of][]{Goi10}. For each quasar image and for the control star S40, we combined as a last step
the individual magnitudes on each night into a single mean value and inferred its error from the 
standard deviation of the mean. 

\begin{figure}
\centering
\includegraphics[width=9cm]{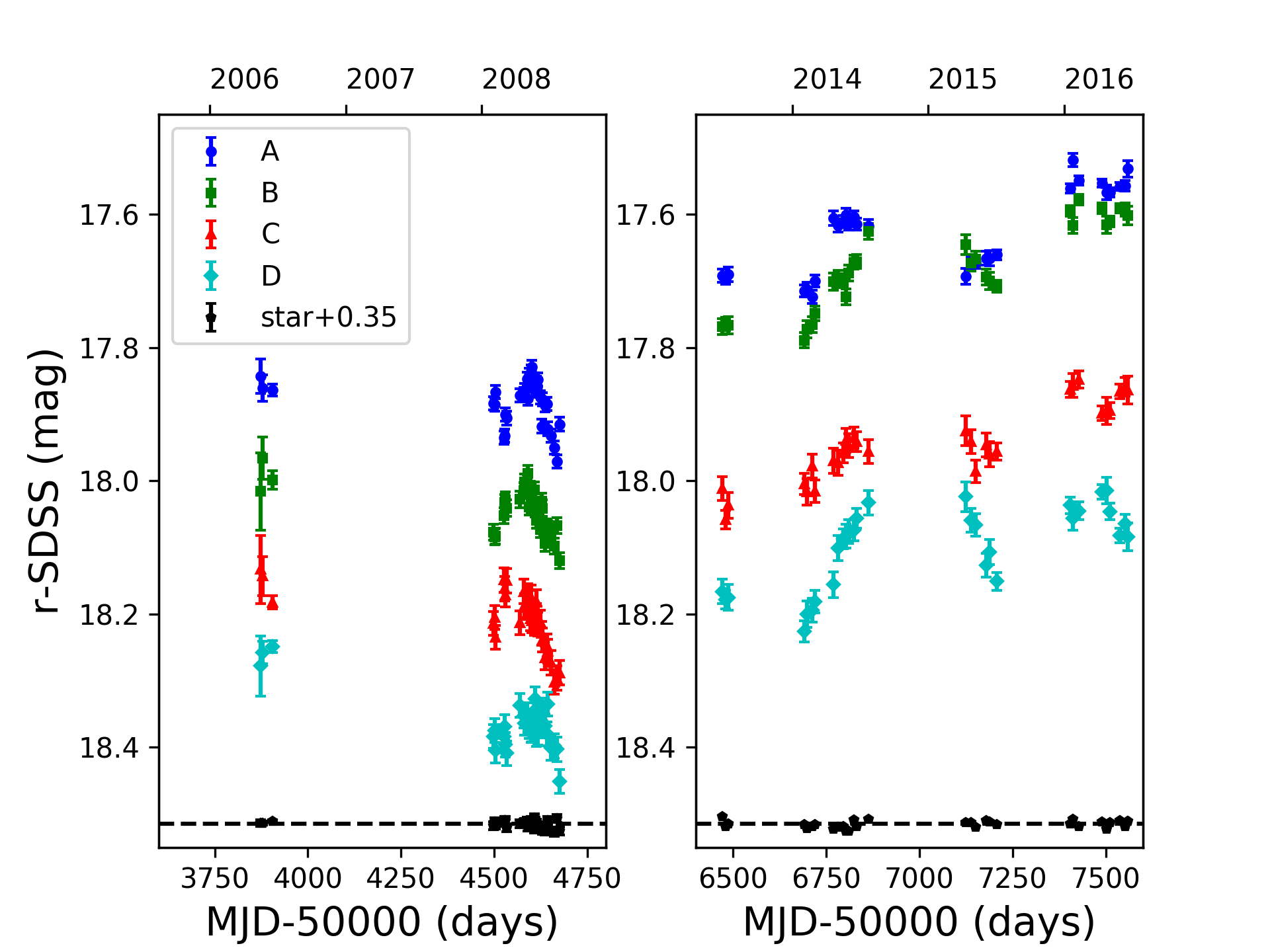}
\caption{LT $r$-band light curves of \object{QSO B1413+117}. The left panel displays RATCam 
photometric data at 3 epochs in May-June 2006 and 33 selected epochs in 2008. The right panel shows 
IO:O data at 30 selected epochs over the period 2013$-$2016. For comparison purposes, we also include 
the brightness record of the field star S40. This object has a magnitude similar to those of the 
quasar images.}
\label{fig:q1413r}
\end{figure}

\begin{figure}
\centering
\includegraphics[width=9cm]{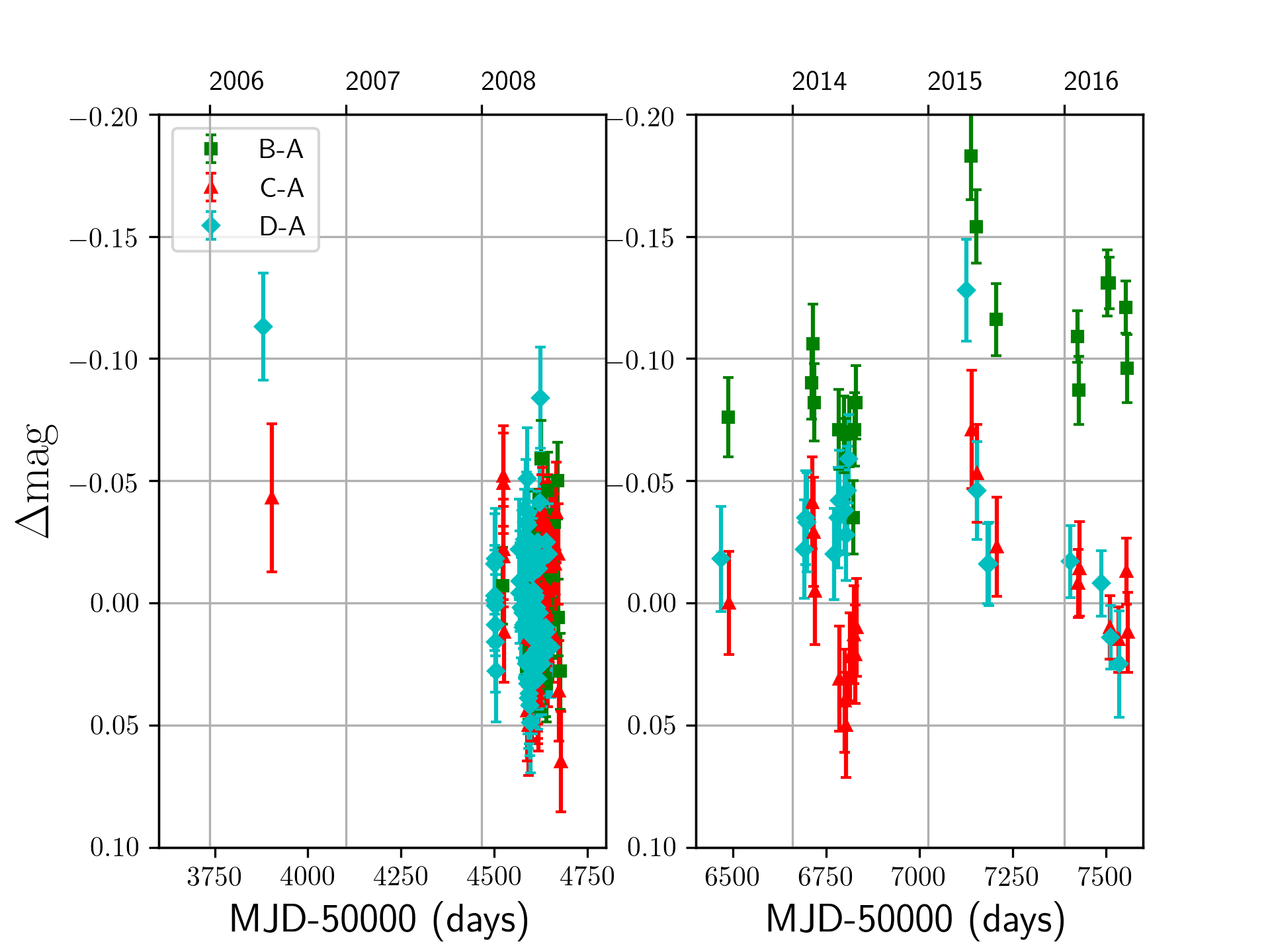}
\includegraphics[width=9cm]{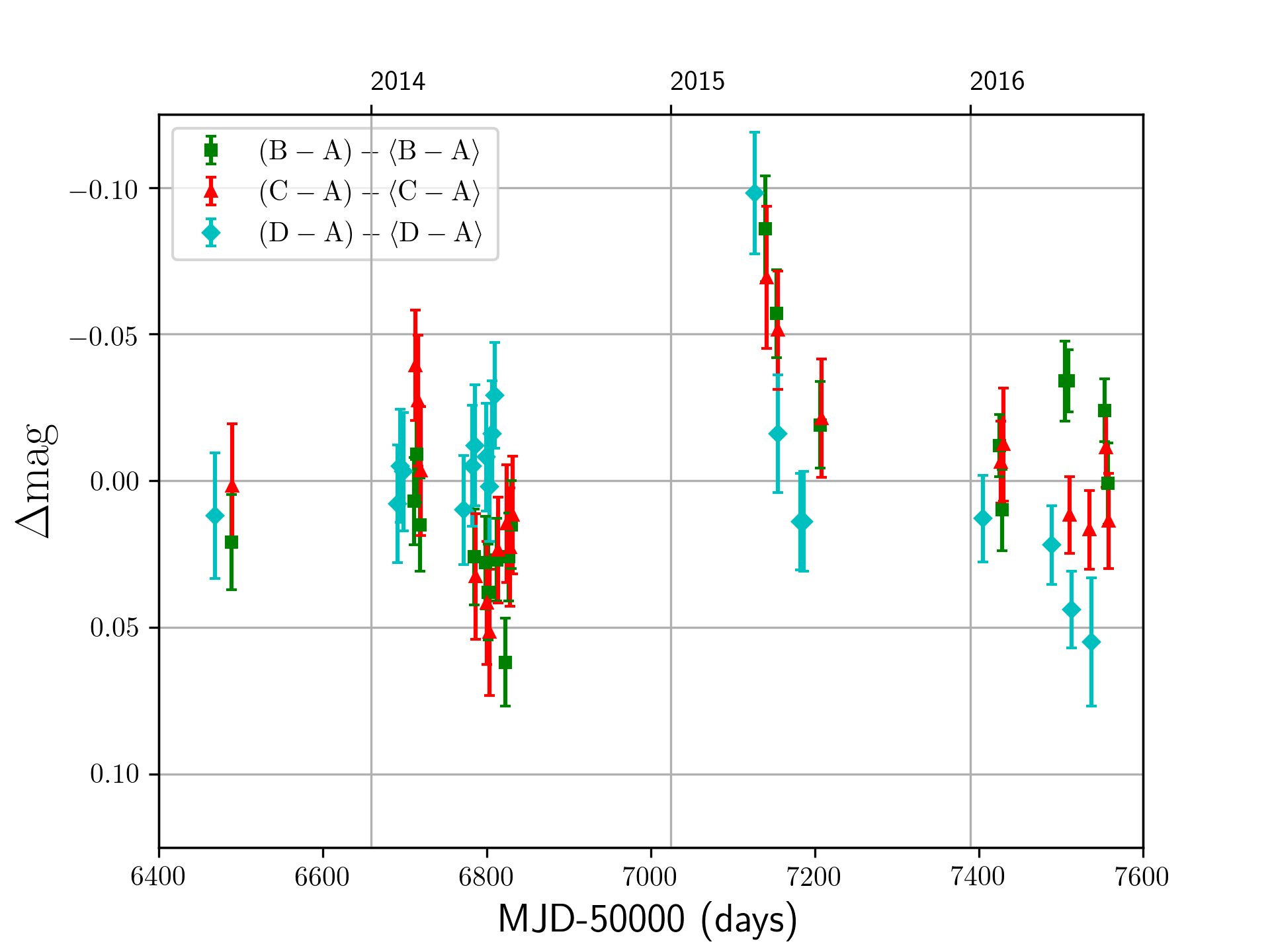}
\caption{DLCs of \object{QSO B1413+117} from LT observations in the $r$ band. To construct the DLCs 
in the top panel, the light curves of the B-D images have been shifted by the time delays and 
magnitude offsets relative to the image A, and the original brightness record of A has later been
subtracted from the shifted curves (only BA, CA, and DA pairs separated by $\leq$ 7 d were considered 
to compute the differences). The bottom panel displays the recent mean-subtracted DLCs.} 
\label{fig:q1413dlc}
\end{figure}      

Our GLQ database also contains acceptable-quality IO:O observations obtained on 59 nights throughout the period 
2013$-$2016. The two consecutive 250 s exposures per observing night were processed separately with 
IMFITFITS, and averaged magnitudes of quasar images (and the star S40) were then computed. To 
construct accurate light curves in 2013$-$2016, the selection criteria were the same as those applied 
to the previous monitoring campaign in 2008: $FWHM <$ 1\farcs5 and $S/N >$ 150. This selection 
procedure removed 29 out of 59 epochs, leaving 30 epochs with high-quality data. We note that the 
statistical properties of the seeing ($\langle FWHM \rangle$ = 1\farcs17 and $\sigma_{FWHM}$ = 
0\farcs15) and the $S/N$ of S40 ($\langle S/N \rangle$ = 207 and $\sigma_{S/N}$ = 42) roughly 
coincide with the corresponding means and standard deviations for the 33 data epochs selected in 2008 
($\langle FWHM \rangle$ = 1\farcs16, $\sigma_{FWHM}$ = 0\farcs17, $\langle S/N \rangle$ = 201 and 
$\sigma_{S/N}$ = 25), so that we have adopted the typical uncertainties in our previous brightness records 
as average photometric errors. We then used weighting factors $\langle S/N \rangle$/$S/N$ to estimate 
errors on every night \citep[e.g.][]{How06}. The $r$-band light curves of A-D covering two months in 
2006 (RATCam) and the period 2013$-$2016 (IO:O; only selected epochs) are available in Table 15 at 
the CDS$^{12}$: Column 1 lists the observing date (MJD$-$50\,000), and Cols. 2$-$3, 4$-$5, 
6$-$7, and 8$-$9 give the magnitudes and magnitude errors of A, B, C, and D, respectively. LT $r$-band 
brightness records of the quadruple quasar \object{QSO B1413+117} and the field star S40 are also 
illustrated in Fig.~\ref{fig:q1413r}, which shows an $\sim$ 0.3 mag intrinsic brightening of the four 
quasar images between the periods 2006$-$2008 and 2013$-$2016. 

We also built DLCs of \object{QSO B1413+117}. Using the time delays and the magnitude offsets between 
the fainter images (B-D) and A \citep{Goi10}, we first derived magnitude- and time-shifted light 
curves of B-D. To obtain the DLCs in the top panel of Fig.~\ref{fig:q1413dlc}, the original light 
curve of image A was then subtracted from these shifted brightness records. We only computed 
magnitude differences from BA, CA, and DA pairs separated by $\leq$ 7 d. In 2006, the DLC for the 
images D and A shows an evident deviation of $\sim$ 0.1 mag from its zero mean level in 2008. 
Although this could be interpreted as a typical microlensing gradient of about 10$^{-4}$ mag d$^{-1}$ 
\citep[e.g.][]{Gay05,Foh07,Sha09}, data in Fig. 7 of \citet{Akh17} make it possible to carry out a 
more detailed analysis. In 2013$-$2016, we also clearly detect an average deviation of $\sim$ 0.1 mag 
in the DLC for the images B and A. Thus, we find evidence of microlensing activity between 2008 and 
2013$-$2016. The recent DLCs, after subtracting their mean values, are plotted in the bottom panel of 
Fig.~\ref{fig:q1413dlc}. A prominent gradient between days 7100 and 7200 is simultaneously observed 
in the three difference curves, which indicates the existence of a significant microlensing variation 
in the image A during the first half of 2015. Unfortunately, there is a long gap around day 7000, so 
we do not have information about the overall shape of this microlensing event. The DLCs only put 
constraints on its amplitude ($\geq$ 0.1 mag) and duration (ranging from one month to one year). 
While previous studies have reported microlensing episodes in the optical continuum of the image D
\citep[e.g.][]{Ost97,Ang08,Slu15,Akh17}, our DLCs demonstrate that other quasar images have also been
affected by microlensing over the last ten years.

\subsubsection{QSO B2237+0305}
\label{sec:q2237}

\object{QSO B2237+0305} has been monitored at optical wavelengths over more than 20 years by several 
large collaborations \citep{Cor91,Ost96,Woz00,Alc02,Sch02,Vak04,Uda06,Eig08}. In 2007, we started our 
own monitoring campaign with the LT. Photometric follow-up observations of the four quasar images 
(A-D) were initially peformed with RATCam in the period 2007$-$2009, and were resumed a few years 
later (from 2013) using IO:O. Regarding the $r$-band data, in the seasons 2007$-$2009, one 300 s 
exposure was taken on most nights. Although a single $r$-band frame per night was also obtained 
during seasons 2013$-$2016, the vast majority of exposures lasted 180$-$200 s. To obtain a better 
perspective on the variability of \object{QSO B2237+0305}, we analysed additional $r$-band frames 
collected at the LT in 2006. These publicly available materials are not incorporated in the current 
version of the GLENDAMA archive and correspond to an independent, short-term LT 
programme\footnote{Proposals ID: CL06A07 and CL06B09, PI: Evencio Mediavilla.}. The star $\alpha$ 
\citep{Cor91} was used to estimate the $S/N$ in all frames, as well as the PSF in many of them. 
However, when the brighter star $\gamma$ was within the field of view and not saturated, we took this 
star to describe the PSF \citep[the star $\gamma$ is located 95\arcsec south from the lens system and 
is also called star 1 in][]{Mor05}.

\begin{figure}
\centering
\includegraphics[width=9cm]{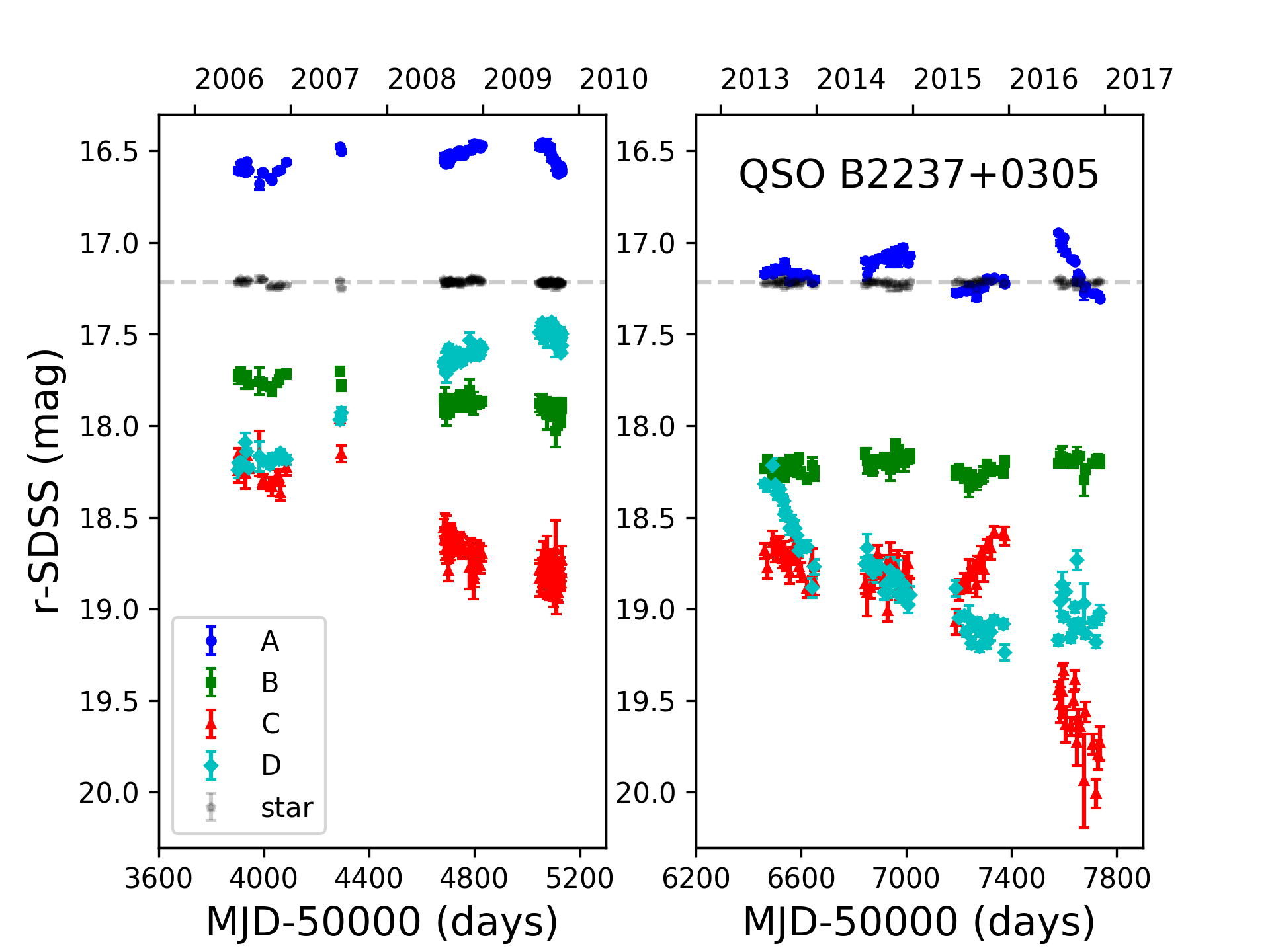}
\caption{LT $r$-band light curves of \object{QSO B2237+0305}. The left panel displays RATCam data at 
110 selected epochs over the period 2006$-$2009. The right panel shows IO:O data at 75 selected 
epochs in 2013$-$2016. Magnitudes of the control star are also shown for comparison (see main text).}
\label{fig:q2237r}
\end{figure}

We performed PSF-fitting photometry on the lens system and field stars. In the strong-lensing region, 
the photometric model consisted of four point-like sources (A-D) and a de Vaucouleurs profile 
convolved with the PSF (lensing galaxy). We set the positions of B-D and the galaxy (relative to A) 
to those derived from HST data in the $H$ band \citep[e.g. Table 1 of][]{Alc02}, and applied the 
IMFITFITS code to the best frames in terms of seeing and $S/N$. The mean values obtained for the 
parameters of the de Vaucouleurs profile ($r_{\rm{eff}}$ = 4\farcs72, $e$ = 0.40 and $\theta_e$ = 
64\degr) were in close agreement with the structure parameters from the best GLITP frames in the $R$ 
band \citep[Table 2 of][]{Alc02}. In a last step, we applied the code to all frames, setting the 
relative positions and the galaxy properties. The individual photometric results were then averaged 
on a nightly basis to calculate $r$-band magnitudes at 185 selected epochs: 16 in 2006, 94 in 
2007$-$2009, and 75 in 2013$-$2016 (including the complete 2016 season, until 14 December). Throughout 
the period from 2007 to 2016, we only discarded 25 epochs (nights) in which our quality requirements 
($FWHM <$ 1\farcs75 and $S/N >$ 100) were not met. Hence, we have robotically observed \object{QSO 
B2237+0305} with an efficiency reaching almost 90\%. Typical errors in the light curves of A-D and 
field stars were estimated following the procedure at the end of Sect. 2 of \citet{Goi10}. This led 
to uncertainties of 0.011, 0.027, 0.053, and 0.026 mag for the A, B, C, and D images, respectively. 
Errors at every epoch were computed by using two weighting factors. Apart from the $\langle S/N 
\rangle / S/N$ ratio \citep[e.g.][]{How06}, we also considered the ratio between the flux of each 
source and its mean value, so the error in a high-flux state is less than that in a low-flux state. 
In more detail, the physically-motivated flux factor was $(\langle FLUX \rangle / FLUX)^{1/2}$, and 
it played a significant role in determining uncertainties in the brightness records of C and D (see 
Fig.~\ref{fig:q2237r}). 

The LT $r$-band dataset covering the period 2006 to 2016 (there is a gap of about 1300 d between the 
end of the first monitoring phase with RATCam and the beginning of the second phase with IO:O) is 
available in Table 16 at the CDS$^{12}$: Column 1 lists the observing date (MJD$-$50\,000), and 
Cols. 2$-$3, 4$-$5, 6$-$7, 8$-$9, and 10$-$11 present the magnitudes and magnitude errors of A, B, 
C, D, and the control star, respectively. These photometric results are also displayed in 
Fig.~\ref{fig:q2237r}. Depending on the star used as PSF tracer and photometric reference ($\gamma$ 
or $\alpha$), we took $\alpha$ or $\beta$ \citep{Cor91} as control star. Therefore, Table 16 (Col. 
10) and Fig.~\ref{fig:q2237r} (black circles) show the magnitudes of the star $\alpha$ and the 
shifted magnitudes of the star $\beta$ (taking an offset $\langle \alpha_r - \beta_r \rangle \sim$ 
0.6 mag into account). Time delays in \object{QSO B2237+0305} are shorter than four days 
\citep{Vak06}, which allows for a direct comparison between the curves of the four quasar images. We 
find sharp, uncorrelated brightness variations that are thought to be due to stars in the central 
bulge of the face-on spiral galaxy acting as a gravitational lens. These individual events and the 
full light curves including additional microlensing variations can be used to prove, among other 
things, the structure of the inner accretion flow in the distant quasar and the average stellar mass 
in the bulge of the nearby spiral galaxy \citep[e.g.][]{Sha02,Koc04}. From 2009 onwards, the LT light 
curves are particularly relevant, since the OGLE collaboration stopped offering photometric data of 
the quadruple quasar at the beginning of the 2009 season. 

\section{Summary and prospects}
\label{sec:final}

We are constructing a publicly available database of ten optically bright GLQs in the northern 
hemisphere. This is fed with materials from a long-term observing programme started in 1999. The 
central idea behind the observational effort is to perform an accurate follow-up of each lens system 
over 10$-$30 years, using mainly the cameras, polarimeters, and spectrographs on the GTC and LT 
facilities at the RMO (see Table~\ref{tab:facil}). Such a long programme is required, among other 
things, to find periods of high microlensing activity in most targets \citep{Mos11}. The database 
currently incorporates $\sim$ 6600 processed frames of 9 GLQs in the period 1999$-$2016 (see 
Tables~\ref{tab:glqs} and \ref{tab:data}), and we intend to reach the 10\,000 frames in the next 
update of the archive (late 2019/early 2020). We remark that this is a singular initiative in the 
research field of GLQs, since other groups do not offer freely
accessible well-structured archives 
with a variety of astronomical materials this rich. In addition
to frames that are ready for astrometric 
and photometric tasks, spectral extractions, or polarisation measurements, this paper also presents 
high-level data products for six of the ten objects in the GLQ sample: \object{QSO B0909+532}, 
\object{FBQS J0951+2635}, \object{QSO B0957+561}, \object{SDSS J1001+5027}, \object{QSO B1413+117,} 
and \object{QSO B2237+0305}. We have published results for two objects in the sample in the last 
year. GTC-LT data provided evidence of two extreme cases of microlensing activity in two double 
quasars at $z \sim$ 2 \citep[\object{SDSS J1339+1310} and \object{SDSS J1515+1511};][]{Goi16,Sha17}. 
In addition, \object{PS J0147+4630} has been discovered very recently \citep{Ber17,Lee17,Rub17} and 
is being monitored with the LT since August 2017. We are also completing a detailed analysis of 
several physical properties of \object{SDSS J1442+4055} (e.g. the time delay between quasar images 
and the redshift of the primary lensing galaxy; see Table~\ref{tab:glqs}), which will be presented in 
a subsequent paper.

Our main results for the six individual objects are listed below.
\begin{itemize}
\item {\it \object{QSO B0909+532}}: We extend previously-published $r$-band brightness records 
\citep[covering the period from 2005 through early 2012;][]{Hai13} by adding new LT magnitudes in the 
period 2012$-$2016. The global LT light curves in the $r$ band are available in Table 4 at the CDS. 
In recent years (2012$-$2016), we detected microlensing variations with an amplitude of $\sim$ 0.1 mag,
which might be useful to improve the Hainline et al. constraint on the size of the $r$-band 
continuum source. We note that our database also contains recent $g$-band frames (see 
Table~\ref{tab:data}).
\item {\it \object{FBQS J0951+2635}}: Table 5 at the CDS incorporates LT-NOT-MAO $r$-band light 
curves covering 2001 to 2016. Despite the relatively low cadence of observations and the 
existence of a gap of $\sim$ 1000 d, these data are critical to trace the long-timescale microlensing 
event in the system \citep{Par06,Sha09}. The single-epoch magnitude differences $A - B$ have evolved 
from a value close to $-$1.1 mag in March 2002 to $-$1.4 mag in January 2016, and this last $r$-band 
difference coincides with the single-epoch flux ratio of the Mg\,{\sc ii} emission line 
\citep{Jak05}.  
\item {\it \object{QSO B0957+561}}: Table 6 at the CDS shows IAC80-LT $r$-band brightness records   
spanning about 21 years (1996$-$2016). These records reveal the presence of an ongoing microlensing 
event, which offers a long-awaited opportunity to simultaneously measure the size of the $g$- and 
$r$-band continuum sources (the database includes frames in both passbands; see 
Table~\ref{tab:data}), as well as the mass distribution in the cD galaxy acting as a gravitational 
lens. We are now starting to use our two-colour light curves of the first GLQ to obtain information 
on the structure of sources and lens. The future constraints on size of sources will be compared to 
current results from microlensing analyses \citep[e.g.][]{Ref00,Hai12} and reverberation mapping 
techniques \citep{Gil12,Goi12}. The main results from LT polarimetric observations are presented in 
Table~\ref{tab:q0957pol}. The optical polarisation degree of the two quasar images is $<$ 1\% over 
the monitoring campaign from late 2011 to early 2017, where the A image has a larger polarisation 
amplitude of $\sim$ 0.5\% \citep[for the B image, the amplitude is consistent with zero; see, 
however, very early measures by][]{Wil80}. In Sect.~\ref{sec:q0957} (Overview), we proposed a scenario 
that may explain this "polarisation excess" in A in addition
to other observed "anomalies". Tables 8$-$9 at
the CDS include LT spectra in 2010$-$2017, while Tables 10$-$11 at the CDS incorporate NOT spectra in 
2010$-$2013. This spectroscopic follow-up allows us to unambiguously confirm that the Mg\,{\sc ii} 
and C\,{\sc iii}] emitting regions do not suffer dust extinction or microlensing effects, 
since the Mg\,{\sc ii} and C\,{\sc iii}] flux ratios over the past 30 years are in very good 
agreement with those obtained at MIR and radio wavelengths (see subsections on spectroscopy and deep
NIR imaging in Sect.~\ref{sec:q0957}). New photometric solutions in the $JHK$ bands are also shown in 
Table~\ref{tab:q0957nirphot}. 
\item {\it \object{SDSS J1001+5027}}: Tables 13$-$14 at the CDS contain LT spectra covering the 
period November 2013 to December 2016. These data along with results in the discovery paper 
\citep{Ogu05}, NIR magnitudes \citep{Rus16} and spectra in the SDSS database favour the presence of a 
compact dusty cloud along the line of sight of B, although other scenarios cannot be completely ruled 
out. The ongoing spectroscopic programme will be continued in the coming years to try to distinguish
contamination, dust extinction, and macro- and micro-lens magnification effects. 
\item {\it \object{QSO B1413+117}}: New LT light curves in the $r$ band (in 2006 and 2013$-$2016) are 
available in Table 15 at the CDS. These complement previous $r$-band magnitudes spanning several 
months in 2008 \citep{Goi10}. We detect microlensing activity between 2006 and 2008, as well as 
between 2008 and 2013$-$2016. Additionally, it is worth mentioning that we find a microlensing event 
in the A image during 2014$-$2015. Such extrinsic variation has an amplitude exceeding 0.1 mag, while 
its duration is between 1 and 12 months.  
\item {\it \object{QSO B2237+0305}}: Table 16 at the CDS shows LT $r$-band brightness records over 
two four-year intervals: 2006$-$2009 and 2013$-$2016, which have plenty of microlensing variability. 
These light curves and additional data in the $g$ band (see Table~\ref{tab:data}) are promising tools
to improve knowledge on the structure of the accretion disc in the distant quasar and the composition 
of the bulge of the spiral lens galaxy at $z <$ 0.1 \citep[e.g.][]{Koc04}. The second data interval 
(2013$-$2016) has a special relevance because there are no OGLE $V$-band magnitudes in recent years.
\end{itemize}

The final GLQ database in the second half of the 2020s will help astronomers to delve deeply into the 
structure of distant active galactic nuclei, the mass distribution in galaxies at different redshifts 
and the cosmological parameters \citep[e.g.][]{Schn92,Schn06}. In addition to frames for the ten targets 
in the GLQ sample, the GLENDAMA global archive will also include observations of binary quasars and 
other non-lensed objects, and even of newly discovered GLQs where one or more images are fainter than 
$r$ = 20 mag (e.g. \object{SDSS J1617+3827}). In addition to the current telescopes at 
the RMO, we will try to use the successor of the LT \citep[LT2;][]{Cop15} and the World Space 
Observatory-Ultraviolet \citep[WSO-UV;][]{Shu14}. These two new facilities will be operational in the 
first half of the next decade, and the UV space telescope may provide details on continuum sources in 
the surroundings of supermassive black holes. So far, the optical polarimetry has mainly been focused 
on a widely separated GLQ (\object{QSO B0957+561}). However, our database also contains broad-band 
polarimetric observations with the LT of three other systems with $\Delta \theta \sim$ 2$-$3\arcsec: 
\object{SDSS J1001+5027}, \object{SDSS J1339+1310,} and \object{QSO B2237+0305}, and we are developing 
a new method for reducing these frames (pixel size of $\sim$ 0\farcs45 and normal seeing conditions). 
We will also try to measure polarisations with better spatial resolution at telescopes other than the 
LT. Whereas $V$-band polarisation degrees of some unresolved GLQs were obtained by \citet{Hut98} and 
\citet{Slu05}, \citet{Cha01} and \citet{Hut10} determined $V$-band polarisations of the four images 
of \object{QSO B1413+117} through observations at high spatial resolution. 

We add a remark on the expected role of our GLQ database in cosmological studies. Seventy percent of the 
targets is being photometrically monitored in an intensive way at certain periods, which will allow 
users to determine delays between images in different time segments throughout the entire duration of 
the project. This procedure is useful for checking time-segment-dependent biases 
\citep[e.g.][]{Tew13,Tie18} and obtaining unbiased measures of time delays for cosmology. As shown by 
\citet{Sha12} \citep[see also][]{Kun97}, chromatic biases are also possible, that is, different time 
delays at different wavelengths, and thus the final archive will incorporate data to check for 
chromaticity in delays of at least three systems. Robustly measured delays from the new database (see
current results in Table~\ref{tab:glqs}) and other ongoing monitoring campaigns (e.g. the COSMOGRAIL 
project\footnote{\url{https://cosmograil.epfl.ch/}}) will shed light on an unbiased value of $H_0$ 
and additional cosmological quantities. Despite this optimistic perspective, some problems remain, and 
they need to be fixed. For example, the spectroscopic redshift of the main lens in \object{QSO 
B1413+117} is unknown. One must also avoid the bias introduced by unaccounted mass along GLQ 
sightlines, since $H_0$ can be noticeably overestimated when line-of-sight deflectors are ignored 
\citep[e.g.][]{Wil17}. Finally, the overall experience gained from ongoing projects will be a basic 
tool to decide on future time-domain observations of large collections of GLQs, which will lead to 
robust estimates of $H_0$, as well as the amount of dark matter and dark energy in the Universe 
\citep[e.g.][]{Ogu10,Tre16}. 

\begin{acknowledgements}
We thank the anonymous referee for carefully reading our long manuscript and for several useful 
comments. We also thank the Universidad de Cantabria (UC) web service for making it possible the 
GLENDAMA global 
archive. We acknowledge A. Ull\'an for doing observations with the Telescopio Nazionale Galileo (TNG) 
and processing raw frames of the Gravitational Lenses International Time Project. We are indebted to 
C.J. Davis, J. Marchant, C. Moss and R.J. Smith for guidance in the preparation of the robotic 
monitoring programme with the Liverpool Telescope (LT). We also acknowledge the staff of the LT for 
their development of the Phase 2 User Interface (which allows users to specify in detail the 
observations they wish the LT to make) and data reduction pipelines. The LT is operated on the island 
of La Palma by Liverpool John Moores University in the Spanish Observatorio del Roque de los 
Muchachos (ORM) of the Instituto de Astrof\'isica de Canarias (IAC) with financial support from the 
UK Science and Technology Facilities Council. We thank the support astronomers and other staff of the 
observatories in the Canary Islands (J.A. Acosta, C. Alvarez, T. Augusteijn, R. Barrena, A. Cabrera, 
R.J. C\'ardenes, 
R.Corradi, J. Garc\'ia, T. Granzer, J. M\'endez, P. Monta\~{n}\'es, T. Pursimo, R. Rutten, M. R. 
Zapatero and C. Zurita, among others) for kind interactions regarding several observing programmes at 
the ORM and the Observatorio del Teide (OT). Based on observations made with the Gran Telescopio 
Canarias, installed at the Spanish ORM of the IAC, in the island of La Palma. This archive is also 
based on observations made with the Isaac Newton Group of Telescopes (Isaac Newton and William 
Herschel Telescopes), the Nordic Optical Telescope and the Italian TNG, operated on the island of La 
Palma by the Isaac Newton Group, the Nordic Optical Telescope Scientific Association and the 
Fundaci\'on Galileo Galilei of the Istituto Nazionale di Astrofisica, respectively, in the Spanish 
ORM of the IAC. We also use frames taken with the IAC80 and STELLA 1 Telescopes operated on the 
island of Tenerife by the IAC and the AIP in the Spanish OT. We also thank the staff of the 
Chandra X-ray Observatory (CXO; E. Kellogg, H. Tananbaum and S.J. Wolk) and Swift Multi-wavelength 
Observatory (SMO; M. Chester and N. Gehrels) for their support during the preparation and execution 
of the monitoring campaign of QSO B0957+561 in 2010. The CXO Center is operated by the Smithsonian 
Astrophysical Observatory for and on behalf of the National Aeronautics and Space Administration 
(NASA) under contract NAS803060. The SMO is supported at Penn State University by NASA contract 
NAS5-00136. This publication makes use of data products from the Two Micron All Sky Survey (2MASS), 
which is a joint project of the University of Massachusetts and the Infrared Processing and Analysis 
Center/California Institute of Technology, funded by the NASA and the National Science Foundation. We 
also used data taken from the Sloan Digital 
Sky Survey (SDSS) database. SDSS is managed by the Astrophysical Research Consortium for the 
Participating Institutions of the SDSS Collaboration. The SDSS web site is www.sdss.org. Funding for 
the SDSS has been provided by the Alfred P. Sloan Foundation, the Participating Institutions, and 
national agencies in the U.S. and other countries. SDSS acknowledges support and resources from the 
Center for High-Performance Computing at the University of Utah. We are grateful to both 
collaborations (2MASS and SDSS) for doing those public databases. The construction of the archive has 
been supported by the GLENDAMA project and a few complementary actions: PB97-0220-C02, 
AYA2000-2111-E, AYA2001-1647-C02-02, AYA2004-08243-C03-02, AYA2007-67342-C03-02, AYA2010-21741-C03-03
and AYA2013-47744-C3-2-P, all them financed by Spanish Departments of Education, Science, Technology 
and Innovation, "Lentes Gravitatorias y Materia Oscura" financed by the SOciedad para el DEsarrollo 
Regional de CANtabria (SODERCAN 
S.A.) and the Operational Programme of FEDER-UE, and AYA2017-89815-P financed by MINECO/AEI/FEDER-UE.
RGM acknowledges grants of the AYA2010-21741-C03-03 and AYA2013-47744-C3-2-P subprojects to develop 
the core software of the database. This archive has been also possible thanks to the support of the 
UC. 

\end{acknowledgements}

\clearpage

\begin{appendix}

\section{Pilot programme for QSO B0957+561: data analysis}
\label{sec:appa}

\subsection{Flux ratio in the $r$ band}
\label{sec:appa1}

Instead of the magnitudes in Table 6, we used the corresponding fluxes (in mJy) and a time delay of 
420 d to study the delay-corrected flux ratio $B/A$ in the $r$ band. To evaluate this ratio at 
different epochs, we compared the fluxes of the B image and the fluxes of the A image shifted by +420 
d. Because the shifted epochs of A generally do not coincide with those of B, we made bins in A 
around the epochs of B. These bins had semisizes $\alpha$ = 1$-$12 d. \citet{Sha12} considered four 
time segments (observing seasons) of B that were called TS1, TS2, TS3, and TS4, and in this paper, we 
extend our previous analysis by incorporating 16 additional segments (see Table~\ref{tab:q0957frat}). 
We note that TS0 corresponds to the season 2005/2006, in which the transition from IAC80 Telescope to 
LT took place.

\begin{table}[!ht]
\centering
\caption{Long-term evolution of the $r$-band flux ratio $B/A$.}
\begin{tabular}{ccccc}
\hline\hline
TS\# (season)\tablefootmark{a} & $\chi^2_0$/dof\tablefootmark{b}
& $B/A$\tablefootmark{c} & $\langle t_{\rm{B}} \rangle$\tablefootmark{d} \\
\hline
TS-9 (1996/1997)                                        & 1.06 & 1.028 $\pm$ 0.010 &    583.1\\
TS-8 (1997/1998)\tablefootmark{$\star$} & 0.72 & 1.038 $\pm$ 0.008 &  885.3\\
TS-7 (1998/1999)                                        & 1.17 & 1.021 $\pm$ 0.005 &   1254.1\\
TS-6 (1999/2000)                                        & 1.06 & 1.022 $\pm$ 0.007 &   1605.1\\
TS-5 (2000/2001)\tablefootmark{$\star$} & 1.00 & 1.039 $\pm$ 0.006 & 1969.4\\
TS-4 (2001/2002)                                        & 0.69 & 1.008 $\pm$ 0.006 &   2394.4\\
TS-3 (2002/2003)                                        & 0.82 & 0.995 $\pm$ 0.012 &   2747.8\\
TS-2 (2003/2004)                                        & 1.34 & 0.992 $\pm$ 0.009 &   3052.8\\
TS-1 (2004/2005)\tablefootmark{$\star$} & 1.00 & 1.011 $\pm$ 0.007 & 3450.1\\
TS0  (2005/2006)                                        & 0.99 & 1.026 $\pm$ 0.009 &   3849.4\\
TS1  (2006/2007)\tablefootmark{$\star$} & 1.00 & 1.022 $\pm$ 0.004 & 4162.6\\
TS2  (2007/2008)                                        & 1.05 & 1.054 $\pm$ 0.004 &   4556.6\\        
TS3  (2008/2009)                                        & 0.89 & 1.016 $\pm$ 0.003 &   4954.2\\
TS4  (2009/2010)                                        & 2.81 & 1.060 $\pm$ 0.004 &   5276.8\\
TS5  (2010/2011)                                        & 0.67 & 1.060 $\pm$ 0.008 &   5656.6\\
TS6  (2011/2012)                                        & 1.18 & 1.086 $\pm$ 0.017 &   6001.2\\
TS7  (2012/2013)                                        & 2.05 & 1.127 $\pm$ 0.010 &   6348.5\\
TS8  (2013/2014)                                        & 0.99 & 1.162 $\pm$ 0.007 &   6727.6\\
TS9  (2014/2015)                                        & 0.93 & 1.170 $\pm$ 0.007 &   7091.8\\
TS10 (2015/2016)                                        & 1.99 & 1.178 $\pm$ 0.010 &   7458.4\\
\hline
\end{tabular}
\tablefoot{
\tablefoottext{a}{Time segment and observing season of B};
\tablefoottext{b}{reduced chi-square for the best fit (dof = degrees of freedom)};
\tablefoottext{c}{formal 2$\sigma$ confidence interval};
\tablefoottext{d}{average epoch of B (MJD$-$50\,000) in the overlapping period between A(+420 d) and 
the time segment}.\\
\tablefoottext{$\star$}{Two bin semisizes lead to fits of similar quality, so we cite the average 
values of $\chi^2_0$/dof, $B/A$ and $\langle t_{\rm{B}} \rangle$ using both semisizes.} 
} 
\label{tab:q0957frat}
\end{table}

We used a $\chi^2$ minimisation to find the flux ratio for each time segment. In 
Table~\ref{tab:q0957frat}, we give the best solutions and their reduced chi-square values. This 
Table~\ref{tab:q0957frat} also contains the 2$\sigma$ intervals for $B/A$, where each interval 
includes all values of $B/A$ satisfying the condition $\chi^2 \leq \chi^2_0$ + 4. We obtained 
$\chi^2_0$/dof $\sim$ 2$-$3 for the segments TS4, TS7, and TS10, and thus the formal uncertainties 
for these periods should be taken with caution. The AB comparisons for the 20 time segments, that is, 
from TS-9 to TS10, are shown in the panels of Fig.~\ref{fig:q0957frat1}. Taking the best solutions of 
$B/A$ to amplify/reduce the time-delay shifted signal A, both A and B signals are compared to each 
other in these panels (A = filled circles and B = open red circles). If we exclusively focus on the LT 
photometry during the last ten years, our simple scenario ($B/A$ is constant within each segment) does 
not work on TS4, TS7, and TS10. In addition to best solutions associated with reduced chi-square values 
ranging from 2 to 2.8, we see some anomalies in these periods. The simple scenario does not 
convincingly explain the observations, since the variations in A seem to be smoother than those in B 
(see Fig.~\ref{fig:q0957frat1}).   

\begin{figure*}
\centering
\includegraphics[width=18cm]{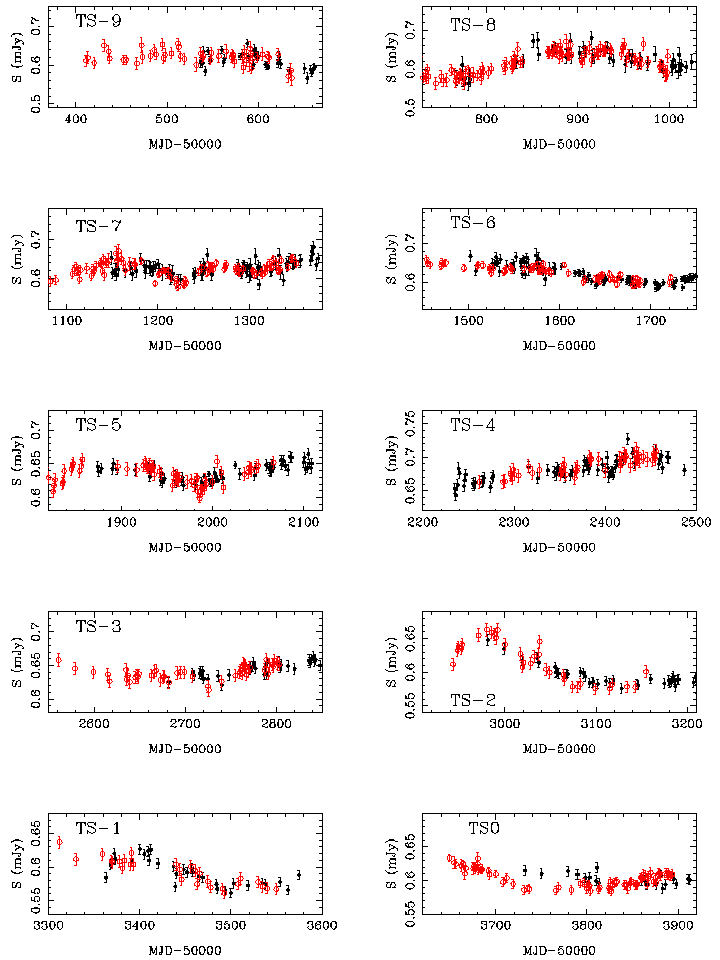}
\caption{AB comparisons in the $r$ band. We show overlapping periods between the flux records of A 
(filled circles) and B (open red circles), where the fluxes of A are shifted by +420 d and properly 
amplified/reduced (see main text).}
\label{fig:q0957frat1}
\end{figure*}

\setcounter{figure}{0}
\begin{figure*}
\centering
\includegraphics[width=18cm]{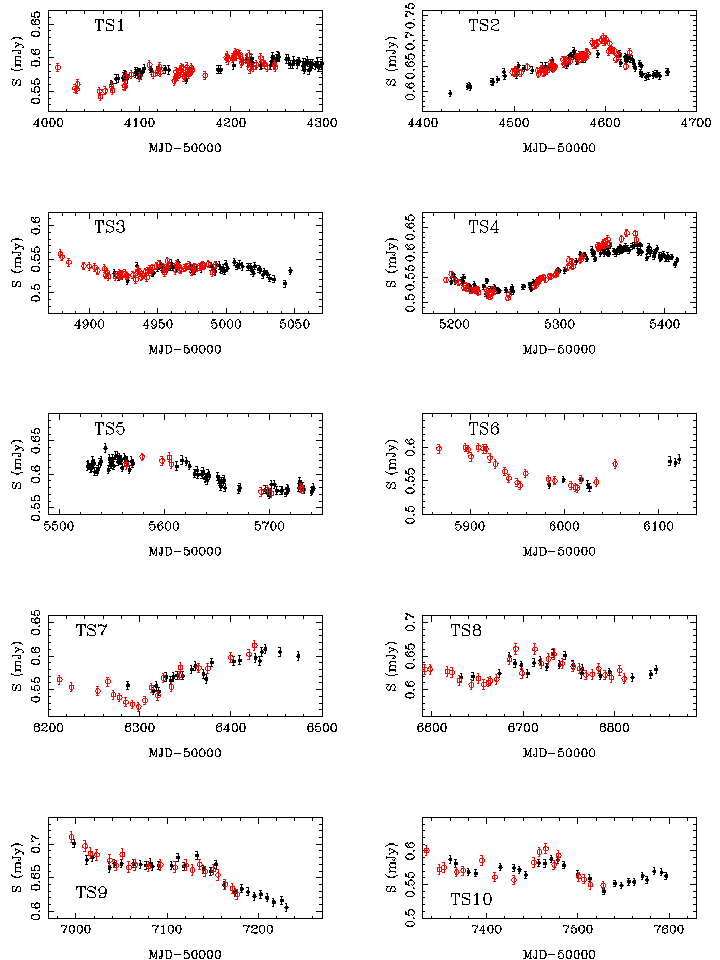}
\caption{({\it continued}) AB comparisons in the $r$ band. We show overlapping periods between the 
flux records of A (filled circles) and B (open red circles), where the fluxes of A are shifted by 
+420 d and properly amplified/reduced (see main text).}
\label{fig:q0957frat2}
\end{figure*}

\newpage

\subsection{Broad-band polarimetric follow-up}
\label{sec:appa2}

To carry out photometric and polarimetric reduction of RINGO2 data (three epochs; see imaging polarimetry 
in Sec.~\ref{sec:q0957}), we firstly extracted instrumental fluxes (in counts) of objects of interest 
in each of the eight stacked frames at each epoch. The fluxes were extracted on the 24 
frames using IMFITFITS. The IMFITFITS software produced PSF fitting photometry of the two quasar 
images and several field stars. Although the RINGO2 re-imaging optics causes a PSF depending on the 
position in the field \citep{Ste17}, this effect does not seem very relevant in our study. The PSF 
star and the fitted objects are separated by only $\leq$ 1\arcmin, and some aperture photometry tests 
with field stars led to polarisation parameters similar to those from PSF fitting. In general, 
aperture photometry is the best method to extract fluxes, but here we study a crowded 
region. In a second step, the eight fluxes per object at each epoch were used to calculate the 
corresponding normalised Stokes parameters ($q = Q/I$, $u = U/I$). We combined 
measurements according to the equations in \citet{Cla02} and \citet{Jer16}. However, these Stokes 
parameters must be corrected from instrumental (device-dependent) biases to obtain the true 
polarisation. For example, \citet{Jer16} and \citet{Ste17} reported on different mean instrumental 
polarisations over four periods of time, and our observations were performed during the third period 
they studied. After comparing the ($q$, $u$) values for the H field star and parameter distributions 
of standard zero-polarised stars, we reasonably assumed that the H star is an unpolarised object. At 
each epoch, the instrumental polarisation ($q_{\rm{H}}$, $u_{\rm{H}}$) was then subtracted from the 
($q$, $u$) values of the quasar images. 

\begin{figure}
\centering
\includegraphics[width=9cm]{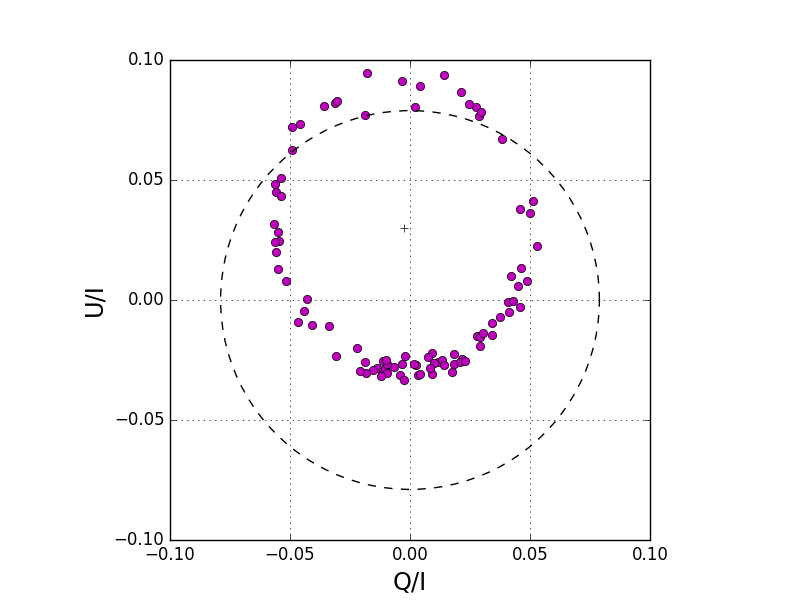}
\includegraphics[width=9cm]{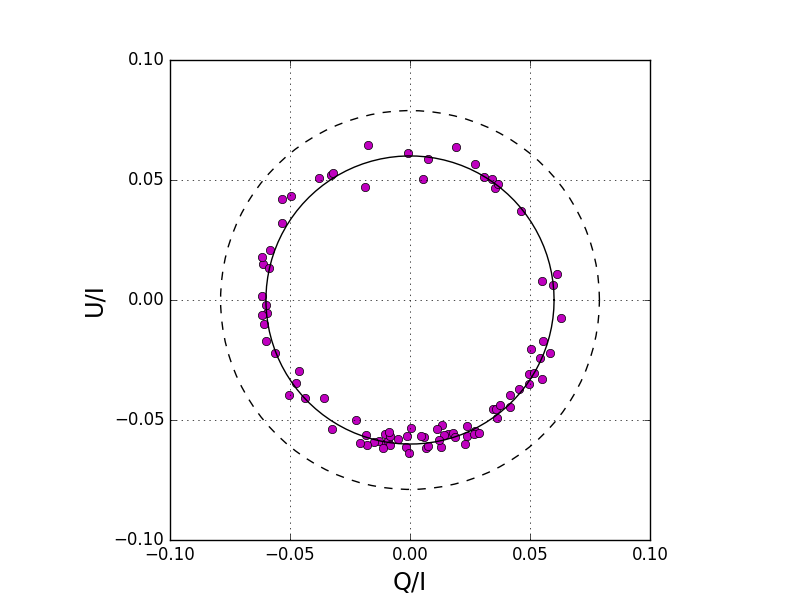}
\includegraphics[width=9cm]{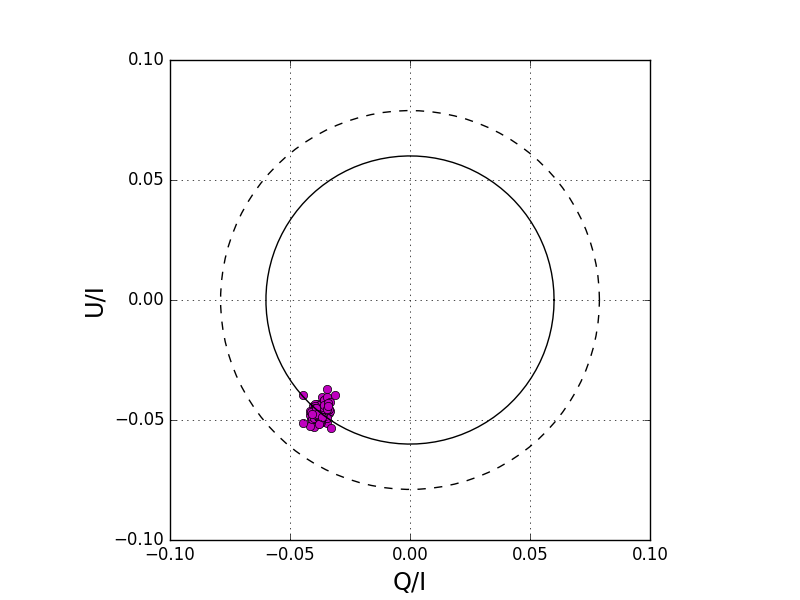}
\caption{RINGO2 data of the polarised star VICyg12. The radius of the dashed open circles is the true 
polarisation degree, and the radius of the solid open circles is the measured polarisation 
degree. The top panel shows instrumental Stokes parameters ($q = Q/I$, $u = U/I$) for 103 values of 
the sky position angle (observations between May 2011 and March 2012). All data points are 
distributed in a ring around the instrumental polarisation (cross). In the middle panel, the centre 
of the distribution is shifted to the origin (0, 0) by subtracting the instrumental polarisation (an 
elliptical distortion is also corrected). In the bottom panel, the data are de-rotated using the sky 
position angles associated with them.}
\label{fig:VICyg12ringo2qu}
\end{figure}

\begin{figure}
\centering
\includegraphics[width=9cm]{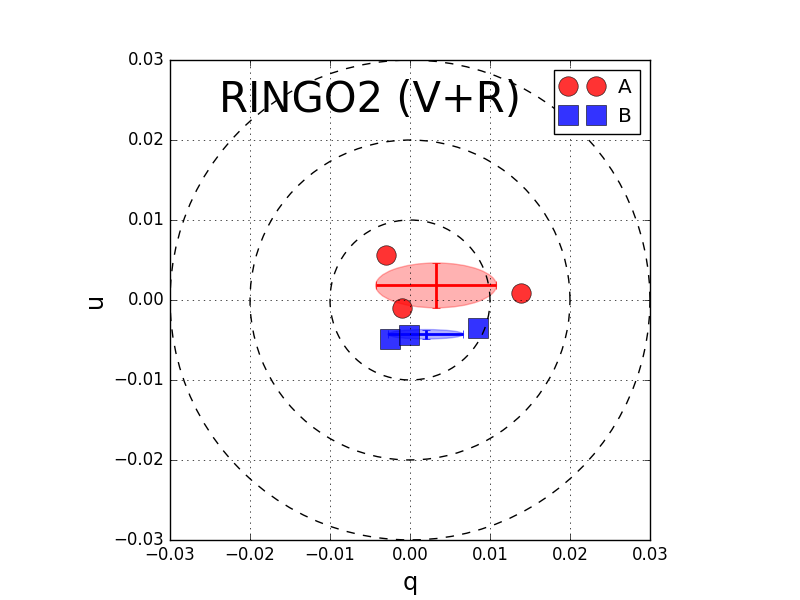}
\caption{RINGO2 data of \object{QSO B0957+561}. The Stokes parameters ($q$, $u$) of A and B at the 
three observing epochs are corrected for instrumental polarisation and depolarisation effects (see 
main text). We also display the means and root-mean-square deviations of both parameters for each 
quasar image (crosses and ellipses), as well as dashed open circles representing three different 
polarisation degrees: 1\%, 2\% and 3\%.}
\label{fig:q0957ringo2qu}
\end{figure}

There is a second main instrumental effect: depolarisation of the signal. To account for this 
additional issue, we analysed RINGO2 data of the standard polarised star VICyg12 
\citep[e.g.][]{Sch92}. Available sets of (eight) frames for different sky position angles ($ROTSKYPA$ 
values) allowed us to plot the $q-u$ diagram in the top panel of Fig.~\ref{fig:VICyg12ringo2qu}. 
After subtracting the instrumental polarisation and removing an elliptical distortion in the 
distribution of shifted ($q$, $u$) values, corrected data were spread along a ring centred at the 
origin of the $q-u$ plane (see the middle panel of Fig.~\ref{fig:VICyg12ringo2qu}). The radius of 
this ring yielded the (measured) polarisation degree $PD_{\rm{meas}}$ for the polarised star and led 
to a depolarisation factor $F = PD_{\rm{meas}}/PD_{\rm{true}}$ = 0.76 (see also the Jermak PhD 
thesis). When rotating a set of frames by an angle $\phi$, the associated polarisation will appear 
rotated through an angle $2\phi$, that is, $q_{\phi} = q \cos(2\phi) + u \sin(2\phi)$ and $u_{\phi} = -q 
\sin(2\phi) + u \cos(2\phi)$. Hence, all data were de-rotated using the known values of $ROTSKYPA$ 
($\phi = -ROTSKYPA$; see the bottom panel of Fig.~\ref{fig:VICyg12ringo2qu}). The measured 
polarisation angle ($PA_{\rm{meas}}$) did not coincide with the true one, and we found 
$PA_{\rm{true}} = PA_{\rm{meas}} + K$, where $K$ = 42\degr\ \citep[e.g.][who estimated $K$ = 41 $\pm$ 
3\degr\ in the period of interest]{Ste17}. As a last step in the reduction of RINGO2 observations of 
\object{QSO B0957+561}, we have corrected the depolarisation bias in our science data (quasar 
images). More specifically, after removing the instrumental polarisation bias, the ($q$, $u$) values 
of A and B were de-rotated through angles $2\phi = -2(ROTSKYPA + K)$, and then divided by $F$ (see 
Fig.~\ref{fig:q0957ringo2qu}). 

\begin{figure}
\centering
\includegraphics[width=9cm]{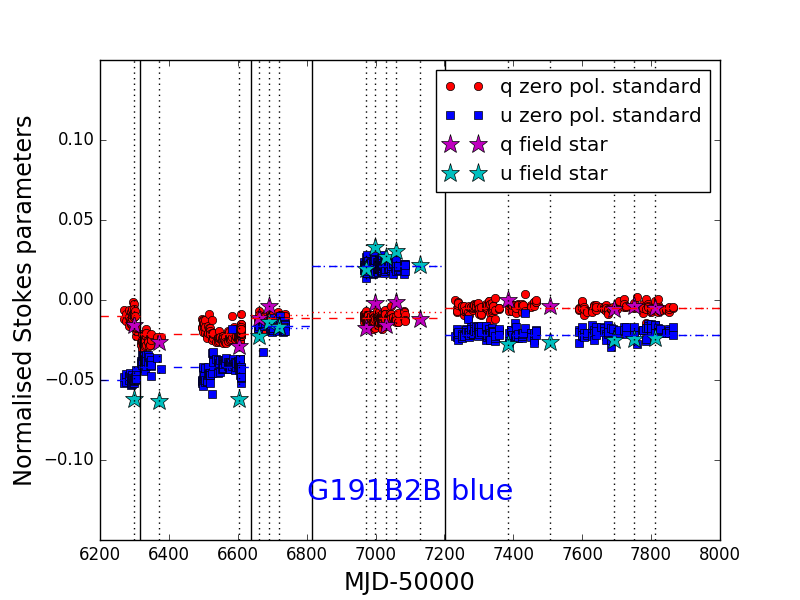}
\includegraphics[width=9cm]{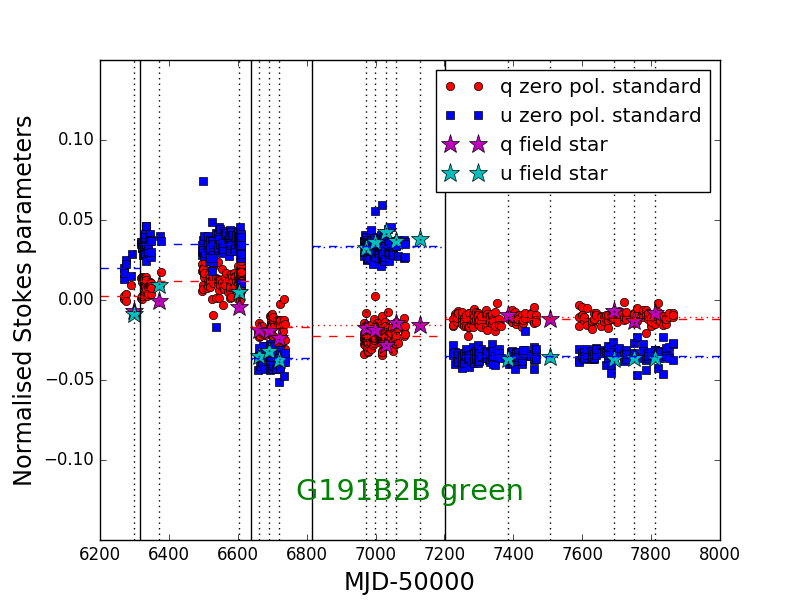}
\includegraphics[width=9cm]{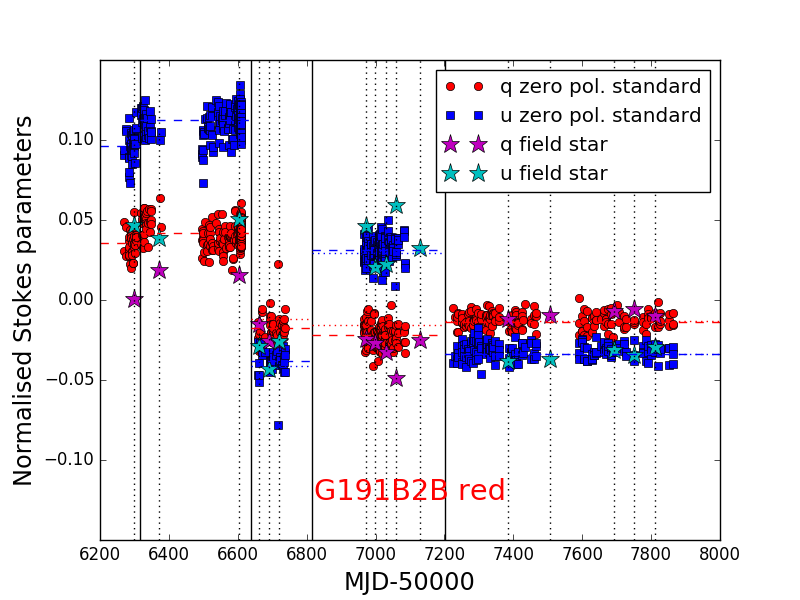}
\caption{RINGO3 data of the standard zero-polarised star G191B2B and the unpolarised field star H. 
The top, middle, and bottom panels show results in the blue, green, and red bands, respectively. We
depict the observing dates (vertical dotted lines) and when hardware updates occurred (vertical solid 
lines), as well as the Stokes parameters for G191B2B (filled circles and squares) and H (filled 
stars). The horizontal dashed lines represent average parameters of G191B2B (these would be equal to 
zero for an ideal instrument), and the horizontal dotted lines are the instrumental polarisation 
offsets reported in \citet{Jer16}.}
\label{fig:zpolringo3}
\end{figure}

The reduction procedures of RINGO3 data were similar to those used to reduce RINGO2 observations. 
However, RINGO3 is a three-band optical polarimeter, so we obtained three sets of eight stacked 
frames at 16 observing epochs (see Sec.~\ref{sec:q0957}). The three bands are labelled B (blue), G 
(green), and R (red), and they differ from standard $ugriz$ and $UBVRI$ passbands. In each optical 
band, the photometric outputs for a given object led to its instrumental Stokes parameters at all 
epochs. Unfortunately, the RINGO3 hardware was changed four times during our polarimetric follow-up 
between 2013 and 2017, which forced us to split the data into five periods and analyse the 
instrumental biases within each individual period. To discuss the instrumental polarisations, we used 
available observations of the standard zero-polarised stars G191B2B and BD+28\degr4211 
\citep[e.g.][]{Sch92}, and our data for the unpolarised field star H. As both standard stars showed a
similar behaviour, we focused on the comparison between G191B2B and H. In Fig.~\ref{fig:zpolringo3},
we illustrate the time evolution of the instrumental polarisation in each band \citep[see 
also][]{Jer16}. The vertical solid lines represent epochs in which there were hardware updates, while 
the vertical dotted lines correspond to the observing dates. In the three last periods, our estimates 
of instrumental polarisations from G191B2B data (red circles and blue squares) agree well with the 
Stokes parameters for H (magenta and cyan stars) and the instrumental polarisation offsets in the 
Jermak PhD thesis (horizontal dotted lines). However, in the two first periods (before fitting a 
depolarising Lyot prism in December 2013), there appear discrepancies between results from the 
G191B2B and H stars. The ($q_{\rm{H}}$, $u_{\rm{H}}$) values are the best tracers of the polarisation 
bias, since the H field star was observed in the same conditions as quasar images. 

\begin{figure}[!ht]
\centering
\includegraphics[width=9cm]{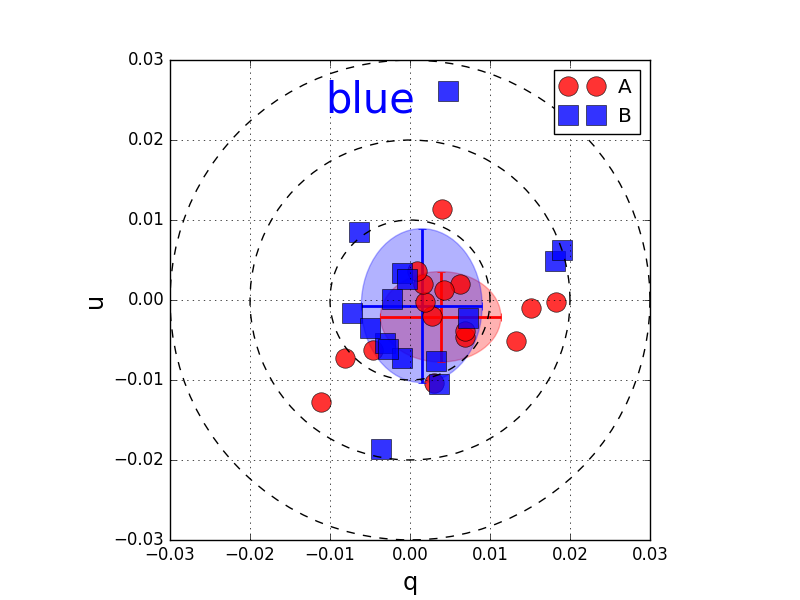}
\includegraphics[width=9cm]{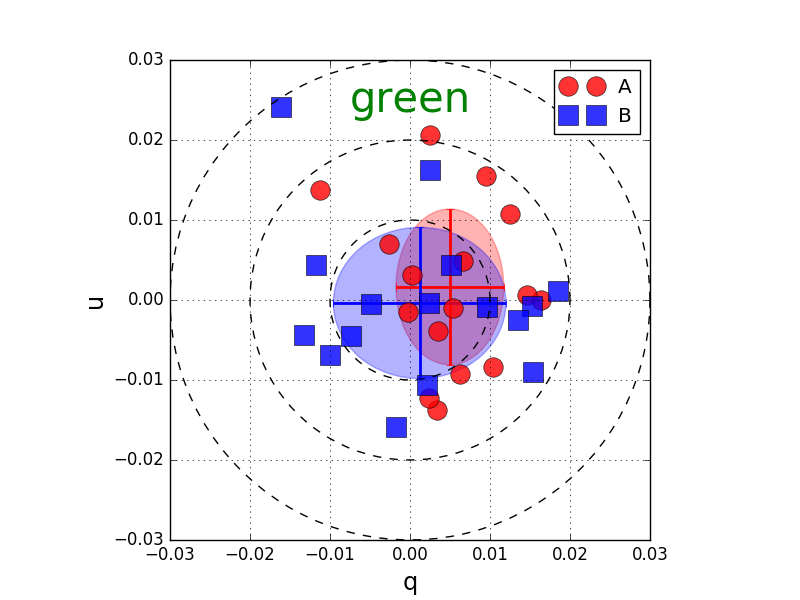}
\includegraphics[width=9cm]{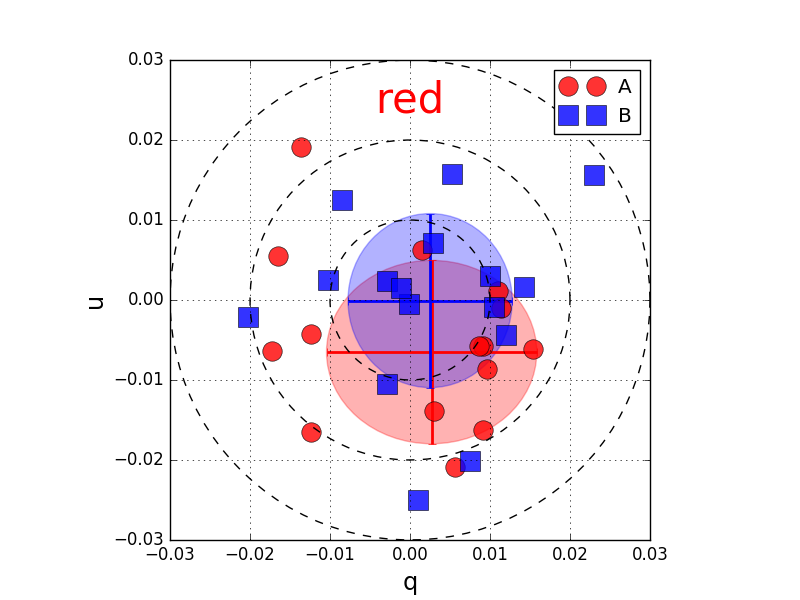}
\caption{RINGO3 data of \object{QSO B0957+561}. The Stokes parameters ($q_{\rm{A}}$, $u_{\rm{A}}$) 
and ($q_{\rm{B}}$, $u_{\rm{B}}$) are corrected for instrumental polarisation and depolarisation 
biases (see main text). In each band, we show the means and root-mean-square deviations of the Stokes 
parameters for each quasar image (crosses and ellipses), and the 1\%, 2\%, and 3\% polarisation rings 
(dashed open circles).}
\label{fig:q0957ringo3qu}
\end{figure}

In order to remove the instrumental 
polarisation bias at a given epoch, we subtracted ($q_{\rm{H}}$, $u_{\rm{H}}$) for each of the three 
bands of the RINGO3 polarimeter
in this epoch from the 
instrumental Stokes parameters of the quasar images. To correct for elliptical distortion, a 
multiplicative factor of 1.14 was also applied to the shifted values of $q_{\rm{A}}$ and 
$q_{\rm{B}}$. In addition, we studied depolarisation factors (and $K$ values) using RINGO3 data of 
standard polarised stars, and we confirmed the Jermak results for the last three periods. 
Therefore, averaging over the three bands, we took $K$ = 55, 115.5, and 125\degr\ in the third, 
fourth, and fifth time segments in Fig.~\ref{fig:zpolringo3}. Averaging over the three bands and these 
three time segments, $F$ was also taken to be equal to 0.96. \citet{Slo16} proved that the values of 
$K$ and $F$ for the two first periods are very similar to those for the third period (using a method 
different from ours; see below), and consequently, we adopted $K$ = 55\degr\ and $F$ = 0.96 in the 
two initial time segments in Fig.~\ref{fig:zpolringo3}. Our results in these initial phases of the 
instrument should be considered with caution. In fact, \citet{Jer16} suggested the instrumental 
polarisation is not constrainable during the two first periods, and she exclusively focused on the 
other periods. We note that \citet{Slo16} discussed Stokes parameters in all time segments, but their 
results are based on a different method. We (and Jermak) used the \citet{Cla02} framework, which 
assumes the same polaroid at eight different angles. However, \citet{Slo16} considered eight 
different polaroids \citep[the n-polarizer method of][]{Spa99}. Finally, the quasar polarisations 
were de-rotated through angles $2\phi = -2(ROTSKYPA + K)$ and divided by $F$ (see 
Fig.~\ref{fig:q0957ringo3qu}). 

\newpage

\subsection{Emission-line fluxes}
\label{sec:appa3}

\begin{figure}[!ht]
\centering
\includegraphics[width=9cm]{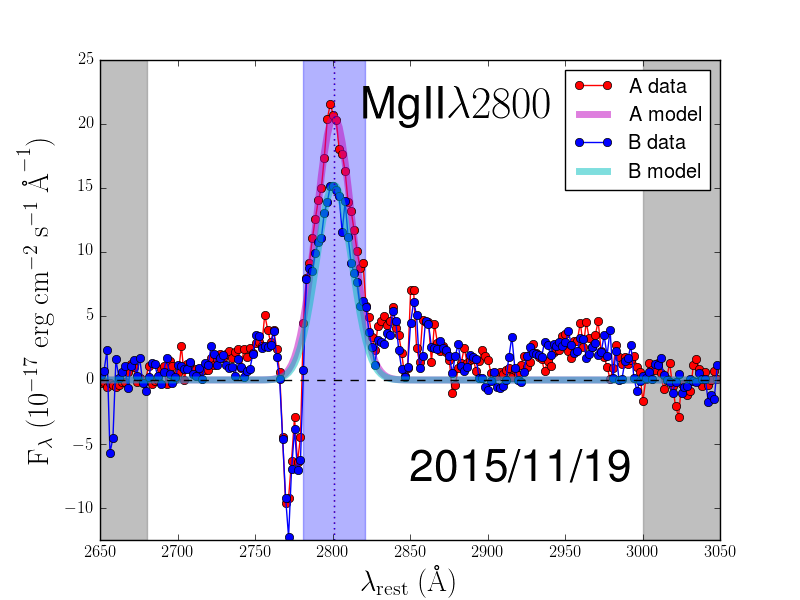}
\caption{Mg\,{\sc ii} emission line profiles on 19 November 2015. We also display Gaussian fits to 
the regions of 40 \AA\ width centred at 2800 \AA\ (blue rectangle), and highlight the continuum windows
to the left and right of the Mg\,{\sc ii} line (grey rectangles). See main text for details.}
\label{fig:q0957MgII}
\end{figure}

We obtained the profiles of the Mg\,{\sc ii} emission line in our FRODOSpec and SPRAT data (see Spectroscopy in 
Sec.~\ref{sec:q0957}) after de-redshifting the spectra to their rest frame ($z$ = 1.414) and 
converting wavelengths from air into vacuum conditions and then subtracting the 
underlying continuum. This underlying signal was determined by linear interpolation between two 
continuum regions to both sides of the line, that is, using a region to the left (2650$-$2680 \AA) and 
another region to the right (3000$-$3050 \AA). The profiles corresponding to SPRAT data on 19 
November 2015 are plotted in Fig.~\ref{fig:q0957MgII}. To avoid a strong Mg\,{\sc ii} absorption 
feature and contamination by other emissions, Mg\,{\sc ii} emission-line fluxes were estimated by 
integrating the profiles over 40 \AA\ intervals centred at 2800 \AA. Central regions with this 40 \AA\ width 
were also fitted to a Gaussian distribution (see Fig.~\ref{fig:q0957MgII}), and the Gaussian 
fits produced fluxes similar to those from direct integrations. Table~\ref{tab:q0957MgIIana} displays
details on the analysis of the Mg\,{\sc ii} emission over this decade, and its last three columns 
describe the Mg\,{\sc ii} emission-line fluxes of A and B, and the single-epoch flux ratios. 

\begin{table*}
\centering
\caption{Mg\,{\sc ii} line fluxes in 2010$-$2017.}
\begin{tabular}{cccccccc}
\hline\hline
Civil date\tablefootmark{a} & Obs. Epoch\tablefootmark{b} & $A_{\rm{cont}}$\tablefootmark{c} & 
$B_{\rm{cont}}$\tablefootmark{c} & $(B/A)_{\rm{cont}}$ & $A_{\rm{Mg\,II}}$\tablefootmark{d} &
$B_{\rm{Mg\,II}}$\tablefootmark{d} & $(B/A)_{\rm{Mg\,II}}$ \\
\hline
101105 & 5506.2 & 38.15 & 46.80 & 1.227 & 1707.8 & 1363.3 & 0.798 \\ 
101109 & 5510.2 & 39.40 & 47.45 & 1.204 & 1854.5 & 1379.7 & 0.744 \\
101223 & 5554.0 & 36.26 & 43.74 & 1.206 & 1603.1 & 1082.5 & 0.675 \\
110109 & 5571.0 & 37.27 & 46.97 & 1.260 & 1768.0 & 1577.0 & 0.892 \\
110109 & 5571.1 & 35.87 & 44.38 & 1.237 & 1839.1 & 1227.4 & 0.667 \\
110224 & 5616.9 & 36.12 & 43.29 & 1.199 & 1511.0 & 1190.1 & 0.788 \\
110301 & 5621.9 & 34.51 & 42.31 & 1.226 & 1586.5 & 1162.7 & 0.733 \\
110410 & 5661.9 & 32.00 & 39.86 & 1.246 & 1423.7 & 1109.7 & 0.779 \\
110530 & 5711.9 & 37.61 & 43.97 & 1.169 & 1666.6 & 1339.6 & 0.804 \\
111024 & 5859.2 & 33.28 & 46.27 & 1.390 & 1711.6 & 1385.6 & 0.810 \\
111101 & 5867.2 & 31.78 & 44.05 & 1.386 & 1725.8 & 1198.8 & 0.695 \\
111205 & 5901.1 & 33.60 & 43.05 & 1.282 & 1791.2 & 1487.9 & 0.831 \\
111218 & 5914.2 & 32.46 & 40.04 & 1.234 & 1598.2 & 1433.5 & 0.897 \\
111221 & 5917.2 & 32.97 & 40.13 & 1.217 & 1744.4 & 1533.2 & 0.879 \\
130220 & 6344.0 & 33.94 & 41.54 & 1.224 & 1539.9 & 1246.1 & 0.809 \\
140104 & 6662.1 & 39.90 & 43.22 & 1.083 & 1441.7 & 1177.4 & 0.817 \\
150315 & 7097.0 & 30.80 & 44.90 & 1.458 & 1414.8 &  975.6 & 0.690 \\
151119 & 7346.2 & 34.88 & 38.45 & 1.102 & 1432.7 & 1114.0 & 0.778 \\
151121 & 7348.1 & 36.04 & 39.97 & 1.109 & 1485.9 & 1088.3 & 0.732 \\
170117 & 7771.1 & 34.21 & 45.15 & 1.320 & 1564.6 & 1087.1 & 0.695 \\
170118 & 7772.1 & 33.68 & 44.33 & 1.316 & 1516.2 & 1078.3 & 0.711 \\
\hline
\end{tabular}
\tablefoot{
\tablefoottext{a}{yymmdd};
\tablefoottext{b}{MJD$-$50\,000};
\tablefoottext{c}{continuum flux at 2800 \AA\ in 10$^{-17}$ erg cm$^{-2}$ s$^{-1}$ \AA$^{-1}$};
\tablefoottext{d}{Mg\,{\sc ii} emission-line flux in 10$^{-17}$ erg cm$^{-2}$ s$^{-1}$} 
} 
\label{tab:q0957MgIIana}
\end{table*}

We also calculated C\,{\sc iii}] and C\,{\sc iv} emission-line fluxes (and single-epoch flux ratios) 
from the available SPRAT and NOT/ALFOSC spectra in Tables 9$-$11. For the C\,{\sc iii}] emission at 
1909 \AA, the underlying continua were obtained through linear interpolations between data at 
1760$-$1830 and 1980$-$2050 \AA, while for the C\,{\sc iv} emission at 1549 \AA, we used the 
1450$-$1460 and 1710$-$1730 \AA\ continuum regions. After subtracting the continua under the C\,{\sc 
iii}] and C\,{\sc iv} emission lines, the line fluxes were computed by integrating the line profiles 
over their central regions of 40 \AA\ width. The results for the C\,{\sc iii}] and C\,{\sc iv} emissions
are presented in Table~\ref{tab:q0957CIII]ana} and Table~\ref{tab:q0957CIVana}, respectively.  

\begin{table*}
\centering
\caption{C\,{\sc iii}] line fluxes in 2010$-$2017.}
\begin{tabular}{cccccccc}
\hline\hline
Civil date\tablefootmark{a} & Obs. Epoch\tablefootmark{b} & $A_{\rm{cont}}$\tablefootmark{c} & 
$B_{\rm{cont}}$\tablefootmark{c} & $(B/A)_{\rm{cont}}$ & $A_{\rm{C\,III]}}$\tablefootmark{d} &
$B_{\rm{C\,III]}}$\tablefootmark{d} & $(B/A)_{\rm{C\,III]}}$ \\
\hline
100328 & 5283.9 & 57.92 & 63.70 & 1.100 & 1992.8 & 1447.8 & 0.727 \\
111218 & 5914.2 & 52.60 & 72.08 & 1.370 & 1894.5 & 1571.5 & 0.830 \\
130314 & 6366.0 & 55.12 & 64.79 & 1.175 & 1597.7 & 1258.7 & 0.788 \\
150315 & 7097.0 & 47.39 & 86.31 & 1.821 & 1990.0 & 1576.1 & 0.792 \\ 
151119 & 7346.2 & 62.80 & 69.82 & 1.112 & 1838.5 & 1364.5 & 0.742 \\
151121 & 7348.1 & 58.31 & 63.77 & 1.094 & 1869.4 & 1590.0 & 0.851 \\
170117 & 7771.1 & 58.41 & 86.23 & 1.476 & 2031.2 & 1467.4 & 0.722 \\
170118 & 7772.1 & 59.87 & 87.73 & 1.465 & 1842.6 & 1299.1 & 0.705 \\
\hline
\end{tabular}
\tablefoot{
\tablefoottext{a}{yymmdd};
\tablefoottext{b}{MJD$-$50\,000};
\tablefoottext{c}{continuum flux at 1909 \AA\ in 10$^{-17}$ erg cm$^{-2}$ s$^{-1}$ \AA$^{-1}$};
\tablefoottext{d}{C\,{\sc iii}] emission-line flux in 10$^{-17}$ erg cm$^{-2}$ s$^{-1}$} 
} 
\label{tab:q0957CIII]ana}
\end{table*}

\begin{table*}
\centering
\caption{C\,{\sc iv} line fluxes in 2010$-$2013.}
\begin{tabular}{cccccccc}
\hline\hline
Civil date\tablefootmark{a} & Obs. Epoch\tablefootmark{b} & $A_{\rm{cont}}$\tablefootmark{c} & 
$B_{\rm{cont}}$\tablefootmark{c} & $(B/A)_{\rm{cont}}$ & $A_{\rm{C\,IV}}$\tablefootmark{d} &
$B_{\rm{C\,IV}}$\tablefootmark{d} & $(B/A)_{\rm{C\,IV}}$ \\
\hline
100328 & 5283.9 & 78.33 &  87.88 & 1.122 & 3955.8 & 3424.2 & 0.866 \\  
111218 & 5914.2 & 75.72 & 102.14 & 1.349 & 4498.0 & 4654.6 & 1.035 \\
130314 & 6366.0 & 71.81 &  84.06 & 1.171 & 4968.5 & 4168.3 & 0.839 \\
\hline
\end{tabular}
\tablefoot{
\tablefoottext{a}{yymmdd};
\tablefoottext{b}{MJD$-$50\,000};
\tablefoottext{c}{continuum flux at 1549 \AA\ in 10$^{-17}$ erg cm$^{-2}$ s$^{-1}$ \AA$^{-1}$};
\tablefoottext{d}{C\,{\sc iv} emission-line flux in 10$^{-17}$ erg cm$^{-2}$ s$^{-1}$} 
} 
\label{tab:q0957CIVana}
\end{table*}

\clearpage

\begin{minipage}[h]{\textwidth}
\section{Analysis of new spectra of SDSS J1001+5027}
\label{sec:appb}

The new FRODOSpec spectra (Table 13) contain the Mg\,{\sc ii} emission line, while the SPRAT spectra
(Table 14) include the C\,{\sc iii}] and C\,{\sc iv} emissions. Thus, the procedure used for 
analysing these data was identical with that used in Sect.~\ref{sec:appa3}. Working in the quasar rest 
frame ($z$ = 1.838), we obtained line profiles by subtracting their underlying continua. In 
Sect.~\ref{sec:appa3}, we describe the continuum regions to perform linear interpolations and estimate 
underlying signals. As a last step, the emission-line fluxes were calculated by integrating the 
profiles over their 40 \AA\ width central regions. Tables~\ref{tab:q1001MgIIana}, 
\ref{tab:q1001CIII]ana}, and \ref{tab:q1001CIVana} show continuum and emission-line fluxes of A and B, 
as well as single-epoch flux ratios.
\end{minipage}

\begin{table*}
\centering
\caption{Mg\,{\sc ii} line fluxes of \object{SDSS J1001+5027}.}
\begin{tabular}{cccccccc}
\hline\hline
Civil date\tablefootmark{a} & Obs. Epoch\tablefootmark{b} & $A_{\rm{cont}}$\tablefootmark{c} & 
$B_{\rm{cont}}$\tablefootmark{c} & $(B/A)_{\rm{cont}}$ & $A_{\rm{Mg\,II}}$\tablefootmark{d} &
$B_{\rm{Mg\,II}}$\tablefootmark{d} & $(B/A)_{\rm{Mg\,II}}$ \\
\hline
131107 & 6604.2 & 19.55 & 13.57 & 0.694 & 1200.5 &  775.3 & 0.646 \\
140208 & 6697.0 & 15.55 & 12.81 & 0.823 & 1106.2 &  983.5 & 0.889 \\
140326 & 6743.0 & 17.68 & 14.63 & 0.828 &  964.4 &  695.8 & 0.721 \\
\hline
\end{tabular}
\tablefoot{
\tablefoottext{a}{yymmdd};
\tablefoottext{b}{MJD$-$50\,000};
\tablefoottext{c}{continuum flux at 2800 \AA\ in 10$^{-17}$ erg cm$^{-2}$ s$^{-1}$ \AA$^{-1}$};
\tablefoottext{d}{Mg\,{\sc ii} emission-line flux in 10$^{-17}$ erg cm$^{-2}$ s$^{-1}$} 
} 
\label{tab:q1001MgIIana}
\end{table*}

\begin{table*}
\centering
\caption{C\,{\sc iii}] line fluxes of \object{SDSS J1001+5027}.}
\begin{tabular}{cccccccc}
\hline\hline
Civil date\tablefootmark{a} & Obs. Epoch\tablefootmark{b} & $A_{\rm{cont}}$\tablefootmark{c} & 
$B_{\rm{cont}}$\tablefootmark{c} & $(B/A)_{\rm{cont}}$ & $A_{\rm{C\,III]}}$\tablefootmark{d} &
$B_{\rm{C\,III]}}$\tablefootmark{d} & $(B/A)_{\rm{C\,III]}}$ \\
\hline
150226 & 7080.0 & 26.90 & 16.37 & 0.608 & 1042.1 & 742.7 & 0.713 \\
151202 & 7359.2 & 26.60 & 15.99 & 0.601 &  944.5 & 709.9 & 0.752 \\
160405 & 7483.9 & 26.90 & 16.90 & 0.628 &  950.5 & 617.9 & 0.650 \\
161206 & 7729.2 & 27.88 & 17.69 & 0.635 &  990.9 & 758.2 & 0.765 \\
\hline
\end{tabular}
\tablefoot{
\tablefoottext{a}{yymmdd};
\tablefoottext{b}{MJD$-$50\,000};
\tablefoottext{c}{continuum flux at 1909 \AA\ in 10$^{-17}$ erg cm$^{-2}$ s$^{-1}$ \AA$^{-1}$};
\tablefoottext{d}{C\,{\sc iii}] emission-line flux in 10$^{-17}$ erg cm$^{-2}$ s$^{-1}$} 
} 
\label{tab:q1001CIII]ana}
\end{table*}

\begin{table*}
\centering
\caption{C\,{\sc iv} line fluxes of \object{SDSS J1001+5027}.}
\begin{tabular}{cccccccc}
\hline\hline
Civil date\tablefootmark{a} & Obs. Epoch\tablefootmark{b} & $A_{\rm{cont}}$\tablefootmark{c} & 
$B_{\rm{cont}}$\tablefootmark{c} & $(B/A)_{\rm{cont}}$ & $A_{\rm{C\,IV}}$\tablefootmark{d} &
$B_{\rm{C\,IV}}$\tablefootmark{d} & $(B/A)_{\rm{C\,IV}}$ \\
\hline
150226 & 7080.0 & 33.78 & 16.57 & 0.491 & 1273.3 & 1102.6 & 0.866 \\
151202 & 7359.2 & 32.11 & 17.37 & 0.541 & 1566.6 & 1000.1 & 0.638 \\
160405 & 7483.9 & 35.31 & 18.30 & 0.518 & 1619.3 & 1304.8 & 0.806 \\
161206 & 7729.2 & 41.07 & 21.35 & 0.520 & 1869.8 & 1542.6 & 0.825 \\
\hline
\end{tabular}
\tablefoot{
\tablefoottext{a}{yymmdd};
\tablefoottext{b}{MJD$-$50\,000};
\tablefoottext{c}{continuum flux at 1549 \AA\ in 10$^{-17}$ erg cm$^{-2}$ s$^{-1}$ \AA$^{-1}$};
\tablefoottext{d}{C\,{\sc iv} emission-line flux in 10$^{-17}$ erg cm$^{-2}$ s$^{-1}$} 
} 
\label{tab:q1001CIVana}
\end{table*}

\end{appendix}

\end{document}